\documentclass[12pt]{article}

\usepackage{epsf,amsfonts,hyperref}
\newcommand\encadremath[1]{\vbox{\hrule\hbox{\vrule\kern8pt 
\vbox{\kern8pt \hbox{$\displaystyle #1$}\kern8pt} 
\kern8pt\vrule}\hrule}}
\def\enca#1{\vbox{\hrule\hbox{
\vrule\kern8pt\vbox{\kern8pt \hbox{$\displaystyle #1$}
\kern8pt} \kern8pt\vrule}\hrule}}

\newcommand\figureframex[3]{
\begin{figure}[bth]
\hrule\hbox{\vrule\kern8pt 
\vbox{\kern8pt \vbox{
\begin{center}
{\mbox{\epsfxsize=#1.truecm\epsfbox{#2}}}
\end{center}
\caption{#3}
}\kern8pt} 
\kern8pt\vrule}\hrule
\end{figure}
}
\newcommand\figureframey[3]{
\begin{figure}[bth]
\hrule\hbox{\vrule\kern8pt 
\vbox{\kern8pt \vbox{
\begin{center}
{\mbox{\epsfysize=#1.truecm\epsfbox{#2}}}
\end{center}
\caption{#3}
}\kern8pt} 
\kern8pt\vrule}\hrule
\end{figure}
}

\makeatletter
\@addtoreset{equation}{section}
\makeatother
\newtheorem{theorem}{Theorem}[section]
\newtheorem{conjecture}{Conjecture}[section]
\newtheorem{remark}{Remark}[section]
\newtheorem{proposition}{Proposition}[section]
\newtheorem{lemma}{Lemma}[section]
\newtheorem{corollary}{Corollary}[section]
\newtheorem{definition}{Definition}[section]
\def\br{\begin{remark}\rm\small}
\def\er{\end{remark}}
\def\bt{\begin{theorem}}
\def\et{\end{theorem}}
\def\bd{\begin{definition}}
\def\ed{\end{definition}}
\def\bp{\begin{proposition}}
\def\ep{\end{proposition}}
\def\bl{\begin{lemma}}
\def\el{\end{lemma}}
\def\bc{\begin{corollary}}
\def\ec{\end{corollary}}
\def\beaq{\begin{eqnarray}}
\def\eeaq{\end{eqnarray}}
\newcommand{\proof}[1]{{\noindent \bf proof:}\par
{#1} $\square$}

\newcommand{\eq}[1]{eq.~(\ref{#1})}

\newcommand{\beq}{\begin{equation}}
\newcommand{\eeq}{\end{equation}}
\newcommand{\bea}{\begin{eqnarray}}
\newcommand{\eea}{\end{eqnarray}}

\renewcommand{\and}{{\qquad {\rm and} \qquad}}

\newcommand{\virg}{{\qquad , \qquad}}

 \newcommand{\Tr}{{\,\rm Tr}\:}
\newcommand{\tr}{{\,\rm tr}\:}

\newcommand{\Res}{\mathop{\,\rm Res\,}}

\newcommand{\td}[1]{{\tilde{#1}}}

\renewcommand{\l}{\lambda}
\newcommand{\om}{\omega}

\newcommand{\ee}[1]{{{\rm e}^{#1}}}

\renewcommand{\d}{{{\partial}}}

\newcommand{\Pint}{{\int\kern -1.em -\kern-.25em}}

\renewcommand{\l}{\lambda}

\newcommand{\ovl}{\overline}

\newcommand{\curve}{{\cal L}}

\newcommand{\tmin}{{t_{\rm min}}}
\newcommand{\tmax}{{t_{\rm max}}}
\newcommand{\domain}{{\cal D}}
\newcommand{\defect}{{\ovl{\cal D}}}

\newcommand{\qq}{{\mathfrak{q}}}

\begin{document}
\sloppy


\pagestyle{empty}
\hfill SPT-09/050
\addtolength{\baselineskip}{0.20\baselineskip}
\begin{center}
\vspace{26pt}
{\large \bf {A matrix model for plane partitions}}
\newline
\vspace{26pt}

{\sl B.\ Eynard}\hspace*{0.05cm}\footnote{ E-mail: bertrand.eynard@cea.fr }\\
\vspace{6pt}
Institut de Physique Th\'{e}orique de Saclay,\\
F-91191 Gif-sur-Yvette Cedex, France.\\
\end{center}

\vspace{20pt}
\begin{center}
{\bf Abstract}
 
We construct a matrix model equivalent (exactly, not asymptotically), to the random plane partition model, with almost arbitrary boundary conditions.
Equivalently, it is also a random matrix model for a TASEP-like process with arbitrary boundary conditions.
Using the known solution of matrix models, this method allows to find the large size asymptotic expansion of plane partitions, to ALL orders. It also allows to describe several universal regimes.
On the algebraic geometry point of view, this gives the Gromov-Witten invariants of $\mathbb C^3$ with branes, i.e. the topological vertex, in terms of the symplectic invariants of the mirror's spectral curve.

\end{center}
%





\vspace{26pt}
\pagestyle{plain}
\setcounter{page}{1}


\section{Introduction}

The statistical physics problem of counting plane partition configurations of some domain, as well as its various equivalent formulations, has become a very active and fascinating area of mathematical physics in the past years, culminating with Okounkov's renowned work. Beyond a beautiful combinatorics problem, it has also many indirect applications, like a tiling problem similar to a discrete version of TASEP, i.e. the simplest model of out of equilibrium statistical physics, and algebraic geometry, as it plays a key role in the computation of Gromov-Witten invariants of some toric Calabi-Yau 3-folds, through the topological vertex method \cite{topvertex}.

The works of Okounkov, Kenyon and Sheffield \cite{KOS}, have brought immense progress, in the understanding of  large size asymptotics behaviors of plane partitions.
It was observed, that in many universal regimes, the statistical properties of large plane partitions, is very similar to that of matrix models, and many works have taken advantage of that similarity.

Here, in this article, we show that there is not only a "similarity" between plane partitions countings and matrix models, in fact we show that plane partitions IS a matrix model, even for finite size.
As a consequence, we may use all the machinery developped for  solving matrix models, and we are able to compute all orders corrections to the large size asymptotics.

Our matrix model, is a multimatrix model, with non-polynomial potentials. It may look very complicated at first sight, and its spectral curve may look rather complicated too. 
However, the solution of matrix models is expressed in terms of symplectic invariants, and up to a symplectic transformation (which does not change the symplectic invariants), our complicated matrix model's spectral curve, is equivalent to the Harnack curve of Kenyon-Okounkov-Sheffield \cite{KOS}.

Moreover, our formulation allows to use the full toolbox of matrix models technology. For instance the method of orthogonal polynomials gives determinantal formulae for correlation functions, the integrable structure, Riemann-Hilbert problem, and much more. And the loop equations method allows to compute the large size expansion order by order \cite{MehtaBook, Mehtamultimat}.

\subsection{Main results of this article}

Our main result is the theorem \ref{thTASEPMM}:

\smallskip
{\noindent \bf Theorem \ref{thTASEPMM}:}
{\em
The tiling model-plane partitions-tilings generating function, can be written as a matrix integral.
(a more precise result is written in theorem \ref{thTASEPMM}).
This identification is exact, it is not asymptotic.
}
\medskip

An immediate consequence is obviously:

\smallskip
{\noindent \bf Corollary:}
{\em
All asymptotic limits of tiling model-plane partitions-tilings, are random matrix limit laws.
}
\medskip

(but it remains to classify all possible random matrix limit laws).

\medskip

Also, since our matrix model is a chain of matrices, classical results of matrix models apply:

\smallskip
{\noindent \bf Corollary:}
{\em
The matrix model is integrable, the generating function is a Tau-function, and for instance correlation functions are given by determinantal formulae of Janossi densities type \cite{eynMehta}.
}
\medskip

We would like also to emphasize that our "Tiling $\leftrightarrow$ matrix model" identification works for very general cases, with almost any possible boundary conditions, we can also give weights to points of the domain to be tiled, some points can be forbidden (defects), or obliged, or just have an arbitrary weight.

\bigskip

The second part of our paper, starting at section \ref{secMMsolution}, consists in "solving" the matrix model.
We do it explicitly only for some not too complicated boundary conditions.
We recover the Harnack curves of Kenyon-Okounkov-Sheffield.

We discuss many examples in sections \ref{secexs} and after,  and in particular we apply our method to the enumeration of TSSCPP's in section \ref{secTSSCPP}.

\tableofcontents

\section{Some statistical physics models}

\subsection{Plane partitions}

Consider 3 integers $N_\l,N_\mu,N_\nu$, and 3 partitions $\l,\mu,\nu$, for example:
\beq
\l={{\mbox{\epsfxsize=0.9truecm\epsfbox{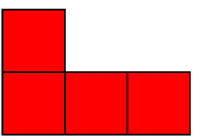}}}}
\virg
\mu={{\mbox{\epsfxsize=0.6truecm\epsfbox{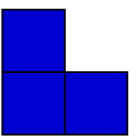}}}}
\virg
\nu={{\mbox{\epsfxsize=0.6truecm\epsfbox{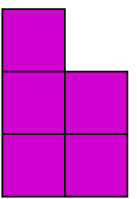}}}}
\eeq
A plane partition $\pi$ with boundaries $\l,\mu,\nu$, and of size $N_\l,N_\mu,N_\nu$, is a piling of cubic boxes in the corner of a room, with boundary conditions given by $\l,\mu,\nu$, for example:
$${\mbox{\epsfxsize=11truecm\epsfbox{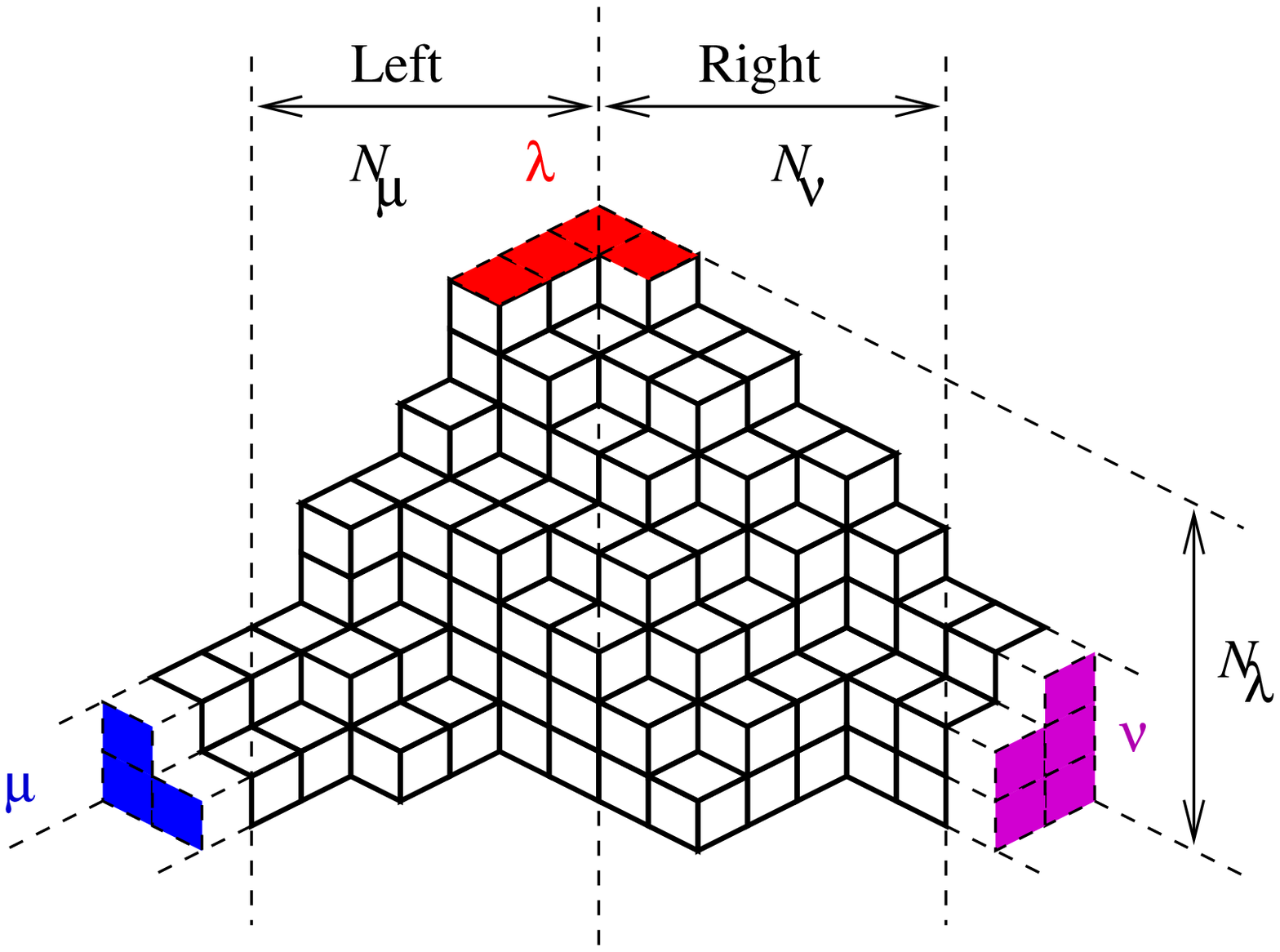}}}$$
$N_\l$ is the height of the plane-partition (height of the cubic boxes piling), $N_\mu$ (resp. $N_\nu$) is the extension towards left (resp. right), so that beyond $N_\mu$ (resp. $N_\nu$), the section is frozen to $\mu$ (resp. $\nu$).

The partition function we would like to compute is:
\beq
Z_{N_\l,N_\mu,N_\nu}(q;\l,\mu,\nu) = \sum_{\pi,\, \partial \pi = (\l,\mu,\nu)}\,\, q^{|\pi|}
\eeq
where $|\pi|$ is the number of boxes, i.e. the volume, it is called the "weight" of $\pi$.

This partition function is the so-called "topological vertex" in topological string theories \cite{topvertex}, it is the building block to compute Gromov-Witten invariants of all toric Calabi-Yau 3-folds \cite{Behrend, bryan, topvertex, mmhouches, nv, nekrasov2, Okounkov2bis, Okounkov3, Takasaki, MOO}.

From the combinatorics point of view, it is the generating function for counting plane partitions with given boundaries and weighted by their volume. 
From the statistical physics point of view, it can be viewed as a model for a growing 3-dimensional crystal in the corner of a room.
All those topics have remained important research areas in physics and mathematics, and it would be difficult to summarize all what has been done.
Let us mention that Kasteleyn \cite{kasteleyn,kasteleyn1} found an explicit expression for the partition function of a domino-tiling, which can be rephrased as plane partition, and since then, the subject has been studied a lot, see for example \cite{Cohn, Kenyon, DFZJZ}.

\subsubsection{Remark: semi-standard tableaux}

If we slice our plane partition $\pi$ at all integer times (time = horizontal coordinate) $t=-N_\mu,\dots,N_\nu$, at each time the slice is a 2-dimensional partition $\l(t)$.

\figureframex{11}{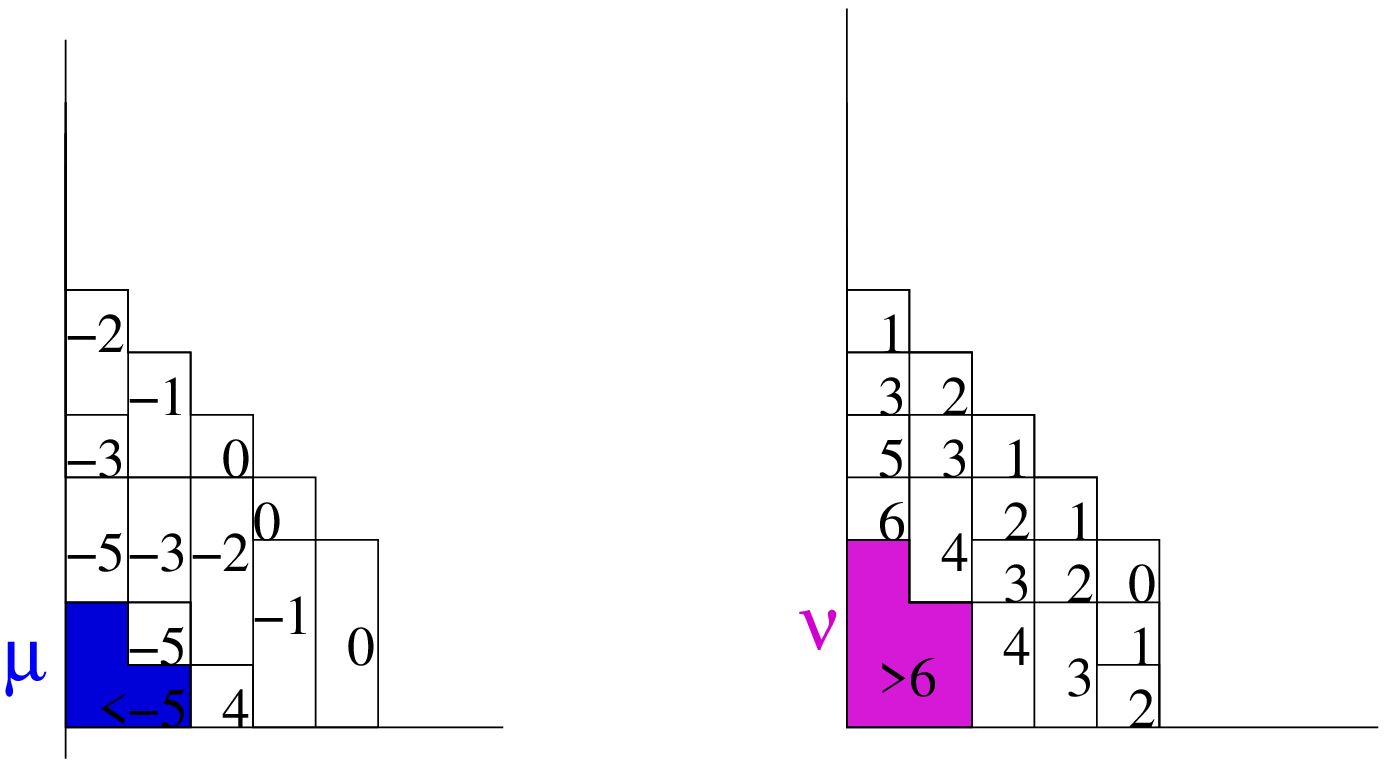}{A plane partition is equivalent to the data of two semi-standard Young tableaux of the same shape $\l(0)$. Each of these two tableaux, is the superposition of growing (or decreasing) partitions $\l(t)$.
\label{figparttime}}

It is then clear that partitions $\l(t)$ are growing from $t=-N_\mu$ to $0$, and then decreasing from $t=0$ to $N_\nu$:
\beq
\forall t<0\, , \quad \l(t)\prec \l(t+1)
\virg
\forall t>0\, , \quad \l(t)\prec \l(t-1)
\eeq
where $\prec$ is the partial ordering of partitions ($\l\prec\mu$ means that $\l$ can be obtained from $\mu$ by removing boxes).
Since we have $\mu\prec \l(-N_\mu)\prec \dots \l(t-1)\prec \l(t) \prec\dots \prec \l(1)\prec \l(0)$, we may draw all $\l(t)$ with $t\leq 0$ inside the Ferrer diagram of $\l(0)$, and we write in each box, the time $t$ at which the box appears for the first time. We do the same for $t\geq 0$, and we have two semi-standard tableaux with the same shape $\l(0)$. A semi-standard tableau is a Ferrer diagram with integer entries decreasing along columns, and strictly decreasing along rows. See fig.\ref{figparttime}.

When $\l=\mu=\nu=\emptyset$, and $N_\l=N_\mu=N_\nu=\infty$, the statistics of the partition $\l(0)$ is the sum over all pairs of semi-standard tableaux of shape $\l(0)$, i.e. it is the Plancherel measure \cite{Robinson, ssuites1, vershik}:
\beq
{\cal P}(\l) = \left({{\rm dim}(\l)\over |\l|!}\right)^2 = {1\over (|\l|!)^2} \,\, (\# ({\rm semi-standard\,tableaux\,of\,shape\,}\l))^2
\eeq

We shall study this limit in section \ref{secplancherel}.

\subsection{Jumping non-intersecting particles}

T.A.S.E.P. means "totally asymmetric exclusion process" \cite{liggett}, it is the simplest model of statistical physics out of equilibrium, it has focused considerable amounts of works \cite{derrida,derrida2, spohn, ferrari1, ferrari2, Mallick}, and it is still intensively studied.
It is a model of self avoiding particles which can either stay at their place, or jump 1 step forward, provided that the next space is unoccupied.
In the dynamics we shall be considering here, time is discrete, and at each unit of time, several particles can jump.

\medskip

It is well known that plane partitions can be rephrased in terms of self-avoiding jumping particles model which is a kind of discrete T.A.S.E.P. \cite{Robinson, ssuites1, vershik, Planchrel1, johansson, BDJ, baik, Aldous}, let us re-explain it here.

\bigskip

Let us draw in red the $N_\l$ non-intersecting lines going through tiles 
{{\mbox{\epsfxsize=0.4truecm\epsfbox{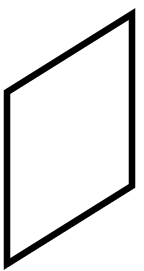}}}} and {{\mbox{\epsfxsize=0.4truecm\epsfbox{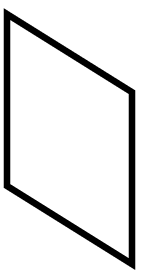}}}}:
\figureframex{11}{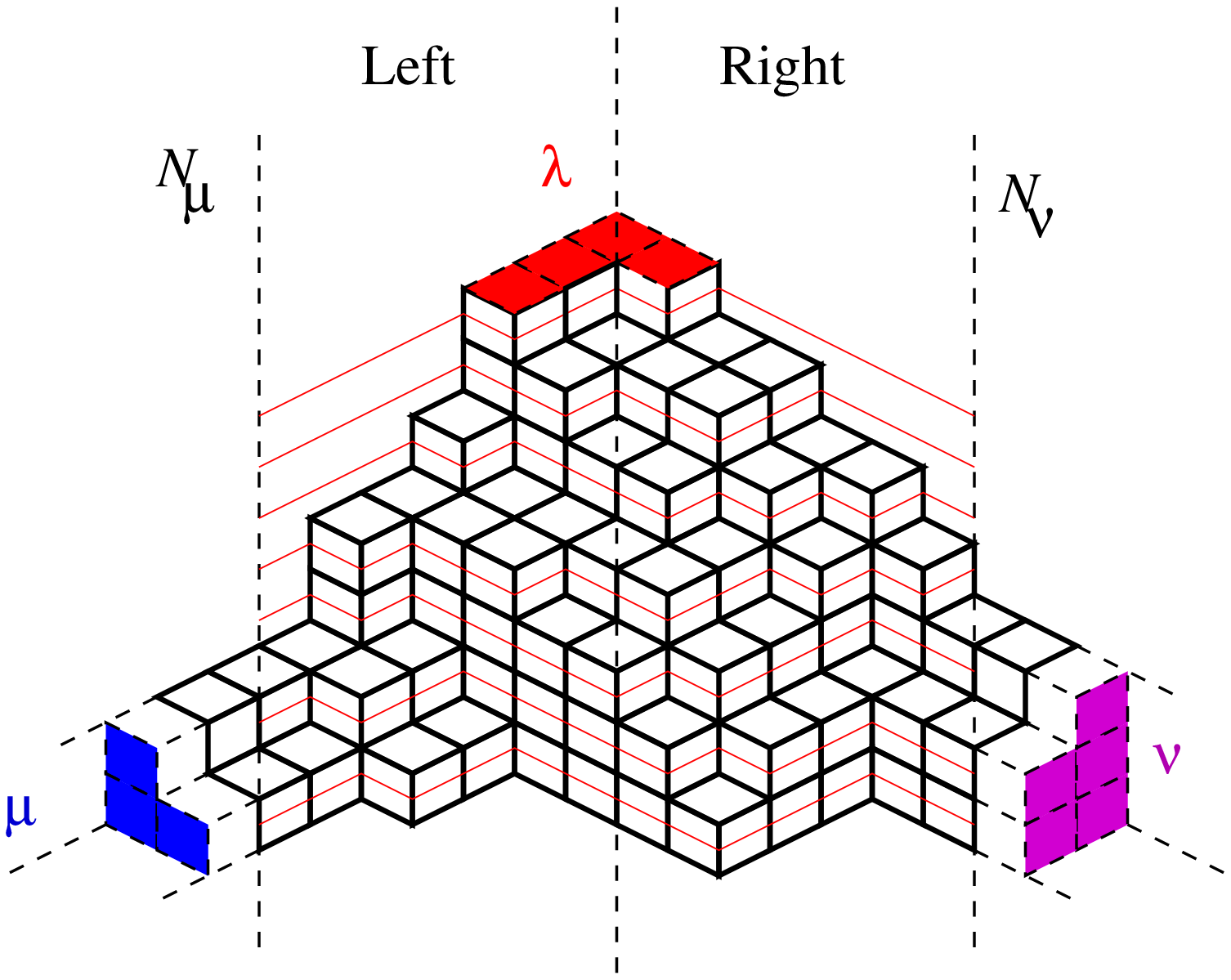}{
The level lines, go only through upright or downright tiles, they form $N_\l$ non-intersecting lines whose slopes are piecewise $\pm {1\over 2}$. 
\label{figtilingh}
}
The tiles {{\mbox{\epsfxsize=0.4truecm\epsfbox{rhombusuprgt.eps}}}} correspond to upright lines with slope $+1/2$,
and the tiles {{\mbox{\epsfxsize=0.4truecm\epsfbox{rhombusuplft.eps}}}} correspond to downright lines with slope $-1/2$.
The intersection of those red lines with integer time lines $t=-N_\mu,\dots,N_\nu$ are interpreted as positions of some particles $h_i(t)$ ($h=0$ is at the top). See figure \ref{figtilingh}.
\figureframex{12}{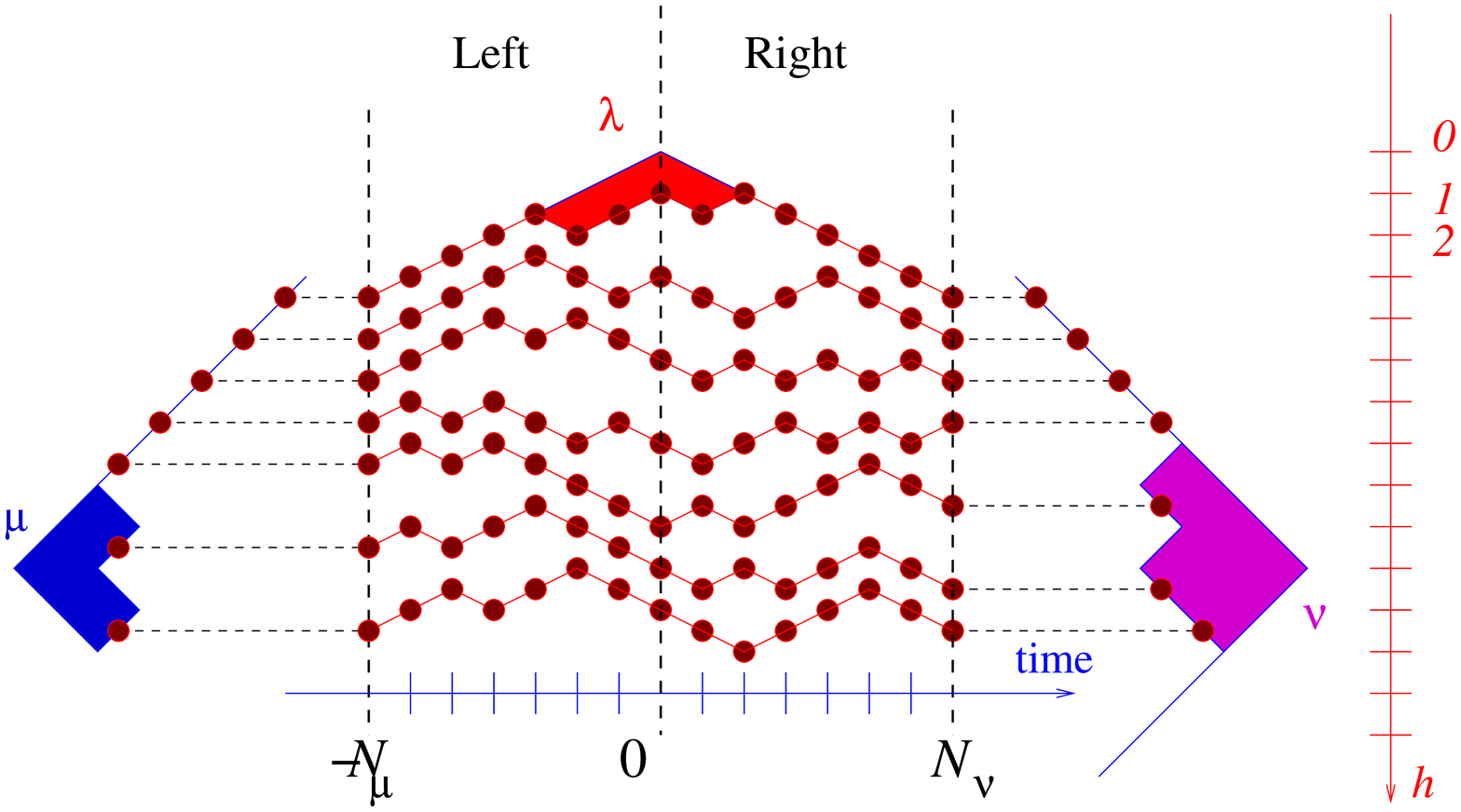}{\label{tilingh2}}

Therefore, a plane partition $\pi$, can also be described as the data of $1+N_\mu+N_\nu$ sets of $N_\l$ variables:
\beq
h_{i}(t) 
\virg t\in\{-N_\mu,\dots,N_\nu\}
\virg i=1,\dots,N_\l
\eeq
such that for every integer $t\in\{-N_\mu,\dots,N_\nu\}$, the $h_i(t)-{1\over 2}|t|$ are non-negative ordered integers:
\beq
h_i(t)-{1\over 2}|t| \in {\mathbb N}
\virg
h_1(t)>h_2(t)>h_3(t)>\dots>h_{N_\l}(t)\geq {1\over 2}|t|+r_t(\l)
\eeq
where $r_t(\l)$ is the profile function of the partition $\l$.

At time $t=-N_\mu$ and time $t=N_\nu$ we have:
\beq
h_i(-N_\mu)=\mu_i-i+N_\l+{1\over 2}N_\mu
\virg
h_i(N_\nu)=\nu_i-i+N_\l+{1\over 2}N_\nu
\eeq
and at each time $t$, we have a partition $\l(t)=(\l_1(t)\geq\l_2(t)\geq\dots\geq \l_{N_\l}(t))$:
\beq
\l_i(t)  = h_i(t)+i-N_\l-{1\over 2}|t|
\eeq
we thus have $\l(-N_\mu)=\mu$ and $\l(+N_\nu)=\nu$.

Moreover we have:
\beq
\forall t \virg h_i(t)-h_i(t+1)=\pm 1/2
\eeq

This is what we call here a discrete T.A.S.E.P. process:

$\bullet$ there are $N$ particles at positons $h_i(t)$, and at each unit of time, they jump by $\pm {1\over 2}$, and they can never occupy the same position
(in the usual formulation of TASEP, particles jump by 0 or 1, and here we have tilted the picture so that they jump by $\pm 1/2$, which is clearly the same thing up to $h_i(t)\to h_i(t)+t/2$).

\subsubsection{Summary plane partition and jumping particles}

A plane partition $\pi$ with boundaries $\l,\mu,\nu$, and of size $N_\l,N_\mu,N_\nu$, is equivalent to $h_i(t)$, $i=1,\dots,N_\l$, $t=-N_\mu,\dots,N_\nu$, such that:
\bea
\bullet\quad && h_i(t)-h_i(t+1)=\pm 1/2  \cr
\bullet\quad && h_1(t)>h_2(t)>h_3(t)>\dots>h_{N_\l}(t)\geq {1\over 2}|t|+r_t(\l)  \cr
\bullet\quad && h_i(-N_\mu)=\mu_i-i+N_\l+{1\over 2}N_\mu  \cr
\bullet\quad && h_i(N_\nu)=\nu_i-i+N_\l+{1\over 2}N_\nu  
\eea

The total numer of boxes in the partition is:
\beq
|\pi| = \sum_{i,t} h_i(t) \quad - {N_\l\over 2}\,(N_\l+N_\mu+N_\nu)
\eeq

\subsubsection{Height function and density}
\label{secdensity}

There is a relationship between the height function of the pile of cubes, and the density of $h_i$'s.
Define the density at time $t$ as the Dirac-comb distribution:
\beq
\rho(h,t) = {1\over N_\l}\,\sum_{i=1}^{N_\l}\, \delta(h-h_i(t)).
\eeq

The profile of the partition at time $t$ is recovered from the integral of $\rho$ as follows.
Define the integrated density:
\beq
I(h,t) = N_\l + N_\l \int_h^{{1\over 2}|t|}\,\, \rho(h',t)\, dh',
\eeq
$I(h,t)$ computes the index $i$ such that $h(x,t)=h_i$.
Then, define the function
\beq
\l(h,t) = h-{1\over 2}|t|+I(h,t)-N_\l.
\eeq
The plot of $\l(h,t)$ against $I(h,t)$ is the shape of the partition is $\l(t)$ at time $t$. See fig.\ref{figprofile}.

\figureframex{12}{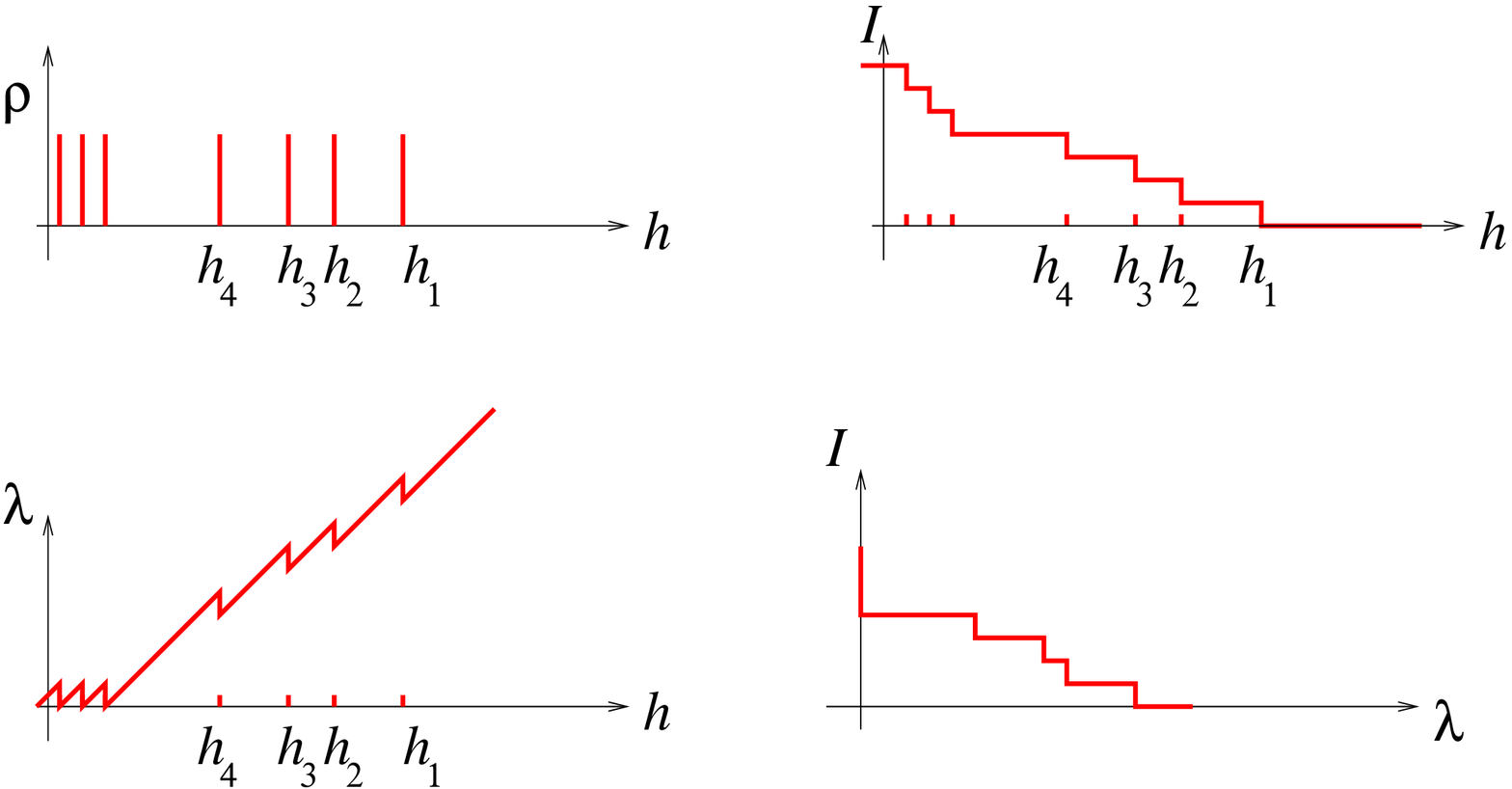}{The density $\rho(h,t)={1\over N_\l}\,\sum_{i=1}^{N_\l}\, \delta(h-h_i(t))$ encodes the profile of the partition $\l(t)$.\label{figprofile}}

The surface of the pile of cubes in ${\mathbb R}^3$ is recovered from plotting the partition $\l(t)$ at all times, i.e. it is given by:
\beq
\left\{\begin{array}{l}
x_1 = \l+{1\over 2}(|t| - t) =I-N_\l+h+{t\over 2} \cr
x_2 = \l+{1\over 2}(|t|+t) =I-N_\l+h-{t\over 2} \cr
x_3 = I 
\end{array}\right.
\eeq


\subsection{Lozenge tilings}

Another representation of plane partitions and the self-avoiding particle model, is with lozenge tilings of the rhombus lattice.

The rhombus lattice is a tiling of the plane, with lozenges  ${{\mbox{\epsfxsize=0.7truecm\epsfbox{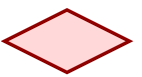}}}}$, whose centers are at the positions $(h,t)$ such that $h-{t\over 2}\in \mathbb Z$ and $t\in\mathbb Z$, see fig \ref{figrhombustiling}.
\figureframex{10}{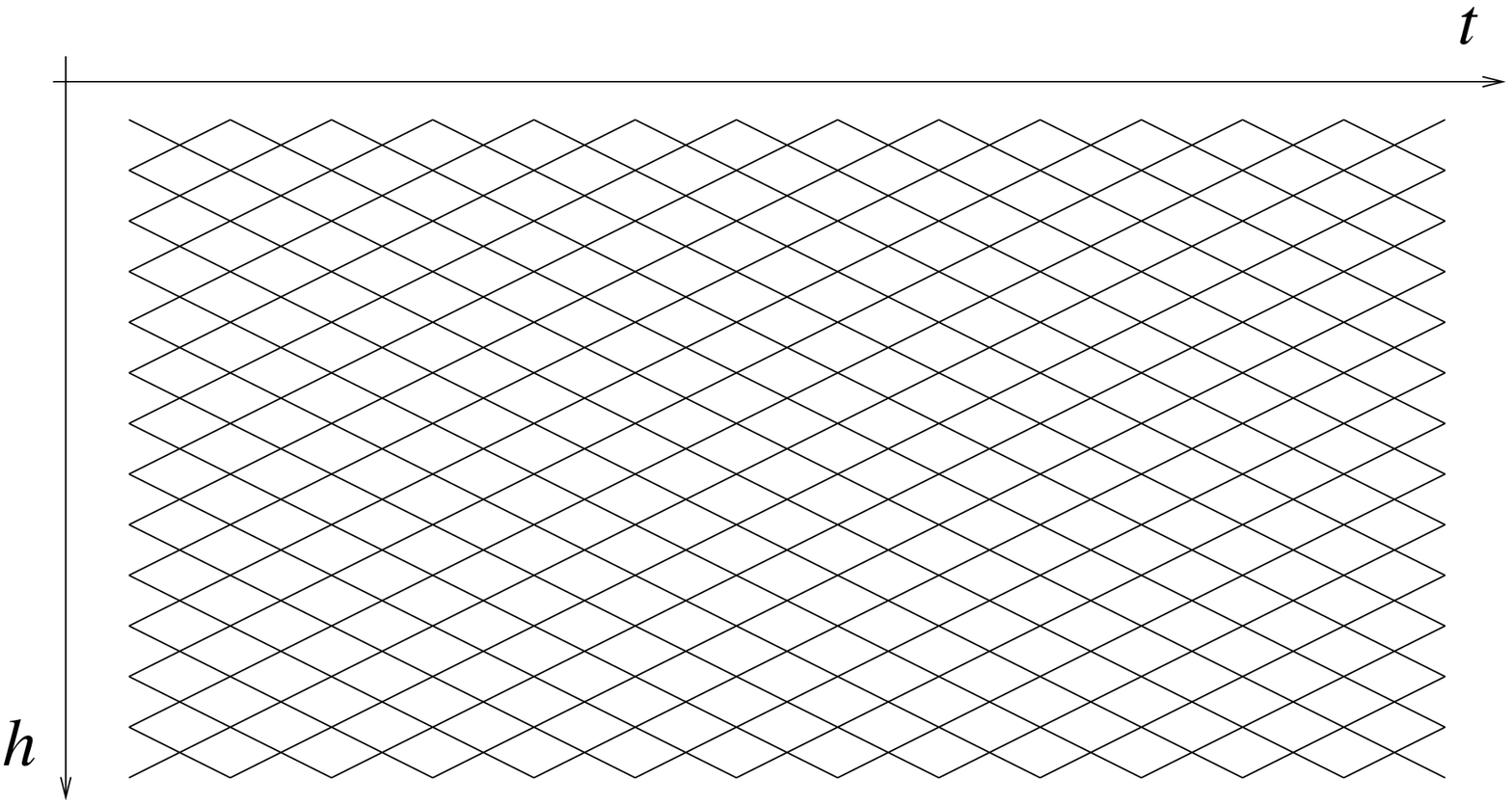}{The rhombus lattice is a tiling of the plane, with lozenges whose centers are at the positions $(h,t)$ such that $h-{t\over 2}\in \mathbb Z$ and $t\in\mathbb Z$. Notice that we orient $h$ from top to bottom.\label{figrhombustiling}}

\figureframex{14}{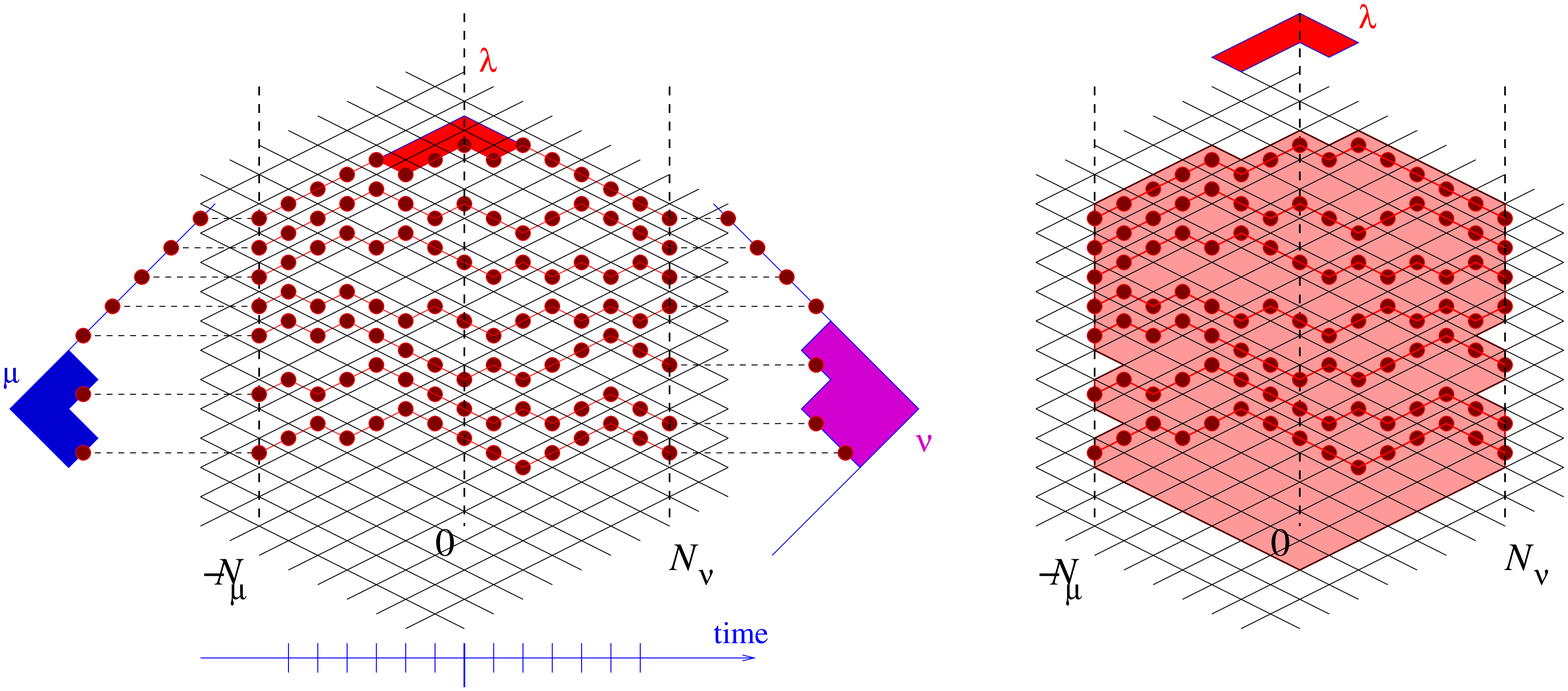}{
A plane partition configuration, can be represented as $N$  self-avoiding walks in some domain of the rhombus tiling of the plane. The white lozenges are forbidden, they are related to the 3 boundary partitions $\l,\mu,\nu$.
\label{tilinhigdomain}}

A plane partition or a self-avoiding particle model configuration can be represented as $N=N_\l$  oriented self-avoiding walks in some domain $\domain$ of the rhombus tiling of the plane. See figure \ref{tilinhigdomain}.
At a given time, each walker moves to the right $t\to t+1$, and either up $h\to h-{1\over 2}$ or down $h\to h+{1\over 2}$.

The  centers of occupied lozenges are at positions $(h_i(t),t)$, $i=1,\dots,N$.

\smallskip

The domain $\domain$ is the domain which contains all possible paths.
$\domain$ is a subdomain of the hexagon defined by the following 6 inequalities:
\bea
\tmin \leq &t& \leq \tmax \cr
h_{N}(\tmin) - {t-\tmin \over 2}\leq & h&  \leq h_1(\tmin) + {t-\tmin \over 2} \cr
h_{N}(\tmax) - {\tmax-t \over 2}\leq &h& \leq h_1(\tmax) + {\tmax-t \over 2} \cr
\eea
Indeed, since at each time, and particularly at $t=\tmin$ and $t=\tmax$, we have $h_N\leq h_i\leq h_1$, no path starting at some $h_i(\tmin)$ at $t=\tmin$ and ending at some $h_j(\tmax)$ at $t=\tmax$ can go out of this hexagon.

\smallskip
Moreover, the domain $\domain$ can be chosen such that some positions are forbidden, in fact we allow $\domain$ to be any arbitrary subset of the maximal hexagon.

\smallskip
For example for plane partitions, the positions corresponding to the boundaries $\l,\mu,\nu$ are forbidden, i.e. $\domain$ has holes corresponding to the 3 partitions $\l,\mu,\nu$ at the boundaries:

$\bullet$ At $t=\tmin=-N_\mu$, we remove from $\domain$, all lozenges which are not of the form $h_i(\mu)$ for some $i$.

$\bullet$ At $t=\tmax=N_\nu$, we remove from $\domain$, all lozenges which are not of the form $h_i(\nu)$ for some $i$.

$\bullet$ At every time $t$ we impose $h\geq {|t|\over 2}+ r_t(\l)$.

See fig \ref{tilinhigdomain}.

\subsubsection{Defects}

It is interresting to generalize our model to include walks on more general domains, not only limitted by 3 boundary partitions.
In particular, we may allow defects and holes at almost any place.

\bigskip

Consider a connected compact domain $\domain$ in the rhombus lattice, but not necessarily simply connected.
\figureframex{9}{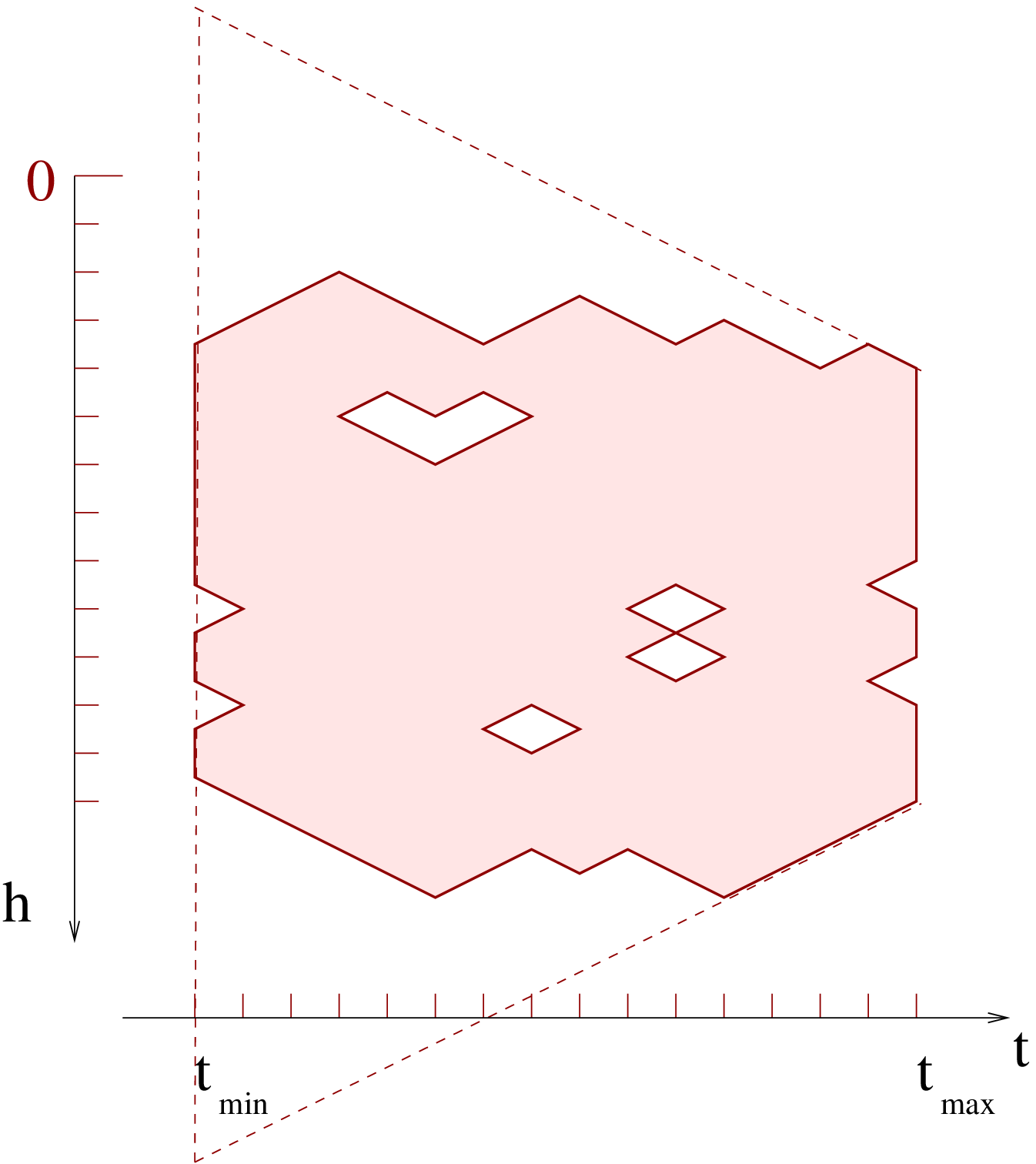}{The maximal domain of $\domain$ is the region comprised in the dashed  line.\label{figmaxdomain}}

{\bf Definitions:}\\
{\it
The maximal domain ${\rm Max}(\domain)$ of $\domain$ is the intersection of $\tmin\leq t\leq \tmax$ with $h_N(\tmax)+{t-\tmax\over 2} \leq h \leq h_1(\tmax) -{t-\tmax\over 2}$ (see the dashed region in figure \ref{figmaxdomain} and figure \ref{figdomainshade}). 

The defect of $\domain$, is  $\defect = {\rm Max}(\domain)\setminus\domain$.
The shadow of $\defect$, is the domain $\widehat\defect\subset {\rm Max}(\domain)$ which is unaccessible to particles moving in $\domain$, with slopes $\pm {1\over 2}$.
We have $\defect\subset\widehat\defect$.

}
Those notions are illustrated in fig \ref{figdomainshade}.

\medskip
\figureframex{14}{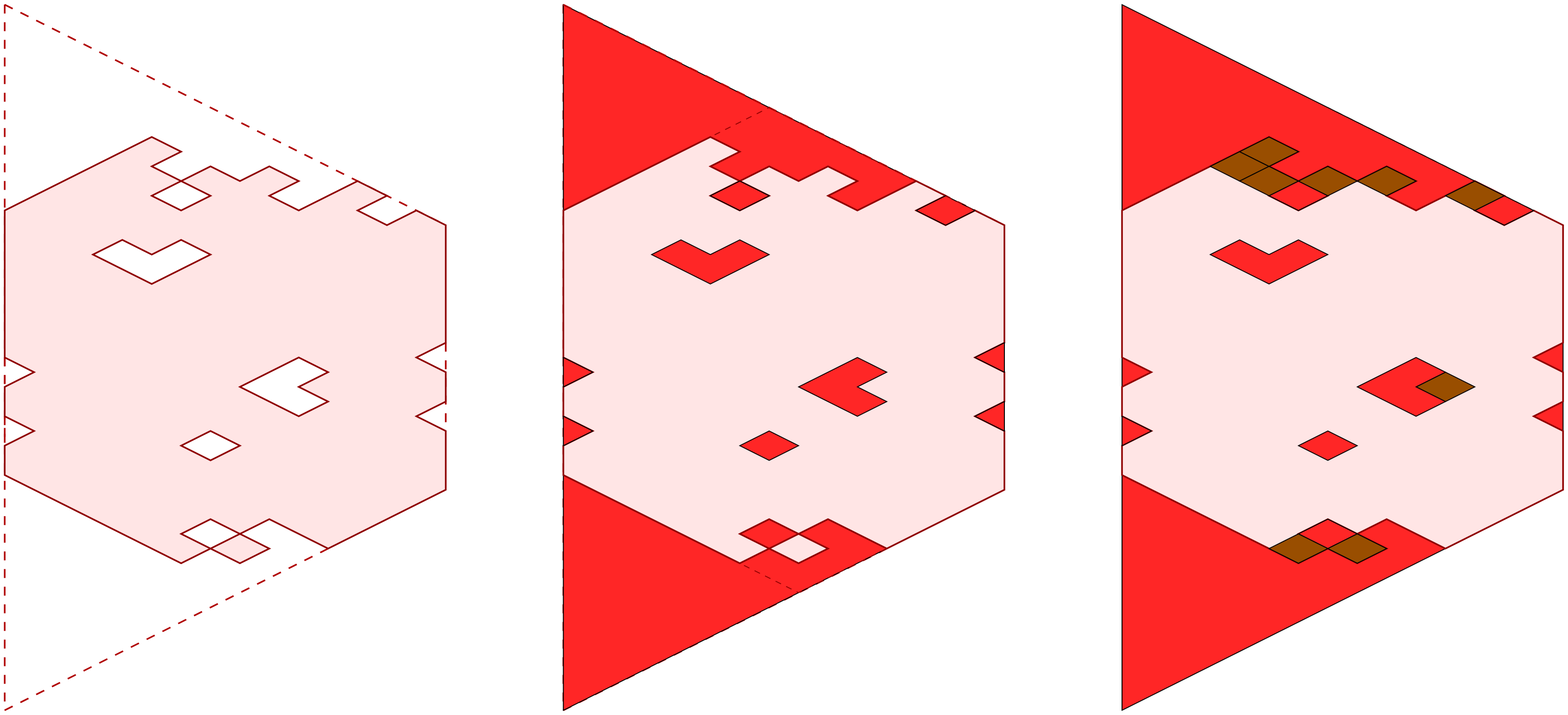}{Consider a given domain $\domain$. Its maximal domain ${\rm Max}(\domain)$ is represented with dashed line on the left figure (it is obtained by lines of slopes $\pm {1\over 2}$ starting from the extremities of the domain at $\tmax$). $\qquad$
In the middle figure we have represented the defect of $\domain$, that is  $\defect = {\rm Max}(\domain)\setminus\domain$ (the red region).$\qquad$
In the right figure we have represented the shadow of $\defect$, that is the domain $\widehat\defect\subset {\rm Max}(\domain)$ which is unaccessible to particles moving in $\domain$, with slopes $\pm {1\over 2}$ (the brown+red region), it contains $\defect$. This means that particles moving in $\domain$ can never enter the brown tiles.
\label{figdomainshade}}

{\bf Definition:}\\
{\it
The minimal defect $\defect_0$ of $\domain$, is the smallest subdomain of $\defect$, such that:
\beq
\widehat\defect_0=\widehat\defect
\eeq
}

\figureframex{14}{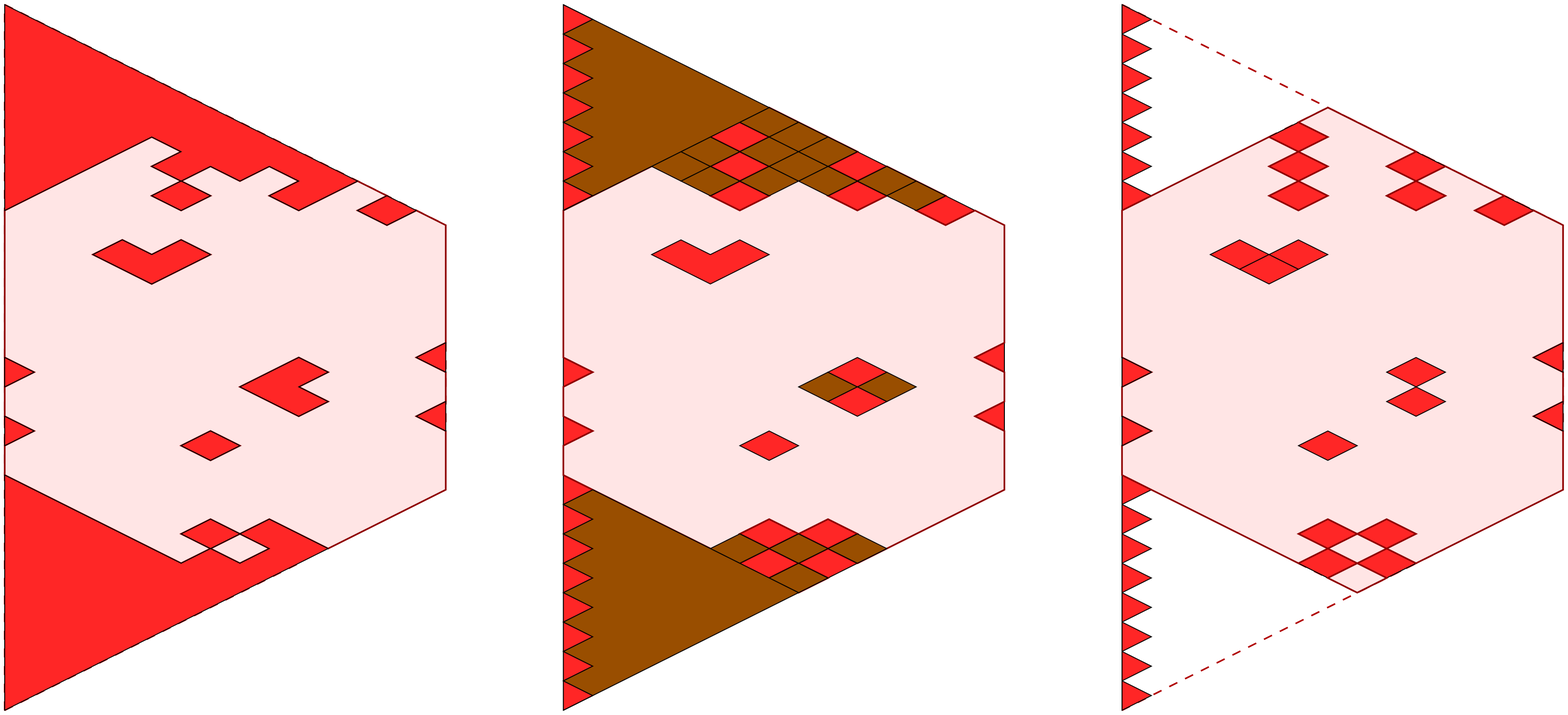}{
The left figure represents the domain $\domain$ and its defect $\defect$ in red.
The right figure represents (in red) the minimal defect $\defect_0$ of $\domain$.
The middle figure shows that both $\defect$ and $\defect_0$ have the same shadow (in brown+red).
\label{figminimaldefect}}

\subsubsection{Domain at time t}

Let us call $\domain(t)$, the slice of the domain $\domain$ at time $t$, i.e. the set of allowed positions of particles at time $t$:
\beq
\domain(t) = \{ h\, / \,\, h-{t\over 2}\in \mathbb Z\, , \,\, (h,t)\in \domain\}
\eeq
Let us call $\defect(t)$ the slice of $\defect$ at time $t$, i.e. the position of holes:
\beq
\defect(t) = \{ h\, / \,\, h-{t\over 2}\in \mathbb Z\, , \,\, (h,t)\in \defect\}
\eeq

Let us assume that the number of particles $N$, is such that:
\beq
\forall t=\tmin,\dots,\tmax
\qquad \quad , \#\domain(t)\geq N,
\eeq
in other words such that it is indeed possible for $N$ avoiding particles to move in $\domain$.

\medskip

\br
Most often we will choose $N=\#\domain(\tmax)$, i.e. the position of all particles at time $\tmax$ are fixed.
Very often we will also choose the domain such that $N=\#\domain(\tmax)=\#\domain(\tmin)$, i.e. the position of all particles are also fixed at time $\tmin$.

But we emphasize that the method we present here works also without those assumptions.

\er

\subsubsection{Filling fractions}

In general, the domain at time $t$, is a union of intervals:
\beq
\domain(t) = \mathop{{\cup}}_{i=1}^{m_t}\,\, [a_i(t),b_i(t)].
\eeq
If the intervals are disconnected, no particle can jump from one interval to another, and thus the number of particles moving in an interval is constant in time, until intervals join. 
We may thus fix the number of particles in each interval.
Call it $n_i(t)=$ filling fraction of the interval $[a_i,b_i]$ at time $t$. We have:
\beq
\sum_{i=1}^{m_t}\, n_i(t)=N.
\eeq
In general, if the domain has $k$ holes, there are $k$ independent filling fractions.

\subsection{Partition function}

The partition function which we wish to compute is:
\beq\label{defZpplanesalphabeta}
Z_N(\domain;q,\alpha,\beta)
= \sum_{h_1(t)>\dots>h_N(t), \, h_i(t)\in \domain(t)}\,\,\, q^{\sum_{i,t} h_i(t)}\,\, 
\prod_{t'=\tmin+{1\over 2}}^{\tmax-{1\over 2}} 
\alpha(t')^{\#({\mbox{\epsfxsize=0.2truecm\epsfbox{rhombusuprgt.eps}}})_{(t')}}
\,\,\beta(t')^{\#({\mbox{\epsfxsize=0.2truecm\epsfbox{rhombusuplft.eps}}})_{(t')}}
\eeq
where at each time $t$ we have $h_1(t)>h_2(t)>\dots>h_N(t)$, 
and where $h_i(\tmin)$ and $h_i(\tmax)$ have fixed values given by two partitions $\mu$ and $\nu$ (encoded by the boundaries of $\domain$):
\beq
h_i(\tmin) = h_i(\mu)
\virg
h_i(\tmax) = h_i(\nu)
\eeq

We count the configuration with a weight $q^{|\pi|}\propto q^{\sum_{t,i} h_i(t)}$, and with weights $\alpha(t+{1\over 2})$ per number of upward jumps {\mbox{\epsfxsize=0.2truecm\epsfbox{rhombusuprgt.eps}}} between time $t$ and $t+1$, and with weights $\beta(t+{1\over 2})$ per number of downward jumps {\mbox{\epsfxsize=0.2truecm\epsfbox{rhombusuplft.eps}}} between time $t$ and $t+1$.

It can be rewritten:
\bea\label{Zhidelta}
Z_N(\domain;q,\alpha,\beta)
&=& \sum_{h_1(t)>\dots>h_N(t), \, h_i(t)\in \domain(t)}\,\,\,
 \prod_{i=1}^N \prod_{t=\tmin}^{\tmax}  q^{h_i(t)}\,\,\cr
&&  \quad     \prod_{t'=\tmin+1/2}^{\tmax-1/2} \prod_{i=1}^{N} \Big[\alpha(t')\delta\left(h_i(t'+{1\over 2})-h_i(t'-{1\over 2})-{1\over 2}\right) \cr
 && +\beta(t')\delta\left(h_i(t'+{1\over 2})-h_i(t'-{1\over 2})+{1\over 2}\right)\Big] \cr
\eea
where here, $\delta$ is the Kroenecker's $\delta$-function.


\subsection{Applications}

$\bullet$ In topological string theory, Gromov Witten invariants of toric Calabi-Yau 3-folds are computed with the topological vertex, which is the following sum of plane partitions \cite{topvertex, ORV}:
\beq
Z_{\infty}(q,1,1;\l,\mu,\nu) = \sum_{\pi, \d\pi = (\l,\mu,\nu)}\,\,\, q^{|\pi|}\,\, 
\eeq

\bigskip

\noindent $\bullet$ When $\l=\mu=\nu=\emptyset$, and $N_\l=N_\mu=N_\nu=\infty$, and $\alpha=\beta=1$, this sum is known, it is the Mac-Mahon formula:
\beq
Z_{\infty}(q,1,1;\emptyset,\emptyset,\emptyset) =  \sum_{\pi}\,\,\, q^{|\pi|}\,\, = \prod_{k=1}^\infty (1-q^k)^{-k}
= 1+q+3q^2+6q^3+13 q^4+\dots 
\eeq

\bigskip

\noindent $\bullet$ The Razumov-Stroganov conjecture \cite{andrew, ZJ3} has put forward a problem of combinatorics of totally symmetric sel-complementary plane partitions (TSSCPP), and the claim is about the relationship with the combinatorics of alternating sign matrices (ASM). There is a 6-vertex matrix model formulation for ASM \cite{PZJ}, and it would be interesting to also have a matrix model for TSSCPP, that's what we address in section \ref{secTSSCPP} below.


%

\section{Matrix model}
\label{secMM}

We are going to represent our self avoiding particle process partition function  \eq{defZpplanesalphabeta} as a multi-matrix integral.

Let us sketch the idea of the next subsection:
we shall introduce $\tmax-\tmin+1$ normal matrices $M_t$ of size $N\times N$, for all integer times $t$ between $\tmin$ and $\tmax$, whose eigenvalues are the $h_i(t)$.
Moreover, we shall Fourrier-transform the $\delta-$functions which enforce $h_i(t+1)=h_i(t)\pm {1\over 2}$:
\bea\label{defLagrangemultri}
&& \Big[\alpha(t+{1\over 2})\,\delta(h_i(t+1)-h_i(t)-{1\over 2}) + \beta(t+{1\over 2})\,\delta(h_i(t+1)-h_i(t)+{1\over 2})\Big]  \cr
&=& \int_{-\infty}^{+\infty} dr_i\,\, \Big[\alpha(t+{1\over 2})\,\ee{2i\pi\,r_i(h_i(t+1)-h_i(t)-{1\over 2})} + \beta(t+{1\over 2})\,\ee{2i\pi\,r_i(h_i(t+1)-h_i(t)+{1\over 2})}\Big]  \cr
&=& \int_{-\infty}^{+\infty} dr_i\,\,\ee{2i\pi\,r_i(h_i(t+1)-h_i(t))}\,\, \Big[\alpha(t+{1\over 2})\,\ee{-i\pi\,r_i} + \beta(t+{1\over 2})\,\ee{i\pi\,r_i}\Big]  \cr
\eea
i.e. we shall introduce some Lagrange multipliers $r_i(t+{1\over 2})$ at all half-integer times, and we will introduce a $N\times N$ hermitian matrix $R_{t+{1\over 2}}$ whose eigenvalues are the Lagrange multipliers $r_i(t+{1\over 2})$, $i=1,\dots,N$, which implement the $\delta-$functions for the jumps between time $t$ and $t+1$. See fig.\ref{figMatrixMR}.

More generally, if we wanted to allow jumps of several steps $h_i(t+1)= h_i(t)+s, s\in \{s_1,\dots,s_k\}$, we would take the Fourrier transform of $\sum_j \alpha_j \ee{2i\pi s_j r_i}$. 

\smallskip
The non-intersecting condition for paths can be realized as a determinant, like Gessel-Viennot formula \cite{GV}, which allows to rewrite
\beq
\prod_i \ee{2i\pi\, r_i\, h_i} \, \to \, \det(\ee{2i\pi\, r_i\, h_j})
\eeq
and we recognize that this expression is the Itzykson-Zuber-Harish-Chandra formula \cite{HC, IZ}:
\beq
\det(\ee{2i\pi\, r_i\, h_j}) = \Delta(R)\Delta(h)\,\, \int_{U(N)}\, dU\,\, \ee{2i\pi \tr R\, U\, h \, U^\dagger}.
\eeq
This is the key to obtain a matrix integral, it introduces angular degrees of freedom in addition to eigenvalues:
\beq
M_t = U\, h(t) \, U^\dagger
\virg h={\rm diag}(h_1(t),h_2(t),\dots,h_N(t)).
\eeq

\figureframex{9}{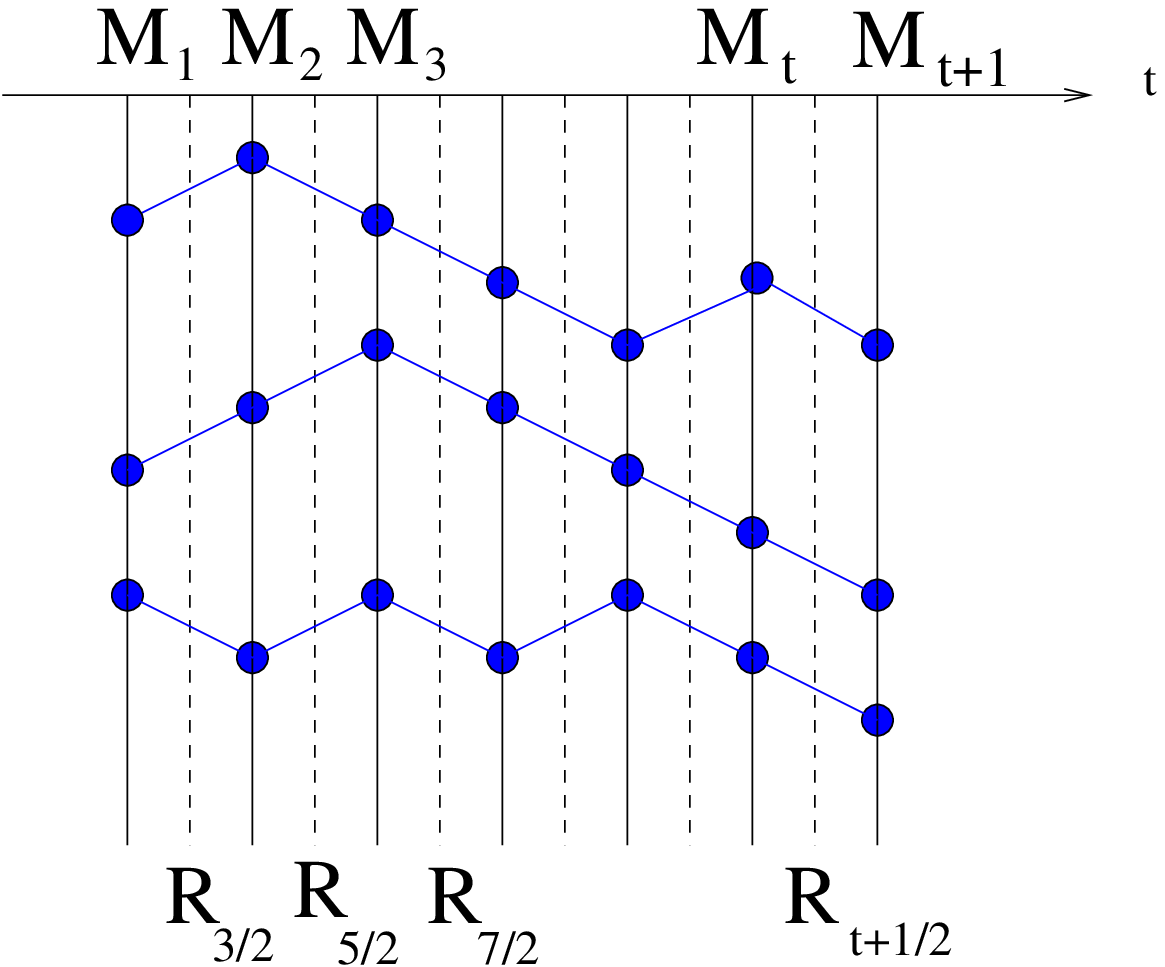}{We introduce a chain of matrices. The eigenvalues of the random matrices $M_t$ with $t$ integer, are the random $h_i(t)$. And the $R_{t'}$ with $t'$ half integer are Lagrange multipliers which enforce the relations $h_i(t+1)-h_i(t)=\pm{1\over 2}$.
\label{figMatrixMR}}
And we shall introduce some potential $V_t$ for the matrices $M_t$, to ensure that their eigenvalues are in $\domain(t)$, and $V_\tmin$ and $V_\tmax$ enforce the initial values at time $t=\tmin$ or $t=\tmax$.

We shall find that the sum over $h_i(t)$'s can be rewritten as a "chain of matrices" matrix model.

So, let us describe the model now.

\subsection{The multi-matrix model}

Let us consider the following multi-matrix integral:
\bea\label{defmatrixmodel}
 {\cal Z}  
&=&  \int_{(H_{N})^{\tmax-\tmin}} \prod_{t=\tmin}^{\tmax-1} dM_t\,\, \int_{(i\, H_{N})^{\tmax-\tmin}} \prod_{t'=\tmin+{1\over 2}}^{t'=\tmax-{1\over 2}}\, dR_{t'} \cr
&&  \prod_{t=\tmin}^{\tmax}\, \ee{-\,\Tr V_t(M_t)} \,\, q^{\Tr M_t}\,
 \prod_{t'=\tmin+{1\over 2}}^{\tmax-{1\over 2}}\, \ee{-\Tr U_{t'}(R_{t'})}\,\, 
 \prod_{t'=\tmin+{1\over 2}}^{\tmax-{1\over 2}}\, \ee{\Tr R_{t'}\,(M_{t'+{1\over 2}}-M_{t'-{1\over 2}} )}\,\, \cr
\eea
Each integral over $M_t$ with $t=\tmin,\dots,\tmax-1$ is over the set $H_{N}$ of hermitian matrices of size $N$, and 
each integral over $R_{t'}$ with $t'=\tmin+{1\over 2},\dots,\tmax-{1\over 2}$ is over the set $i H_{N}$ of anti-hermitian matrices of size $N$.
And there is no integration over $M_{\tmax}$, which is a fixed external field, which we choose equal to:
\beq
M_{\tmax}= \,{\rm diag}(h_1(\nu),\dots,h_{N}(\nu))
\eeq

The potentials $V_t$ or $U_{t'}$ are defined as follows: 

$\bullet$ for $t'$ half-integer:
\beq\label{defpotU}
\ee{-U_{t'}(r)} = \alpha(t')\,\ee{-{r\over 2}} + \beta(t')\,\ee{{r\over 2}}
\eeq
this potential for $R_{t'}$ is the Fourrier transform of the jumps, see \eq{defLagrangemultri}, we have just rescaled $r$ by $2i\pi$.

$\bullet$ and for $t$ integer, $\tmin<t<\tmax$, we choose the potential $V_t(x)$ such that:
\beq\label{defVtchardefectt}
\forall x\in ({t\over 2} + \mathbb Z) \cap {\rm Max}(\domain)\,\, , \qquad \,\,
\ee{-V_t(x)} = 
\left\{\begin{array}{l}
1\,\, {\rm if}\,\, x\in  \domain(t)/\widehat\defect  \cr
0\,\, {\rm if}\,\, x\in  \defect_0(t)  \cr
{\rm arbitrary \,\, otherwise}
\end{array}\right.
\eeq
$\ee{-V_t}$ can be more or less interpreted as the characteristic function of the domain $\domain(t)$, or more precisely,  the complementary domain of the defect $\defect(t)$.

However, this definition does not define a unique potential $V_t$, and
many potentials $V_t$ may have the property \eq{defVtchardefectt}.
In particular, we see that the value of $V_t(x)$ in the shadow of $\domain$, more precisely in $\widehat\domain/\defect_0$, is undetermined.
We show below some rather canonical examples of $V_t$ satisfying those constraints.

We will see below that the value of the partition function ${\cal Z}$ does not depend on the choice of $V_t$.

\smallskip

$\bullet$ and for $t=\tmin$, we choose the potential $V_\tmin(x)$ such that:
\beq\label{defVtchardefecttmin}
\forall x\in ({\tmin\over 2} + \mathbb Z) \cap {\rm Max}(\domain)\,\, , \qquad \,\,
\ee{-V_\tmin(x)} 
\left\{\begin{array}{l}
\neq 0\,\, {\rm if}\,\, x=h_i(\mu)   \cr
=0\,\, {\rm if}\,\, x\neq h_i(\mu)  \cr
{\rm arbitrary \,\, otherwise}
\end{array}\right.
\eeq
In other words, we do not even require that $\ee{-V_\tmin}=1$ in the domain, we only require that it is $\neq 0$.

\medskip

{\bf Examples of $V_t$}:

Since our domain is compact, we have to find a potential $V_t$ with prescribed values at a finite number of points, and a possibility is to choose $\ee{-V_t}$ to be the Lagrange interpolating polynomial going through the prescribed values.

\smallskip

For example, if our domain at time $t$ is contained in the interval
\beq\label{eqktsupbnd}
\domain(t) \subset [k_{t-},k_{t+}],
\eeq
we may choose the potential $V_t$ as the Lagrange interpolating polynomial:
\beq\label{defVtkfini}
\ee{-V_t(x)} =  1- \,\,\sum_{i\in \defect(t)}\, {P_{k_{t-},k_{t+}}(x)\over (x-i)\,P'_{k_{t-},k_{t+}}(i)} 
\virg
P_{k_-,k_+}(h) = \prod_{j=k_-}^{k_+} (h-j).
\eeq

Another possibility is to take the limit $k_+\to\infty$:
\beq\label{defVtkinfty}
\ee{-V_t(x)} =  1- \,\,\sum_{i\in \defect(t)}\, {(-1)^{i}\over i!\,(x-i)\,\Gamma(-x)} .
\eeq
or we may also choose:
\beq
\ee{-V_t(x)} = {\ee{i\pi x}\,\sin{\pi x}\over \pi}\,\, \sum_{i\in \domain(t)}\, {1\over x-i}.
\eeq

Depending on the type of applications we are interested with, it is sometimes more convenient to work with a potential of type \eq{defVtkfini} or a potential of type \eq{defVtkinfty}, or also their $q$-deformations, or sometimes other potentials having property \eq{defVtchardefectt}.

In general, we see that $-V_t(x)$ must have logarithmic singularities on $\defect_0(t)$, and thus we may write:
\beq\label{Vtdefect0sum}
V'_t(x) = - \sum_{i\in \defect_0(t)} {1\over x-i}  \quad + f'(x) + g'(x-{t\over 2})
\eeq
where $g(x)$ can be any arbitrary entire function which vanishes on $x\in \mathbb Z$,
and where $f(x)$ is an analytical function such that if $i\in \domain(t)$ we have $V_t(x)=0$.
Since fixing the values of $f$ and $g$ does not fix the values of $f'$ and $g'$ at those points, we see that what characterizes $V'_t$ is that it has simple poles with residue $-1$ in $\defect_0(t)$, plus an almost arbitrary analytical function.

\subsubsection{Diagonalization}

Let us diagonalize all matrices in \eq{defmatrixmodel}.
One needs to know that any normal matrix (and in particular hermitian matrices) can be diagonalized by a unitary transformation:
\beq
M_t = {\cal U}_t\, X_t\, {\cal U}_t^\dagger
\virg
{\cal U}_t\in U(N)
\,\, , \, X_t={\rm diag}(X_1(t),\dots,X_{N}(t))
\eeq
and the matrix measures are:
\beq
dM_t = {1\over N!}\,\, d{\cal U}_t \, dX_t\,\,\,\Delta(X_t)^2
\eeq
where $d{\cal U}_t$ is the Haar measure on $U(N)$ and $\Delta(X_t)$ is the Vandermonde determinant:
\beq
\Delta(X_t) = \prod_{i>j} (X_i(t)-X_j(t))
\eeq

Similarly for the matrices $R_{t'}$:
\beq
R_{t'} = \td{\cal U}_{t'}\, Y_{t'}\, \td{\cal U}^\dagger_{t'}
\virg
\td{\cal U}_{t'}\in U(N)
\,\, , \, Y_{t'}={\rm diag}(Y_1(t'),\dots,Y_{N}(t'))
\eeq

Therefore we may rewrite \eq{defmatrixmodel} as:
\bea\label{defmatrixmodeldiag}
&& (N !)^{2(\tmax-\tmin)+1}\,\, {\cal Z}  \cr
&=&  \int_{\mathbb R} \prod_{t=\tmin}^{\tmax-1} dX_t\,\, \int_{i{\mathbb R}} \prod_{t'=\tmin+{1\over 2}}^{t'=\tmax-{1\over 2}}\, dY_{t'} \,\, \ee{-\Tr V_\tmax(X_\tmax)} \, q^{\Tr X_\tmax}\,\,\cr
&&  \prod_{t=\tmin}^{\tmax-1}\, \ee{-\Tr V_t(X_t)} \, q^{\Tr X_t}\,\, \Delta(X_t)^2
 \prod_{t'=\tmin+{1\over 2}}^{\tmax-{1\over 2}}\, \ee{-\Tr U_{t'}(Y_{t'})}\,\, \Delta(Y_{t'})^2\,\,\cr
&&  \prod_{t'=\tmin+{1\over 2}}^{\tmax-{1\over 2}}\, I(X_{t'+{1\over 2}},Y_{t'}) \, I(-X_{t'-{1\over 2}},Y_{t'}) \cr
\eea
where $I(X,Y)$ is the Itzykson-Zuber integral
\beq
I(X,Y) = \int_{U(N)}\, d{\cal U}\,\, \ee{\Tr X {\cal U} Y {\cal U}^\dagger}
\eeq
It is well known that \cite{HC,IZ}:
\beq
I(X,Y) = {\det{(\ee{ X_i Y_j})}\over \Delta(X)\,\Delta(Y)}
\eeq
and therefore \cite{Mehtamultimat}, we get:
\bea\label{defmatrixmodeldiagIZ}
&& (N !)^{2(\tmax-\tmin)+1}\,\,{\cal Z}  \cr
&=& {1\over \Delta(X_{\tmax})}\, \,\, \int_{\mathbb R} \prod_{t=\tmin}^{\tmax-1} dX_t\,\, \int_{i{\mathbb R}} \prod_{t'=\tmin+{1\over 2}}^{t'=\tmax-{1\over 2}}\, dY_{t'} \,\,\, \Delta(X_{\tmin})\cr
&& \,\, \prod_{t=\tmin}^{\tmax}\, \ee{-\Tr V_t(X_t)}\, q^{\Tr X_t} \,\,
 \prod_{t'=\tmin+{1\over 2}}^{\tmax-{1\over 2}}\, \ee{-\Tr U_{t'}(Y_{t'})}\,\,\cr
&&  \prod_{t'=\tmin+{1\over 2}}^{\tmax-{1\over 2}}\, 
\det{(\ee{ X_i(t'+{1\over 2})\,Y_j(t')}  )}\,\,\det{(\ee{- \, X_i(t'-{1\over 2})\,Y_j(t')}  )} \cr
\eea

\subsection{Relation between the matrix model and the self-avoiding particles model}



\subsubsection{Integrals over Lagrange multipliers}

Then let us perform the integral over $Y_j(t')$, we have:
\bea
&& \int dY_{t'}\,\, \det{(\ee{ X_i(t'+{1\over 2})\,Y_j(t')}  )}\,\,\det{(\ee{-\, X_i(t'-{1\over 2})\,Y_j(t')}  )}\,\, \ee{-\Tr U_{t'}(Y(t'))} \cr
&=& \sum_{\sigma_{t'},\td\sigma_{t'}} (-1)^{\sigma_{t'}\td\sigma_{t'}}\,\,\prod_i \int dY_i(t')\,\,\ee{Y_i(t')\,(X_{\sigma_{t'}(i)}(t'+{1\over 2})-X_{\td\sigma_{t'}(i)}(t'-{1\over 2}) )}\,\, \ee{-U_{t'}(Y_i(t'))} \cr
&=& \sum_{\sigma_{t'},\td\sigma_{t'}} (-1)^{\sigma_{t'}\td\sigma_{t'}}\,\,\prod_i 
\Big[
\alpha(t') \delta(X_{\sigma_{t'}(i)}(t'+{1\over 2})-X_{\td\sigma_{t'}(i)}(t'-{1\over 2})-{1\over 2}) \cr
&& + \beta(t') \delta(X_{\sigma_{t'}(i)}(t'+{1\over 2})-X_{\td\sigma_{t'}(i)}(t'-{1\over 2})+{1\over 2}) \Big] \cr
&=& N!\,\,\det
\Big[
\alpha(t') \delta(X_i(t'+{1\over 2})-X_j(t'-{1\over 2})-{1\over 2}) \cr
&& + \beta(t') \delta(X_i(t'+{1\over 2})-X_j(t'-{1\over 2})+{1\over 2}) \Big] \cr
\eea
This term implies that there must exist some permutation such that
\beq
X_i(t'+{1\over 2})=X_{\sigma_{t'}(i)}(t'-{1\over 2}) \pm {1\over 2}
\eeq
with respective probabilities $\beta(t'),\alpha(t')$.
And in particular, since this is true for $t=\tmax$, we have $\forall t$:
\beq
{X_i(t)}-{t\over 2} \in {\mathbb Z}\cap [h_{N}(\tmax)-{\tmax-t\over 2},h_{1}(\tmax)+{\tmax-t\over 2}]
\eeq
In particular that implies that $X_i(t)\in ({t\over 2}+\mathbb Z)\cap {\rm Max}(\domain)$, and thus $\ee{-V_t(X_i(t))}=1$ if $X_i(t)\in \domain$ and $\ee{-V_t(X_i(t))}=0$ if $X_i(t)\in\defect_0(t)$. 
In particular, the matrix integral is indeed independent of the choice of $V_t$, provided that it satisfies \eq{defVtchardefectt}.

Then, since the quantities we are summing are symmetric, we can assume, up to a multiplication by $N!$, that all $X_i(t)$ are ordered:
\beq
X_1(t)>X_2(t)>\dots>X_{N}(t)
\eeq
and this yields a factor $(N!)^{\tmax-\tmin}$, which we shall discard because we consider the partition function up to global trivial constants.

In other words, the result of the integral is a sum over ${X_i(t)}=h_i(t)$, where $h_i(t)-{t\over 2}$ are ordered integers:
\bea\label{matrixmodesumXi}
  {\cal Z}  
&=&  {1\over \Delta(h_\nu)}\,\, 
\sum_{h_i(t), i=1,\dots,N, t=\tmin+1,\dots,\tmax-1}
\, \Delta(h_i(\tmin))) \quad
\prod_{i,t} \ee{-V_t(h_i(t))}\, q^{h_i(t)}\cr
&&  \prod_{i,t'} \Big[
\alpha(t') \delta_{h_i(t'+{1\over 2})-h_i(t'-{1\over 2})-{1\over 2}} 
 + \beta(t') \delta_{h_i(t'+{1\over 2})-h_i(t'-{1\over 2})+{1\over 2}} \Big] \cr
\eea
where the sum over $h_i(t)$, is such that:
\beq
h_i(t)-{t\over 2}\in {\mathbb Z}
\quad , {\rm and}\quad
h_1(t)>h_2(t)>\dots > h_{N}(t)
\eeq
and
\beq
h_i(\tmax) = h_i(\nu)
\eeq

Then, notice that
$\ee{-V_t(h_i(t))}=1$ if $h_i(t)\in \domain$ and $\ee{-V_t(h_i(t))}=0$ if $h_i(t)\in\defect_0(t)$. 
In other words, the sum is only over $h_i(t)$'s such that 
\beq
h_i(t)-{t\over 2} \notin \defect_0(t)
\eeq
In other words, we have a self-avoiding particle process on the rhombus lattice, which avoids a prescribed domain $\defect$, i.e. we recover our self-avoiding particles partition function:

\medskip
\bt\label{thTASEPMM}

The self-avoiding particles model partition function $Z_N({\cal D},q,\alpha,\beta)$ in a domain $\domain$, is proportional to the matrix integral ${\cal Z}$:
\beq\label{eqTASEPMM}
{\cal Z} = C_{N,\tmax-\tmin}\,\, {\Delta(h_\mu)\over \Delta(h_\nu)}\,\ee{-\sum_i V_{\tmin}(h_i(\mu))}\,\, Z_N({\cal D},q,\alpha,\beta)
\eeq
where the constant $C_{N,T}$ depends only on $N$ and $T$, and nothing else. It contains the normalization factors, such as the volumes of unitary groups, and the powers of $N!$.

\et

\bigskip

This theorem implies immediately, as a tautology, that whatever limit we consider (for instance large size limit, $q\to 1$ limit, bulk regime, behavior near edges, $\dots$), the asymptotic statistical properties {\bf are always matrix models limit laws} !

This explains why one finds sine kernel laws in the bulk, Tracy-Widom laws near some boundaries, Pearcey laws, and many more...


\subsection{Determinantal formulae}

Theorem \ref{thTASEPMM} allows to apply to the self-avoiding particles model and plane partitions, all the technology developed for matrix models, in particular the methods of orthogonal polynomials \cite{Mehtamultimat, MehtaBook}.

\medskip

Consider a "chain of matrices" integral:
\beq\label{Zgenchain2}
{\cal Z}=\int_{\prod H_N({\cal C}_i)} \prod_{i=1}^{p} dM_i\,\,\, \ee{-Q \Tr [\sum_{i=1}^p V_i(M_i) + \sum_i c_i M_i M_{i+1}]}
\eeq
where $V_i$ are some potentials, where $H_N({\cal C}_i)$ is the set of $N\times N$ normal matrices having their eigenvalues on contour ${\cal C}_i$, and where $M_{p+1}$ is not integrated upon.

\medskip

The Eynard-Mehta theorem  \cite{eynMehta} shows that correlation functions of densisities of eigenvalues $\rho_i(x) = \Tr \delta(x-M_i)$, are determinants.
Namely, there exist some kernels $H_{i,j}(x,x')$ such that:
\bea
\left< \rho_{k_1}(x_1)\dots \rho_{k_n}(x_n)\right> = \det\left( H_{k_i,k_j}(x_i,x_j)\right),
\eea
where the kernels $H_{i,j}$ are Christoffel-Darboux kernels for some families of biorthogonal polynomials.

\smallskip

From the point of view of self-avoiding particles model and plane partitions, this should allow to recover many determinantal formulae in the TASEP literature, see for instance \cite{Krat1, bogo, OR}.

\medskip

Another consequence of the``orthogonal polynomials method'' \cite{MehtaBook}, is that matrix integrals of type \ref{Zgenchain2} are Tau-functions for the Integrable Toda hierarchy \cite{AVM}.
This should allow to recover  many differential equations in the TASEP literature.

\section{Matrix model's topological expansion}
\label{secMMsolution}

The good thing about theorem \ref{thTASEPMM}, is that the general expansion of matrix integrals of the chain of matrices type is known to all orders.

\subsection{Generalities about the expansion of matrix integrals}

See appendix \ref{appchainmat} for a more detailed description.

\medskip

Consider a "chain of matrices" integral:
\beq\label{Zgenchain1}
{\cal Z}=\int_{\prod H_N({\cal C}_i)} \prod_{i=1}^{p} dM_i\,\,\, \ee{-Q \Tr [\sum_{i=1}^p V_i(M_i) + \sum_i c_i M_i M_{i+1}]}
\eeq
where $V_i$ are some potentials, where $H_N({\cal C}_i)$ is the set of $N\times N$ normal matrices having their eigenvalues on contour ${\cal C}_i$, and where $M_{p+1}$ is not integrated upon.

\smallskip
In some good cases (depending on the choice of potentials $V_i$ and paths ${\cal C}_i$), such an integral has a large $Q$ expansion of the form:
\beq\label{gentopexpQ}
\ln{{\cal Z}} \sim \sum_{g=0}^\infty Q^{2-2g}\, {\cal F}_g.
\eeq
Such an expansion does not always exist. It exists only if the paths ${\cal C}_i$ which support the eigenvalues, are "steepest descent paths" for the potentials $V_i$ (see e.g. \cite{EOreview, eynform}). Finding the steepest descent paths associated to given potentials is an extremely difficult problem.

\smallskip
Fortunately, many applications of random matrices regard combinatorics, i.e. they are formal series in some formal parameter, and very often, the corresponding so-called ``formal matrix integrals'' do have a large $Q$ expansion almost by definition\footnote{For formal matrix integrals, the integration paths ${\cal C}_i$ for eigenvalues, are most often not known explicitely, they can be determined so that a large $Q$ power series expansion does exist.}, and \eq{gentopexpQ} holds order by order in the formal parameter (a formal parameter which is not necessarily $Q$).
Here, we are considering applications to statistical physics, our partition functions are formal series, and we shall assume that such an expansion exists (order by order in a suitable formal parameter).

\smallskip
The problem is then to compute the coefficients ${\cal F}_g$.
\smallskip

The answer was found in \cite{EPrats}, by using loop equations (i.e. Schwinger-Dyson equations in the context of matrix models), and which just correspond to integrations by parts.

\medskip

The solution proceeds in two steps (which we explain below):

{\bf 1)} Compute the "spectral curve" ${\cal S}$ of the matrix model. The spectral curve ${\cal S}=(x,y)$ is a pair of two analytical functions $x(z),y(z)$ of a variable $z$ living on a Riemann surface. The spectral curve is obtained from the "classical limit" of the integrable system whose tau-function is the matrix integral. 
Roughly speaking, if we eliminate $z$, the function $y(x)$ is more or less the equilibrium density of eigenvalues of the first matrix of the chain.
We explain in appendix \ref{appchainmat} how to find the spectral curve of a general chain of matrices.
We emphasize that associating a spectral curve ${\cal S}=(x,y)$ to a given matrix model, is something already done in the matrix models literature.

{\bf 2)} Then compute the symplectic invariants $F_g({\cal S})$ of that spectral curve (symplectic invariants of an arbitrary spectral curve ${\cal S}$ were first introduced in \cite{EOFg}, they are rather easy to compute, and we recall their definition in appendix \ref{appspinv}), and the main result of \cite{EPrats} is that ${\cal F}_g=F_g({\cal S})$, i.e.:
\beq
\ln{{\cal Z}} = \sum_{g=0}^\infty Q^{2-2g} F_g({\cal S})
\eeq

\subsection{Spectral curve of the self-avoiding particles matrix model}
\label{secrecipespcurve}

We recall in appendix \ref{appchainmat} the main results of \cite{EPrats}, i.e. how to compute the spectral curve of an arbitrary chain of matrices.
Here, in this section, we merely apply the general recipe of \cite{EPrats} (see appendix \ref{appchainmat}) to our matrix integral \eq{defmatrixmodel}, and we give a "ready to use recipe".
\smallskip

\medskip
{\bf Recipe for finding the spectral curve} of the matrix integral  \eq{defmatrixmodel}:
\medskip

$\bullet$ Find $2(\tmax-\tmin+1)$ analytical functions of a variable $z$ ($z$ belongs to a Riemann surface ${\curve}$). There is one such analytical function for each matrix of the chain, plus one additional function at the end of the chain. Let us call them:
\beq
\hat{X}(z,t)\, , \,\, t=\tmin,\dots,\tmax
\virg
\hat{Y}(z,t')\, , \,\, t'=\tmin-{1\over 2},\dots,\tmax-{1\over 2}.
\eeq
Those functions are completely determined by the following constraints:
\smallskip

{\it 
\begin{enumerate}

\item Those functions must obey the following system of equations $\forall\, z$:
\beq\label{spcurveeq}
\left\{\begin{array}{l}
\hat X(z,t'+{1\over 2})-\hat X(z,t'-{1\over 2}) = U'_{t'}(\hat Y(z,t'))
\,\, , \quad
\forall t'=\tmin+{1\over 2},\dots,\tmax-{1\over 2} \cr
\cr
\hat Y(z,t+{1\over 2})-\hat Y(z,t-{1\over 2}) = \ln{q} - V'_{t}(\hat X(z,t)) 
\,\, , \quad
\forall t=\tmin,\dots,\tmax-1
\end{array}\right.
\eeq

\item There exists a point in ${\cal L}$, which we call $\infty\in {\cal L}$, such that $\hat X(z,\tmin)$ has a simple pole at $z\to\infty$, and we have:
\beq
\hat Y(z,\tmin-{1\over 2}) \sim {N\over \hat X(z,\tmin)}.
\eeq

\item
There exists points $\zeta_i\in {\cal L}$, such that $\hat X(\zeta_i,\tmax)$ is an eigenvalue of $M_\tmax$:
\beq\label{spcurveeqef}
\hat X(\zeta_i,\tmax) = \, h_i(\nu),
\eeq
and $\hat Y(z,\tmax-{1\over 2}) $ has simple poles at the points $\zeta_i$ and behaves like:
\beq\label{spcurveeqef1}
\hat Y(z,\tmax-{1\over 2}) \sim {1\over \hat X(z,\tmax)-\, h_i(\nu)}
\eeq

\item Define for $t=\tmin,\dots,\tmax$:
\beq\label{defWxtResYdx}
W(x,t) = \Res_{z\to \infty} {\hat Y(z,t-{1\over 2})\, d\hat X(z,t)\over x-\hat X(z,t)},
\eeq
and call it "the resolvent" of the matrix $M_t$. 
We require that $\forall\,t$, $W(x,t)$ is analytical in a vicinity of $x\to\infty$, and behaves like:
\beq\label{norminftyWxt}
W(x,t) \mathop{{\sim}}_{x\to\infty} {N\over x},
\eeq
and $W(x,t)$ can be analytically continued to 
\beq
\mathbb C \setminus \domain(t)
\eeq
where we recall that $\domain(t)\subset \mathbb R$ is a compact region of $\mathbb R$.

\item
Typically, $W(x,t)$ may have branchcuts or isolated singularities, like poles or log singularities.
The set of points at which $W(x,t)$ is not analytical is called the support:
\beq
{\rm supp}(t) = \{ x\,\, , \,\,  W(x,t)\, {\rm not\, analytical}\, \}.
\eeq
The interior of ${\rm supp}(t)$ is called "the liquid region" (it contains the cuts, it excludes the isolated singularities).
We have ${\rm supp}(t)\subset \domain(t)$, and 
\beq
\stackrel{\circ}{{\rm supp}}(t)\subset \domain(t)
\eeq
If $\domain(t) = \cup_i [a_i(t),b_i(t)]$, we require that $\forall\, i$:
$\stackrel{\circ}{{\rm supp}}(t)\cap [a_i(t),b_i(t)]$ is connected.
 
\item If the domain $\domain(t)$ at time $t$ is a disconnected union $\domain(t) = \cup_{i=1}^{m_t} [a_i(t),b_i(t)]$, we require that $\forall\, i=1,\dots,m_t$:
\beq\label{defholesff}
{1\over 2i\pi} \oint_{ [a_i(t),b_i(t)]} W(x,t)\,dx = n_i(t),
\eeq
where the integration contour surrounds the interval $[a_i(t),b_i(t)]$ in the $\hat X(z,t)$ plane, in the clockwise direction.



\end{enumerate}
}

Finding  functions satisfying all those requirements for a general domain, with general weights $\alpha(t'),\beta(t')$ is a difficult problem.
But for not too complicated domains and weights, some simplifications may occur, and we will see many examples of explicit solutions below.

From now on, let us assume that we have found the functions $\hat X$ and $\hat Y$ satisfying all the requirements.
%
%
%
Once we have found a solution to this problem, i.e. found the functions $\hat X(z,t)$ and $\hat Y(z,t')$, we define the spectral curves:

\bd
The spectral curve at time $t$ is the pair of functions:
\beq
{\cal S}_t = (\hat X(.,t),\hat Y(.,t-{1\over 2})).
\eeq

\ed

\br
Because of \eq{spcurveeq}, the following 2-forms in $T^*\mathbb C\wedge T^*\mathbb C$, restricted to the spectral curve, are equal:
\bea
d\hat X(z,t) \wedge d\hat Y(z,t-{1\over 2}) 
&=& d\hat X(z,t) \wedge d\hat Y(z,t+{1\over 2}) \cr 
&=& d\hat X(z,t+1) \wedge d\hat Y(z,t+{1\over 2}) \cr 
\eea
and therefore, the spectral curves ${\cal S}_t$ and ${\cal S}_{t+1}$ are symplecticaly equivalent:
\beq\label{StsympeqStt}
{\cal S}_t \equiv {\cal S}_{t+1}.
\eeq

\er

\subsection{Symplectic invariants and topological expansion}

The symplectic invariants $F_g({\cal S})$ were introduced in \cite{EOFg}.
To any spectral curve ${\cal S}$, one can associate, by simple algebraic computations, an infinite sequence of complex numbers $F_g({\cal S})$, $g=0,1,2,3,\dots$.
We recall their definition in appendix \ref{appspinv}.

One of their main properties, is that if two spectral curves ${\cal S}$ and $\td{\cal S}$ are symplectically equivalent, then we have $F_g({\cal S}) = F_g(\td{\cal S})$.

In our case, because of \eq{StsympeqStt}, we have:
\beq
F_g({\cal S}_t)=F_g({\cal S}_{t+1}),
\eeq
and therefore $F_g({\cal S}_t)$ is independent of $t$, the $F_g({\cal S}_t)$'s are conserved quantities.

\bigskip

It was proved in \cite{EPrats}, for any chain of matrices, and here we apply it to our case, that:

\bt\label{thZtopchmattasep}
\beq
\ln{{\mathcal Z}} = \sum_{g=0}^\infty F_g({\cal S}_t)
\eeq
where the right hand side is independent of $t$.

\et

This theorem holds order by order in some appropriate formal large parameter expansion.
We will see examples below, where the formal parameter can be the size of the system, or $\ln q$, or $\alpha$,...etc.

\subsubsection{Arctic circle}

For most interesting applications, there is some "large parameter $Q$" in our problem (typically the size of the domain $\domain$, or $Q=1/\ln q$, or sometimes other parameters), such that the spectral curve  scales like $Q$, typically:
\beq
{\cal S} = Q\,  {\cal S}_\infty + o(Q) 
\eeq
(where we write $\l{\cal S}=(x,\l y)$ for a spectral curve ${\cal S}=(x,y)$).

The large $Q$ spectral curve ${\cal S}_\infty$ was already computed in many works and in particular by Kenyon-Okounkov-Sheffield \cite{KOS} who found the limit shape of the liquid region.

From the homogeneity property of $F_g$'s  (see \cite{EOFg} and appendix \ref{appspinv}) we have $ F_g({\cal S})=Q^{2-2g}\,\, F_g(Q^{-1}\,{\cal S})$, i.e. we find a large $Q$ expansion:
\beq
\ln{{\mathcal Z}} = \sum_{g=0}^\infty Q^{2-2g}\,\, F_g(Q^{-1}\,{\cal S})
\eeq
Such an expansion is not very useful if $F_g(Q^{-1}\,{\cal S})$ depends on $Q$.

In fact, in many examples related to TASEP and plane partitions, we find that the spectral curve $Q^{-1}\,{\cal S}$ depends on $Q$, but up to a symplectic transformation we have miraculously (this happens for instance in the matrix model considered in \cite{eynLP}):
\beq
Q^{-1}\,{\cal S} \equiv {\cal S}_\infty \qquad {\rm modulo\,\, symplectomorphisms}.
\eeq
which implies:
\beq
F_g(Q^{-1}\,{\cal S}) = F_g({\cal S}_\infty),
\eeq
and in that case, $F_g(Q^{-1}\,{\cal S})=F_g({\cal S}_\infty)$ is independent of $Q$.

This miracle is deeply related to the structure of the self-avoiding particles model partition function, and with the so called "arctic circle phenomenon", i.e. the fact that the system freezes beyond a certain size \cite{Jarctic}.
This can also be related to the fact that we have some arbitrariness in choosing the potentials $V_t$, and we could choose some potentials $V_t$ which depend on $Q$ in an appropriate way such that the spectral curve $Q^{-1}\,{\cal S}$ would not depend on $Q$. Although it is doable in theory, finding the corresponding $V_t$'s seems horrendous.
A rather explicit example of this phenomenon was discussed in \cite{eynLP}.

\smallskip

Only in the case where we have this "arctic circle phenomenon", we have:
\beq
\ln{{\mathcal Z}} = \sum_{g=0}^\infty Q^{2-2g}\,\, F_g({\cal S}_\infty),
\eeq
where the spectral curve ${\cal S}_\infty$  is the curve derived from the Harnack curve of Kenyon-Okounkov-Sheffield \cite{KOS}.

In that case, we can compute the large $Q$ expansion of our self-avoiding particles model model, to all orders in $Q$, not only the large $Q$ leading order limit.

\subsection{Reduced matrix model}
\label{secreducedMM}

For a fully general domain $\domain$, there are as many equations \eq{spcurveeq} to solve as the size of the domain, but, when the domain has very few defects, many of those equations simplify considerably, and we may consider a reduced problem.
\smallskip

Notice, that everytime $\defect(t)=\emptyset$, we may choose $V_t(x)=0$ for all $x$, and thus $V'_t=0$.
Therefore, it is possible to simplify dramatically the equations determining the spectral curve.

Notice also, that there can be many times $t$ at which $\defect(t)\neq \emptyset$, but since particles can only follow lines with slopes $\pm {1\over 2}$, many places (the shadow of $\domain$) are never visited. In other words, we may replace the defects $\defect$ by the minimal defect $\defect_0$ of $\domain$, and since in general $\defect_0$ is a very small subset of $\defect$, this allows to simplify the problem.

Only the minimal defect $\defect_0(t)$ needs to be specified.
\figureframex{14}{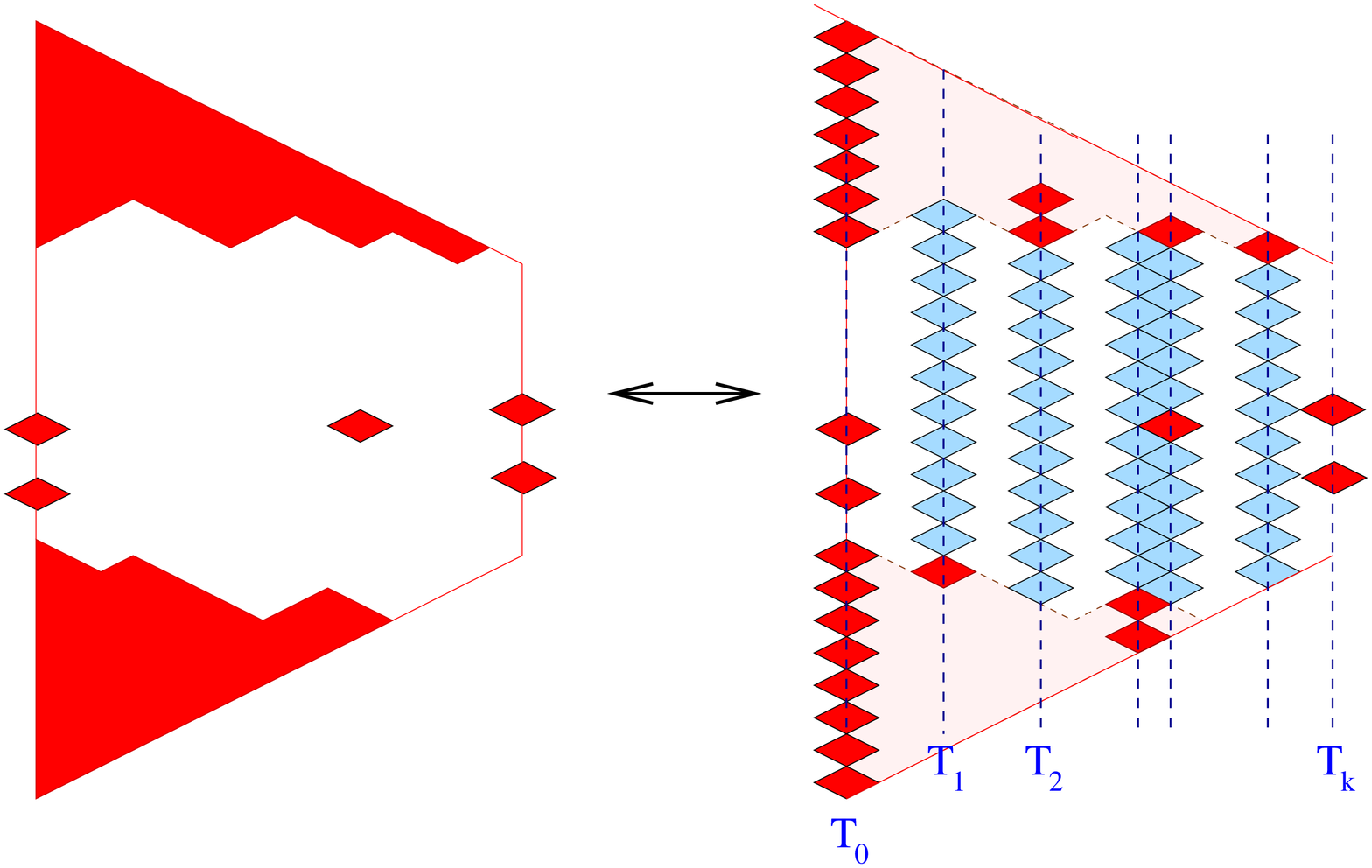}{
The left figure represents the domain $\domain$ and its defect $\defect$ in red.
The right figure represents the minimal defect $\defect_0$ in red. If there are $k+1$ times $T_0,\dots, T_k$ containing red lozenges, then the reduced matrix model is a $2k+1$ matrix model.
The potentials of the reduced model must be such that $\ee{-V_{T_i}}=0$ at the center of red lozenges, and $\ee{-V_{T_i}}=1$ at the blue ones. The values of $\ee{-V_{T_i}}$ elsewhere are arbitrary.
\label{figdomain12}}

Let the times at which $\defect_0(t)\neq \emptyset$ be called:
\beq
\{T_0,T_1,\dots,T_k\}\subset \{\tmin,\dots,\tmax\},
\eeq
among those $T_j$'s we include the extremities $\tmin$ and $\tmax$:
\beq
T_0=\tmin
\virg
T_k=\tmax.
\eeq

The spectral curve equations \eq{spcurveeq} can be rewritten:
\beq\label{spcurveeqred}
\left\{\begin{array}{l}
\hat X(z,T_{i+1})-\hat X(z,T_{i}) = \sum_{t'=T_i+{1\over 2}}^{T_{i+1}-{1\over 2}} U'_{t'}(\hat Y(z,T_{i+1}-{1\over 2})+(t'-T_{i+1}+{1\over 2})\ln q)
\cr
\cr
\hat Y(z,T_{i+1}-{1\over 2})-\hat Y(z,T_{i}-{1\over 2}) = (T_{i+1}-T_{i})\ln{q} - V'_{T_i}(\hat X(z,T_i)) 
\end{array}\right.
\eeq
We thus define:
\beq
\hat X_i(z)= \hat X(z,T_i)
\virg
\hat Y_i(z) = \hat Y(z,T_i-{1\over 2})
\eeq
and the potentials
\beq
\td{U}'_{i}(y) = \sum_{t'=T_{i-1}+{1\over 2}}^{T_{i}-{1\over 2}} U'_{t'}(y+(t'-T_i+{1\over 2})\ln q)
\eeq
i.e.
\beq\label{deftdU}
\ee{-\td{U}_{i}(y)} = \prod_{t'=T_{i-1}+{1\over 2}}^{T_{i}-{1\over 2}} \left(\alpha(t')\,q^{-{1\over 2}(t'-T_i+{1\over 2})}\, \ee{-{y\over 2}} + \beta(t')\,q^{{1\over 2}(t'-T_i+{1\over 2})}\,\ee{y\over 2}\right).
\eeq

The loop equations \eq{spcurveeqred} are equivalent to:
\beq\label{spcurveeqred1}
\left\{\begin{array}{ll}
\hat X_{i+1}(z)-\hat X_i(z) = \td U'_{i+1}(\hat Y_{i+1}(z))
& \virg
\forall i=0,\dots,k-1 \cr
& \cr
\hat Y_{i+1}(z)-\hat Y_i(z) = (T_{i+1}-T_{i})\ln{q} - V'_{T_i}(\hat X_i(z)) 
& \virg
\forall i=0,\dots,k-1
\end{array}\right.
\eeq
One recognizes that these equations are exactly the equations of the spectral curve of another matrix model, which is a chain of matrices, but with only $2k+1$ matrices instead of $2(\tmax-\tmin)-1$, and we get the following result:

\bt\label{thTASEPMMreduced}
The matrix model partition function can be rewritten:
\bea\label{defmatrixmodelred}
 {\cal Z}  
&=& \int_{H_{N}({\cal C})} dM_{0}\,\, \int_{(H_{N})^{k}} \prod_{i=1}^{k-1} dM_{i}\,\, \int_{(i\, H_{N})^{k}} \prod_{i=1}^{k}\, dR_{i} \cr
&&  \prod_{i=0}^{k-1}\, \ee{-\Tr V_{T_i}(M_{i})} \,\, q^{(T_{i+1}-T_{i})\Tr M_{i}}\,
 \prod_{i=1}^{k}\, \ee{-\Tr \td{U}_{i}(R_{i})}\,\, 
 \prod_{i=1}^{k}\, \ee{\Tr R_{i}\,(M_{i}-M_{i-1} )}\,\, \cr
\eea
i.e. a chain of $2k+1$ matrices, instead of a chain of $2(\tmax-\tmin)-1$ as we had before.
\et

\medskip


Notice that knowing $\hat X_i$, we can reconstruct $\hat X(z,t)$ for all times $t\in [T_i,T_{i+1}]$:
\beq\label{XtzXi}
\hat X(z,t) = \hat X_i(z) + \sum_{t'=T_i+{1\over 2}}^{t-{1\over 2}} U'_{t'}(\hat Y(z,T_{i+1}-{1\over 2})+(t'-T_{i+1}+{1\over 2})\ln q),
\eeq
and if $t'\in [T_i+{1\over 2},T_{i+1}-{1\over 2}]$
\beq\label{YtzYi}
\hat Y(z,t') = \hat Y_{i+1}(z) + (T_{i+1}-{1\over 2}-t')\, \ln q.
\eeq

\section{Liquid  region}

Our spectral curve is thus given by a collection of functions $\hat X(z,t)$ and $\hat Y(z,t+{1\over 2})$, defined for all integer times $t$.
We assume that the potentials $V_t$ are non-zero only for times $t\in \{T_1,T_2,\dots,T_k\}$, and the domain at time $T_i$ is the union of $m_i$ intervals:
\beq
\domain(T_i) = \mathop{{\cup}}_{j=1}^{m_i} [a_{i,j},b_{i,j}],
\eeq
with filling fractions $n_{i,j}$, i.e. we require that there are $n_{i,j}$ particles in $[a_{i,j},b_{i,j}]$.

This implies that the potentials $V(x,T_i)$ are of the form:
\beq
V'(x,T_i) = f'(x,T_i)+\sum_{j=1}^{m_i} \psi(x-a_{i,j})-\psi(x-b_{i,j}) ,
\eeq
where $f'(x,T_i)$ is some analytical function, and $\psi=\Gamma'/\Gamma$ is the digamma function. We used the well known property of $\psi$ that:
\beq\label{psisum1}
\sum_{i=1}^n {1\over x-i} = \psi(x)-\psi(x-n).
\eeq

\subsection{Interpolation to real times}

It is in fact better to enlarge these definitions to all times (not necessarily integers or half-integers) by interpolation:
\beq
\forall\,\,t'\in ]T_{i-1},T_i], \qquad \quad \hat Y(z,t') = \hat Y(z,T_i-{1\over 2}) + (T_i-{1\over 2}-t')\ln q
\eeq
and we write:
\beq
y(z,t) = \ee{\hat Y(z,t)}
\eeq
i.e.:
\beq
\forall\,\,t'\in ]T_{i-1},T_i], \qquad \quad y(z,t') = y(z,T_i-{1\over 2})\,\,q^{T_i-{1\over 2}-t'} = y_i(z) \,\,q^{T_i-{1\over 2}-t'}
\eeq
The function $y(z,t')$ is discontinuous at time $T_i$:
\beq
{y_i(z)\over q^{T_i-T_{i-1}}\, y_{i-1}(z)} = \ee{-V'(\hat X(z,T_i),T_i)}
\eeq

\medskip
Similarly we define for $t\in ]T_{i-1},T_i]$:
\beq\label{Xztalltq}
\hat X(z,t) = \hat X(z,T_{i}) - {1\over 2}\,\sum_{t'=t+{1\over 2}}^{T_i-{1\over 2}} {q^{t'}\alpha(t')/\beta(t')-y_i(z)\,\,q^{T_i-{1\over 2}}\over q^{t'}\alpha(t')/\beta(t') +  y_i(z)\,\,q^{T_i-{1\over 2}}}
\eeq

If $\alpha/\beta$ is constant over some interval, then we can find explicitly the interpolating function:
\beq
\hat X(z,t) = \hat X(z,T_{i}) + {t-T_i\over 2} + \psi_q(-{\alpha\over \beta}\,{1\over y_i(z)}\,q^{t+1-T_i})- \psi_q(-{\alpha\over \beta}\,{q\over y_i(z)})
\eeq
where $\psi_q(x)$ is defined as (see appendix \ref{appGammaq}):
\beq
\psi_q(x) = x\,{g'(x)\over g(x)} = \sum_{n=1}^\infty {q^n\over x -q^{n}}.
\eeq

If $q=1$, we have:
\beq\label{Xztalltq1}
\hat X(z,t) = \hat X(z,T_{i}) + {t-T_i\over 2}\,\,{{\alpha\over \beta}-y_i(z)\over {\alpha\over \beta}+y_i(z)},
\eeq
Notice that this expression depends linearly on $t$.

\subsection{Densities of particles}

Consider the density of particles at time $t$ (in section \ref{secdensity}, we have seen that these densities are related to the profile of the plane partitions):
\beq
\bar\rho(x,t) = \sum_{i=1}^N \left< \delta(x-h_i(t)) \right> = \left< \Tr \delta(x-M_t) \right>.
\eeq
We also consider their Stieltjes transforms, i.e. the resolvents:
\beq
\bar W(x,t) = \sum_{i=1}^N \left< {1\over x-h_i(t)} \right> = \left< \Tr {1\over x-M_t} \right> = \int_{\mathbb R} {\bar\rho(x',t)\,dx'\over x-x'}.
\eeq
Let us assume that our model has a formal large $Q$ expansion in some parameter $Q$, it follows from the general solution of the chain of matrices \cite{EPrats}, that one can write:
 \beq\label{largeQexpWrho}
 \bar W(x,t) = \sum_{g=0}^\infty Q^{1-2g}\, W^{(g)}(x,t)
 \virg
 \bar \rho(x,t) = \sum_{g=0}^\infty Q^{1-2g}\, \rho^{(g)}(x,t).
 \eeq
where the first term $W^{(0)}(x,t)=W(x,t)$ is given by \eq{defWxtResYdx}:
\beq
W^{(0)}(x,t)=W(x,t) = \Res_{z\to \infty} {\hat Y(z,t-{1\over 2})\, d\hat X(z,t)\over x-\hat X(z,t)},
\eeq
and the density is the discontinuity of $W(x,t)$ across ${\rm supp}(t)$:
\beq
{W(x-i0,t)-W(x+i0,t)\over 2i\pi} = \rho(x,t).
\eeq

All the higher corrections $W^{(g)}(x,t)$ for $g\geq 1$ were also computed in \cite{EPrats}, and they are the correlators $W_1^{(g)}$ of the symplectic invariants (see appendix \ref{appspinv} or \cite{EOFg}) of the spectral curve ${\cal S}_t$.
In other words, if we know the spectral curve ${\cal S}_t$ at time $t$, there is a simple recursive algebraic algorithm to compute all the corrections, of any correlation function, to all orders in $Q$.
We shall not detail this here.

\smallskip

\br
Let us mention, that if the arctic circle property does not hold (this may be the case for complicated domains), then the spectral curve ${\cal S}_t$ may depend on $Q$, and thus all the  $W^{(g)}$ depend on $Q$. In other words, \eq{largeQexpWrho} is true as an equality of formal series, but it is not really a large $Q$ expansion. In particular, $W^{(0)}$ is not necessarily the large $Q$ limit.
When the arctic circle property holds, then, the $W^{(g)}$ are independent of $Q$, and we really have a large $Q$ expansion.
We will see examples where this holds, below in section \ref{secexs}.
\er

\bigskip

Let us now study the leading term.

\medskip


$W(x,t)$ can have various sorts of singularities, it can have isolated singularities, like poles, or log singularities, and it can have branchcuts. In terms of the density $\rho(x,t)$,
the poles of $W(x,t)$ correspond to $\delta$-Dirac distributions, and cuts correspond to smooth positive densities. Log singularities of $W(x,t)$ correspond to discontinuous jumps of the density.

Notice that \eq{norminftyWxt} implies that $\rho(x,t)$ is normalized, with total weight $N$:
\beq
\int_{\mathbb R} \rho(x,t)\,dx = N,
\eeq
but if we exclude the isolated singularities, i.e. if we integrate only in the liquid region (the interior of the support), we have:
\beq
\int_{\stackrel{\circ}{{\rm supp}(t)}} \rho(x,t)\,dx \leq  N.
\eeq

Notice also that if $\domain(t)$ is disconnected, i.e. a union of intervals $\domain(t) = \cup_{i=1}^{m_t} [a_i(t),b_i(t)]$, we have required that the filling fraction is fixed:
\beq
\int_{a_i(t)}^{b_i(t)} \rho(x,t)\,dx = n_i(t).
\eeq

\subsection{The Envelope of the liquid region}

The liquid region, is the interior of the support of the density, i.e. it excludes all the isolated singularities of $W(x,t)$, it contains only the cuts.

The complement of the liquid region in $\domain$, is called the "solid" region, it contains the isolated singularities of $W(x,t)$, and possibly their accumulation points.

The boundary of the liquid region is called the envelope of the liquid region. It is given by the branchpoints.

\figureframex{12}{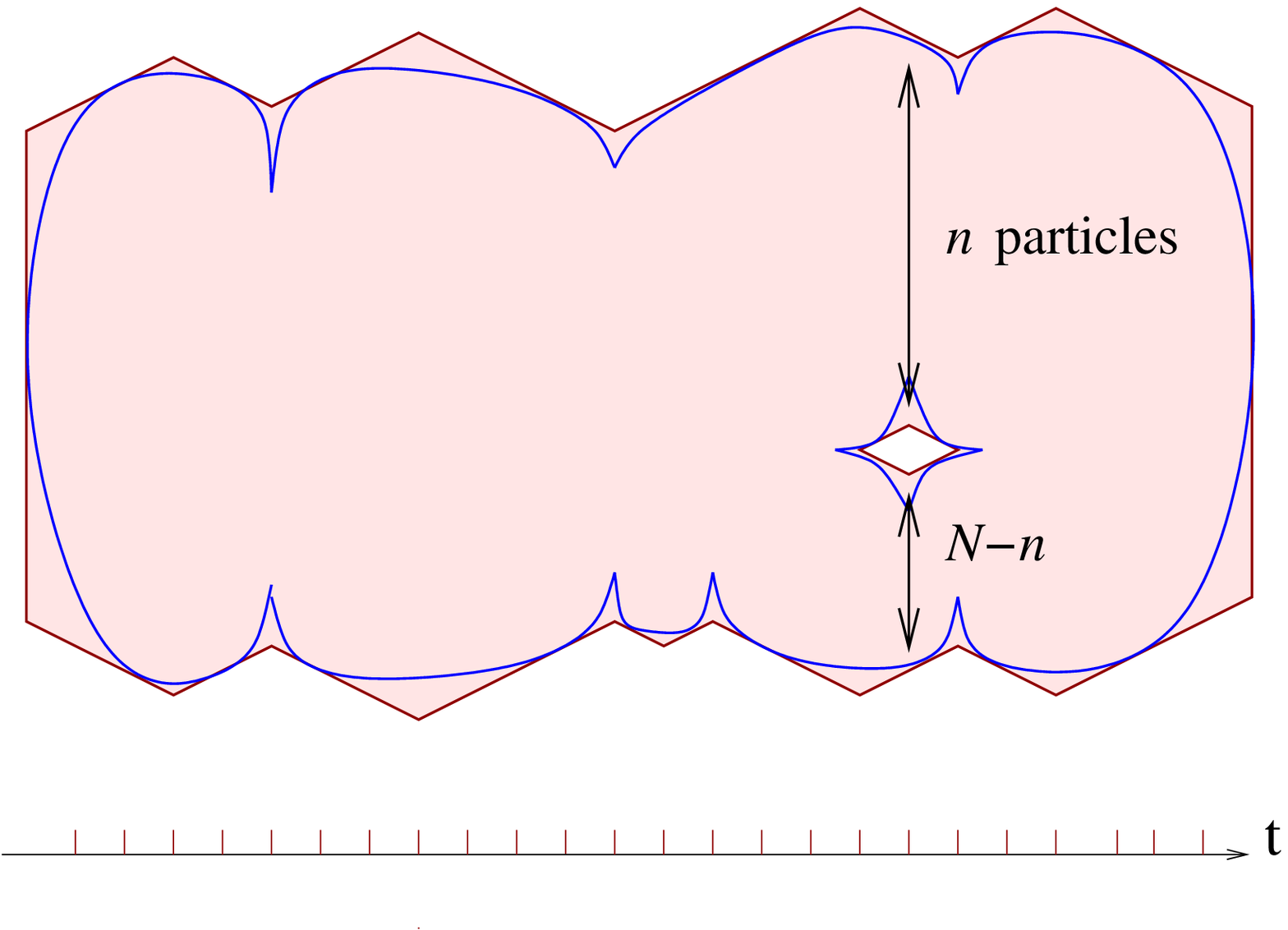}{Given a domain $\domain$, we compute its spectral curve ${\cal S}_t$ for all times $t$. From the spectral curves we compute the densities $\rho(x,t)$, whose supports are contained in $\domain(t)$. The interior of the support is called the "liquid region".
The liquid region is delimited by the envelope, which is a curve $X_c(t)$ contained in $\domain$.
The points on the envelope are branchpoints of the functions $\hat X(z,t)$, i.e. zeroes of $d\hat X(z,t)$
We shall see below that the envelope has some special properties of tangency to the domain $\domain$.
\label{figdomainpoly}}

The branchpoints $z_c(t)$ at time $t$, are zeroes of the differential form $d \hat X$, i.e. they are the points where the density has a vertical tangent:
\beq
d \hat X(z_c(t),t)=0
\eeq
where $d$ is the differential with respect to $z$.
Not all branchpoints are boundaries of the liquid region, let us consider only those which are on the envelope.
The boundary of the liquid region is thus at $X_c(t)$ such that:
\beq
X_c(t) = \hat X(z_c(t),t).
\eeq
By definition, the envelope $X_c(t)$ is a curve contained in $\domain$. Let us study some of its properties.

\subsubsection{Tangency}
\label{sectangents}

\bt
If a point of the envelope has a tangent of slope $\pm 1/2$, then this point must be on the boundary of the shadow of $\domain$.

\et

\figureframex{6}{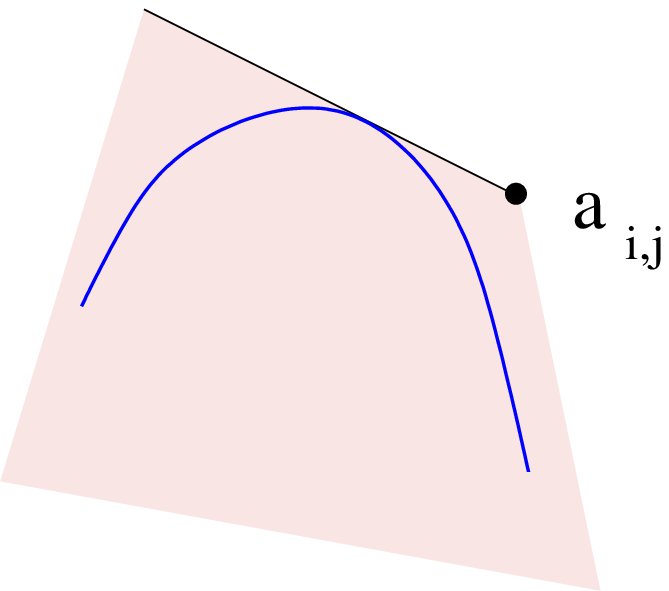}{
A point of the envelope at which the tangent has a slope $\pm 1/2$, is necessarily on the boundary of the shadow of $\domain$, this means that there is a point $a_{i,j}$ of the defect $\defect$ on the tangent.
\label{figenvelopetangent}}

\proof{

Let $\bar t\in ]T_{i-1},T_i]$ be such that the point $(\bar t,X_c(\bar t))$ is a point of the envelope at which the tangent has slope $+1/2$ (resp. $-1/2$).
Let $\alpha = z_c(\bar t)$, i.e. we have $X_c(\bar t)=\hat X(\alpha,\bar t)$.

Since on the envelope we have $d \hat X=0$, we see that at all times $t$ we have:
\beq
{d X_c(t)\over dt} = \left.{\partial \hat X(z,t)\over \partial t}\right|_{z=z_c(t)}
\eeq
and therefore we must have:
\beq
{\partial \hat X(\alpha,t)\over \partial t} = {1\over 2}\quad ({\rm resp.}\, -{1\over 2}),
\eeq
and, from \eq{Xztalltq}, this implies that
\beq
y_i(\alpha) = 0\qquad ({\rm resp.}\, y_i(\alpha)=\infty).
\eeq
Then, if $y_i(\alpha)=0$ (resp. $\infty$), we see from \eq{Xztalltq}, that $\forall\,t\in ]T_{i-1},T_i]$:
\beq
\hat X(\alpha,t) = X_i(\alpha) + {t-T_i\over 2} \qquad ({\rm resp.}\, X_i(\alpha)-{t-T_i\over 2}).
\eeq
and in particular at $t=\bar t$, we have:
\beq
X_c(t) = X_i(\alpha) + {\bar t-T_i\over 2} \qquad ({\rm resp.}\, X_i(\alpha)-{\bar t-T_i\over 2}),
\eeq
which means that the tangent to the envelope at point $\bar t$, passes through the point $a=\hat X(\alpha,T_i)$ at time $T_i$:
\beq
a=\hat X(\alpha,T_i).
\eeq

Then, notice that $y_i(\alpha) = \ee{\hat Y(\alpha,T_i-{1\over 2})}$ can be equal to $0$ or $\infty$, only if $\hat Y(z,T_i-{1\over 2})$ has a singularity at $z=\alpha$.
Because of \eq{spcurveeqred1}, we see that, singularities of $\hat Y(z,T_i-{1\over 2})$ can occur only if $V'(a,T_i)$ has a singularity.

Remember that $V'(x,t)$ has no singularity on $\domain(t)$, it has singularities only at the defects $\defect(t)$, and thus, this means that the point $(a,T_i)$ belongs to the defect $\defect$.
But we know that the point $(X_c(\bar t),t)$ belongs to $\domain$, and the slope between the points $(a,T_i)$ and $(X_c(\bar t),t)$ is equal to $\pm {1\over 2}$, i.e. the point $(X_c(\bar t),t)$ is in the shadow of the defect $(a,T_i)$.

}

\subsubsection{Local Convexity}

In the classical case, where $q^t\alpha(t)/\beta(t)$ is constant, we have a convexity property:
\bt
If $q^t\alpha/\beta$ is constant in time, the liquid region is locally convex.
\et

\proof{
The density at time $t$ must have a real support, starting at a branchpoint, i.e. at $X_c(t)$.
Consider that, up to a reparametrization of $z$, and up to trivial translations in $t$ and $\hat X$, we can write localy, near the branchpoint point $X_c(0)$ we write:
\beq
\hat X(z,t) = X_c +  X_t t+ {1\over 2}X_{z,z} z^2  + X_{z,t} z t + \dots
\eeq
and up to a rescaling of $z$, we assume $X_{z,z}=1$.
Notice from \eq{Xztalltq1}, that the assumption $q^t\alpha/\beta$ constant guarantees that $\hat X(z,t)$ is linear in $t$, there is no $t^2$ term.

The branchpoint at time $t$ is at $z_c(t)$ given by $d\hat X(z,t)/dz=0$, i.e.:
\beq
z_c(t) = - X_{z,t}\,\,t + \dots
\eeq
i.e., the envelope $X_c(t)=\hat X(z_c(t),t)$ is, up to order $2$:
\beq
X_c(t) = X_c + X_t t -{X_{z,t}^2\over 2}\,\,t^2 + O(t^2)
\eeq
Our assumption on the reality of the supports, implies that $X_c(t)\in \mathbb R$ for all $t$, and therefore  $X_t$ and $X_{z,t}^2$ must be real quantities.

The point $\bar z$ such that $X(z,t)=X(\bar z,t)$ is:
\beq
\bar z =  2 z_c(t) -z + \dots 
\eeq

The function $y(z)$ is then given by solving \eq{Xztalltq}:
\bea
y(z,t') &=& q^{t'-T_i+{1\over 2}}\, {\alpha(t')\over \beta(t')}\,\, {1-2(X_t+X_{z,t}z+\dots)\over 1+2(X_t+X_{z,t}z+\dots)} \cr
&=& q^{t'-T_i+{1\over 2}}\, {\alpha(t')\over \beta(t')}\,\,{1-2X_t\over 1+2 X_t}\,\,\left(1-{4X_{z,t}\, z\over 1-4X_t^2} + \dots\right)
\eea

and therefore
\beq
{y(z,t')\over y(\bar z,t')} \sim 1- {8\,X_{z,t}\over 1-4 X_t^2}\,\,(z-z_c(t)) + \dots
\eeq
and the density $\rho(x,t) = {1\over 2i\pi}\,\ln{{y(\bar z)\over y(z)}}$, with $x=\hat X(z,t)$ is, up to order $2$:
\beq
\rho(x,t) 
= {4\,X_{z,t}\over i\pi\,(1-4 X_t^2)}\,\,(z-z_c(t)) + \dots
= {4\,\sqrt{2}\over \pi\,(1-4 X_t^2)}\,\,\sqrt{X_{z,t}^2\,(X_c(t)-x)} + \dots
\eeq
This implies that $X_{z,t}^2\,(X_c(t)-x)\geq 0$ in the liquid region.
If the liquid region is above $X_c(t)$, we must have $X_{z,t}^2<0$, and if the liquid region is below $X_c(t)$, we must have $X_{z,t}^2>0$. 
In all cases, the liquid region is locally convex.
}

\br
If we are not in the classical case, i.e. if $q^t\alpha/\beta$ is not constant, then $\hat X(z,t)$ is not linear in $t$, there maybe a second derivative $X_{t,t}\neq 0$, and the envelope is locally:
\beq
X_c(t) = X_c + X_t t +{X_{t,t}-X_{z,t}^2\over 2}\,\,t^2 + O(t^2)
\eeq
and although $X_{z,t}^2$ has a constant sign, $X_{t,t}-X_{z,t}^2$ can have any sign, and the envelope is not necessarily convex.
\er

\subsubsection{Cusps}

Notice that $\hat Y(z,t)$ is discontinuous at $t=T_i$, i.e. there is a left value and a right value.
This implies that $dX_c/dt$ has a right tangent and left tangent, and thus generically, the envelope has cusps at $t=T_i$.

\figureframex{6}{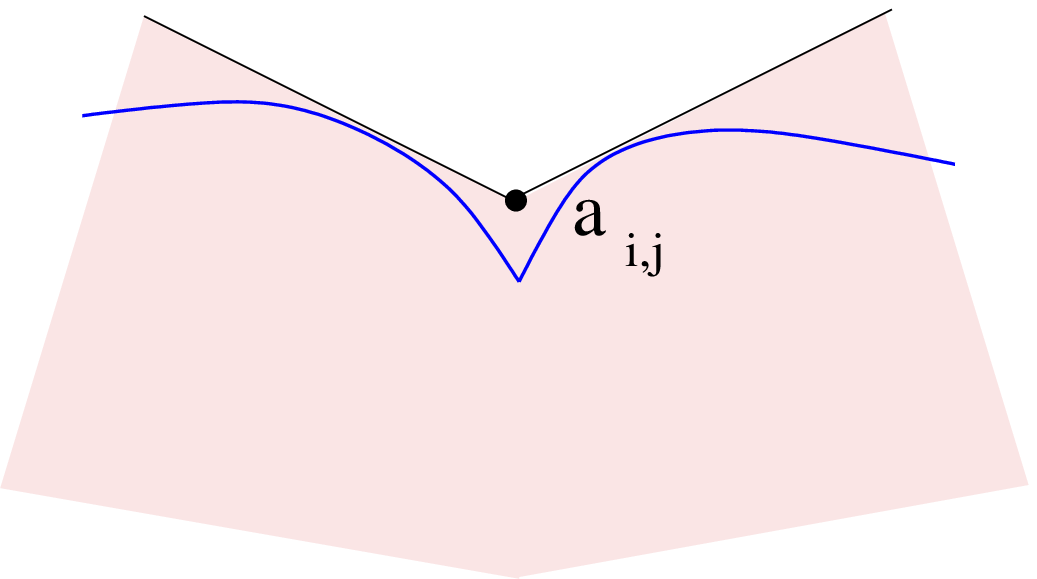}{The envelope has a cusp at $t=T_i$.
\label{figenvelopetangent2}}

\subsubsection{Genus and holes}

Condition 6 of section \ref{secrecipespcurve}, i.e. \eq{defholesff} tells us that the spectral curve must have at least as many non trivial cycles as the number of holes in our domain, i.e. the genus of our spectral curve is at least the number of holes, and generically, it coincides with the number of holes.

In particular, if we consider a simply connected domain $\domain$ with no holes, we can expect to have genus $0$, which means that the functions $\hat X(z,t)$ and $\hat Y(z,t)$ are analytical functions of a complex variable $z$. The Riemann surface in which $z$ lives can be chosen as a domain of the complex plane.

\section{Asymptotic regimes}

Of course, our model depends on so many parameters (shape of the domain, coefficients $\alpha(t'), \beta(t')$, potentials $V_t$, ...) that it is almost impossible to classify all possible asymptotic regimes.
However, a few asymptotic regimes are more relevant for applications in statistical physics or algebraic geometry.

We classify them into 2 kinds:

$\bullet$ macroscopic asymptotic regimes, which describe the behavior of the partition function, i.e. the statistics of our self-avoiding particles model at the scale of the size of the domain.
Typically, we shall consider asymptotic behaviors for large size, of for small $\ln{q}$.

$\bullet$ microscopic asymptotic regimes, which describe the statistics of our self-avoiding particles model particles in a very small region of the domain, typically, the behavior in the bulk, or near the edges, especially near special points, like near the envelope, or near cusps of the envelope.

\smallskip

Let us comment a few of them, and let us emphasize that our method, works for all possible asymptotic regimes, and it gives not only the leading order asymptotics, it gives the full asymptotic expansion to all orders.

\subsection{Classical case q=1, and large size asymptotics}

Classical means that we choose $q=1$.
Here, we shall assume that the weights $\alpha(t')=\beta(t')=1$ are constant, although the general case is doable by the same methods.

Consider a domain $\domain$ whose size is $T=\tmax-\tmin$.
Assume that there are $k$ defects, and $k$ does not depend on $T$:
\beq
k=O(1).
\eeq
The defects are at times $T_i=T \tau_i$, where $\tau_i$ are independent of $T$, and the defect at time $T_i$ is the union of $m_i$ intervals:
\beq
\defect(T_i) = \cup_{j=1,\dots,m_i} [T a_{i,j},T b_{i,j}]
\eeq
where again $a_{i,j}$ and $b_{i,j}$ don't depend on $T$.

Assume also that the number of particles $N$ scales like $T$:
\beq
N = {\mathcal N}\,T,
\eeq
and the number of particles $n_{i,j}$ in each interval $[a_{i,j},b_{i,j}]$ scales like $T$:
\beq
n_{i,j} = {\mathcal N}_{i,j}\, T,
\eeq
and of course $\forall\, i$
\beq
\sum_j {\mathcal N}_{i,j} = {\mathcal N}.
\eeq
See fig \ref{figdomainpoly1}.

\figureframex{12}{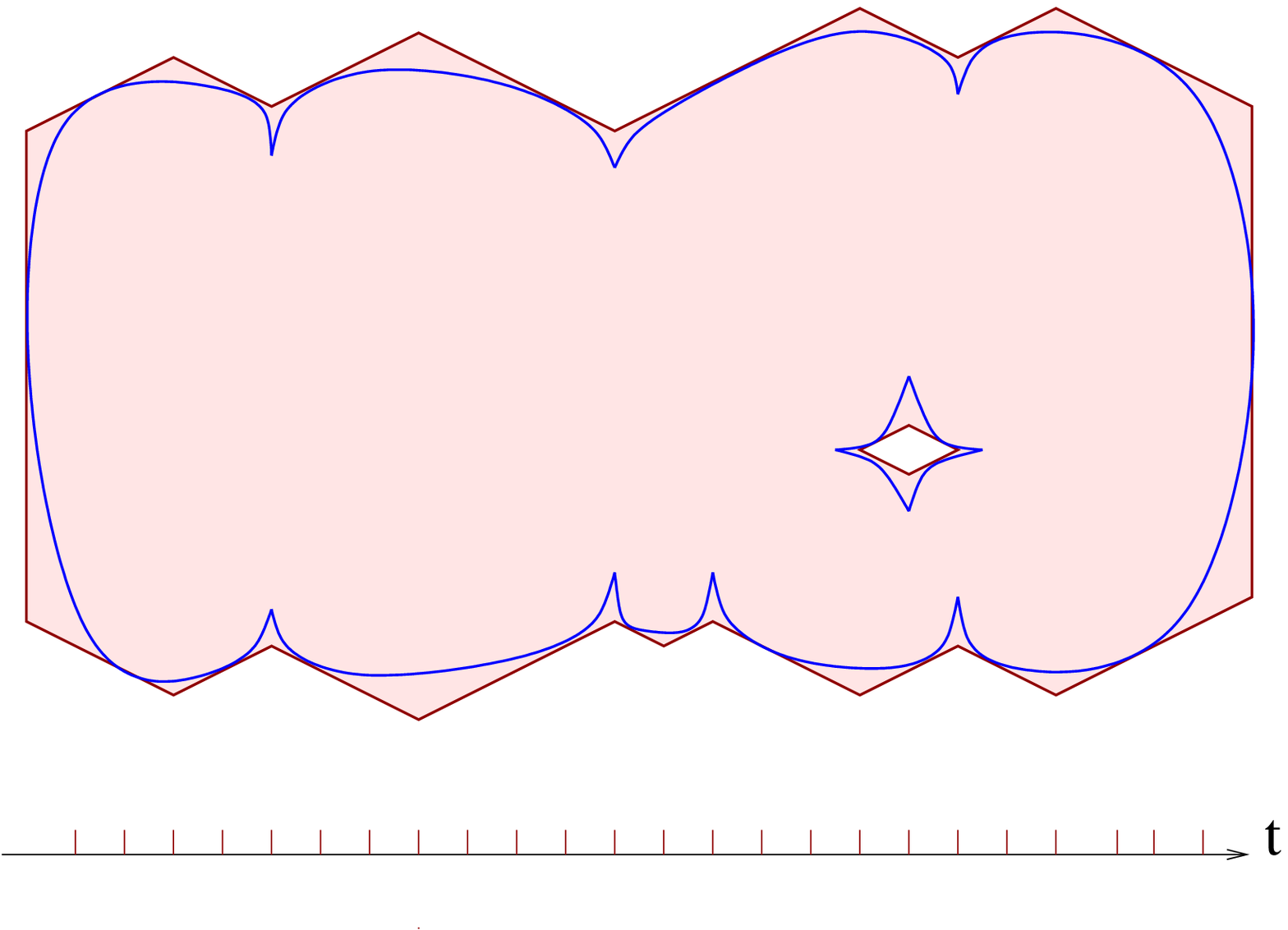}{A polygonal domain with $k$ defect at times $T_i=T\tau_i$.
The envelope of the spectral curve, is the algebraic curve of smallest degree, tangent to all boundaries.
\label{figdomainpoly1}}

\subsubsection{The spectral curve}

The spectral curve equations \eq{spcurveeqred1} are:
\beq\label{loopeqcllimit1}
\left\{\begin{array}{l}
\hat X_{i}-\hat X_{i-1} = -{T\over 2}\,(\tau_{i}-\tau_{i-1})\, \tanh{\hat Y_i\over 2}
\cr
\hat Y_{i+1}-\hat Y_i =  - V'_{i}(\hat X_i) 
\end{array}\right.
\eeq
where we choose the potentials $V_i$ according to \eq{Vtdefect0sum}:
\beq
V'_i(X) =  f'(X)+\sum_{j=1}^{m_i} \psi(X-T a_{i,j})-\psi(X-T b_{i,j}) 
\eeq
where $f'$ has no singularity, and $\psi=\Gamma'/\Gamma$ is the digamma function. We used the well known property of $\psi$ that:
\beq\label{psisum}
\sum_{i=1}^n {1\over x-i} = \psi(x)-\psi(x-n).
\eeq

Moreover, the last matrix of the chain $M_k=M(\tmax)={\rm diag}(\l_j)$ is not integrated upon. Its spectrum is $\domain(\tmax)$, i.e. it is delimited by the intervals $[T a_{k,j},T b_{k,j}]$.
Then, equation \eq{spcurveeqef1} reads:
\beq
\hat Y_k \,\,\,\mathop{{\sim}}_{X_k\to {\rm e.v.\, of\,}M_k}  \quad \sum_{j=1}^{m_k} \sum_{i=T a_{k,j}}^{T b_{k,j}} {1\over X_k-i}
= \sum_{j=1}^{m_k} \psi(X_k-T a_{k,j})-\psi(X_k-T b_{k,j}) 
\eeq

\medskip

\subsubsection{Rescaling}

Since our model is defined as a formal power series, and we are looking for the large $T$ expansion, we should try to find the spectral curve in the large $T$ expansion.
In that purpose we rescale the variables:
\beq
\hat X_i = T x_i
\virg
\hat Y_i = \ln y_i,
\eeq
and we use Stirling's asymptotic formula for the $\Gamma$ and $\psi$ functions, see appendix \ref{appGammaq}:
\beq
\psi(x) \sim \ln{x} - {1\over 2x} - \sum_{n=1}^\infty {B_{2n}\over 2n\, x^{2n}} ,
\eeq
where $B_n$ are the Bernouilli numbers.

That gives:
\beq\label{cllimyiyi1}
{y_{i+1}\over y_i} = \prod_{j=1}^{m_i}\, {x_i-b_{i,j}\over x_i-a_{i,j}}\,\, (1+ {\rm regular} + O(1/T))
\eeq
where the $O(1/T)$ term is, to each power of $T^{-1}$, a rational function of $x_i$ with poles at $a_{i,j}$ or $b_{i,j}$, plus possibly an arbitrary analytical function with no singularity.
We also have:
\beq\label{cllimyiyik}
y_k \sim \prod_{j=1}^{m_k}\, {x_k-a_{k,j}\over x_k-b_{k,j}}\,\, (1+ {\rm regular}+ O(1/T))
\eeq
where the $O(1/T)$ term is, to each power of $T^{-1}$, a rational function of $x_k$ with poles at $a_{k,j}$ or $b_{k,j}$, plus possibly an arbitrary analytical function with no singularity.

We also have from \eq{loopeqcllimit1}:
\beq\label{eqxiyiredasympcl}
x_{i}-x_{i-1} = {\tau_{i}-\tau_{i-1}\over 2}\, \, {1-y_i\over 1+y_i}
\eeq
i.e.
\beq\label{eqyixiredasympcl}
y_i = {\tau_i-\tau_{i-1} -2 (x_i-x_{i-1})\over \tau_i-\tau_{i-1} + 2 (x_i-x_{i-1})}.
\eeq

\subsubsection{Singularities}

We see from \eq{cllimyiyi1},and \eq{cllimyiyik}, that if $x_i=a_{i,j}$, the ratio $y_i/y_{i+1}$ has a zero, and thus either $y_i$ has a zero, or $y_{i+1}$ has a pole, or both.

In fact, since $y_i$ and $y_{i+1}$ are multivalued function of $x_i$, there might be several points on the spectral curve such that $x_i=a_{i,j}$.
In that case, it may happen that $y_i=0$ corresponds to $x_i=a_{i,j}$ in one sheet, and $y_{i+1}=\infty$ corresponds to $x_i=a_{i,j}$ in another sheet. 
Or if both $y_i$ has a zero, and $y_{i+1}$ has a pole at the same point, this means that $x_i-a_{i,j}$ has a double zero, i.e. $dx_i$ has a zero, and this means that we are at a branchpoint, i.e. the point where two sheets meet.

Therefore, we shall assume that generically both possibilities occur, and we conclude that:

$\bullet$ $y_i$ has a zero when $x_i=a_{i,j}$ or $x_{i-1}=b_{i-1,j}$, and has a pole when $x_i=b_{i,j}$ or $x_{i-1}=a_{i-1,j}$.


\medskip

Since the $x_i$'s and $y_i$'s have only meromorphic singularities, it is  natural to look for an algebraic solution, i.e. such that $x_i$ and $y_i$ are  meromorphic functions on an algebraic curve ${\cal L}$.
Indeed, our freedom to choose the potential $V_t$ can be used to eliminate all the singularities except those described above.

\medskip

The problem then consists in finding algebraic functions $x_i$, $y_i$, such that:

$\bullet$ $x_{i}-x_{i-1} = {\tau_{i}-\tau_{i-1}\over 2}\, \, {1-y_i\over 1+y_i}$,

$\bullet$ $y_i$ has a simple zero when $x_i=a_{i,j}$ or $x_{i-1}=b_{i-1,j}$, and has a simple pole when $x_i=b_{i,j}$ or $x_{i-1}=a_{i-1,j}$.

$\bullet$ the density measures ${1\over 2i\pi}\,\ln{\left({y_i(\bar z)\over y_i(z)}\right)}\, dx_i(z)$ (where $x_i(z)=x_i(\bar z)$ are the two points on each side of the support) must have their supports in $\domain$ (in particular the supports are real).
Those reality  conditions for an algebraic curve, are closely related to the conditions which defines the  Harnack curve in Kenyon-Okounkov-Sheffield \cite{KOS}.

\subsubsection{The large size limit: Harnack curve}

So, we have to find a spectral curve, and its envelope, satisfying many equations (which give the poles and zeroes), as well as many reality conditions.
Since we have only meromorphic types of singularities (poles), it is natural to look for real algebraic spectral curves.

It is also natural to look for minimal degree algebraic curves, i.e. having just the number of poles and zeroes implied by our equations and no other poles. In fact introducing other poles would most probably break the condition on the homology of steepest descent paths.
Also we can again use our freedom to choose the potential $V_t$ in order to reduce the degree, so that we eliminate all the singularities except those described above.

\medskip

An envelope and spectral curve having all the required properties can be found from the work of Kenyon-Okounkov-Sheffield \cite{KOS}, and they proved that their envelope is indeed the limit shape of the liquid domain to leading order at large $T$.

Their spectral curve is a Harnack curve, and the envelope $x_c(\tau)$ is an algebraic curve given by a polynomial equation
\beq
P(x_c,\tau)=0
\eeq
where $P$ is a real polynomial, which has some very special properties.

In particular, it satisfies all the properties we want for our envelope, and also, being a maximal Harnack curve means that the area enclosed by the envelope is maximal.
Another property, is that the genus of $P$, is exactely the number of holes of the domain in the envelope.

\subsubsection{Recipe for an algebraic spectral curve}
\label{recipeclassicalspcurve}

Therefore, let us make the assumption that the spectral curve is the algebraic curve of smallest degree satisfying all the constraints. 
We shall discuss the consistency of that assumption in section \ref{secLSAE} below.

The problem we have to solve in order to find the spectral curve is then:

\begin{itemize}

\item Find a Riemann surface whose genus is equal to the number of holes in $\domain$.

\item Find 2 meromorphic functions $u(z)$ and $v(z)$, with only simple poles, and let us denote the set of poles as $\infty_{i,j}$, $i=0,\dots,k$, $j=1,\dots,m_i$.

\item We denote 
\beq
x(z,\tau) = u(z) + \tau v(z),
\eeq
and $\forall\, i$:
\beq
x_i(z) = u(z) + \tau_i v(z).
\eeq
and we require  that $\forall\, i,j$ (cusp condition):
\beq
\Res_{\infty_{i,j}} x_i(z) =0
\eeq

\item The meromorphic functions $u(z)$ and $v(z)$ must be such that $\forall\, i,j$:
\beq
{\rm if}\, i<k\quad \exists\, z\qquad {\rm such\, that}\qquad
x_i(z)=a_{i,j}
\qquad {\rm and} \qquad
v(z)=-{1\over 2},
\eeq
\beq
{\rm if}\, i>0\quad \exists\, z\qquad {\rm such\, that}\qquad
x_i(z)=a_{i,j}
\qquad {\rm and} \qquad
v(z)=+{1\over 2},\eeq
\beq
{\rm if}\, i<k\quad \exists\, z\qquad {\rm such\, that}\qquad
x_i(z)=b_{i,j}
\qquad {\rm and} \qquad
v(z)=+{1\over 2},\eeq
\beq
{\rm if}\, i>0\quad \exists\, z\qquad {\rm such\, that}\qquad
x_i(z)=b_{i,j}
\qquad {\rm and} \qquad
v(z)=-{1\over 2}.
\eeq

\item and we have the filling fractions:
\beq
{1\over 2i\pi} \, \int_{[a_{i,j},b_{i,j}]} {x_i(z)\over y(z)}\, dy(z) = {\cal N}_{i,j}
\eeq
where
\beq
y(z) = {1-2v(z)\over 1+2v(z)},
\eeq
and where we assume that the number of particles in $[a_{i,j},b_{i,j}]$ is $n_{i,j}=T\,{\cal N}_{i,j}$.

\end{itemize}

\bigskip

Explicit examples for given domains (hexagon, cardioid) are treated in section \ref{secexs}.

\subsubsection{Envelope of the liquid region}

The envelope of the liquid region corresponding to the spectral curve above, is obtained as follows:

the branchpoints $z_c(\tau)$ are solutions of $dx(z_c,\tau)=0=du(z_c)+\tau dv(z_c)$, and in fact, it is much easier to compute $\tau$ as a function of $z_c$, namely:
\beq
\tau = - \,{du(z_c)\over dv(z_c)}.
\eeq
Then, one computes $x_c(\tau)=x(z_c(\tau),\tau)$, and again, it is easier to parametrize this equation by $z_c$, that gives:
\beq
x_c(\tau) \,\,\, = \,\,\,
\left\{\begin{array}{l}
\tau = - \,{du(z_c)\over dv(z_c)} \cr
x_c = u(z_c) -  v(z_c)\, {du(z_c)\over dv(z_c)}
\end{array}\right. \, .
\eeq
We see that the envelope is an algebraic curve, and from section \ref{sectangents},we know that it is an algebraic curve tangent to the boundary of the shadow of $\domain$, with cusps at $\tau=\tau_i$, and whose genus is the same as the number of holes of $\domain$.

\medskip

Again, explicit examples of envelopes are given in section \ref{secexs}. For example, the envelope of an hexagonal domain is the ellipse tangent to all sides of the hexagon, see fig \ref{fighexagon}.

\bigskip
{\bf Recovering the spectral curve from the envelope:}
\medskip

Assume that we know the envelope $x_c(\tau_c)$, i.e. the algebraic curve tangent to the boundaries of $\domain$.
Let us explain how to recover the full spectral curve $x(z,\tau)$.

\medskip

Given the equation of the envelope $x_c(\tau_c)$ (which is a multivalued algebraic function), one can choose locally $z=\tau_c$ as a local parameter.
One finds that the spectral curve is (at least in the domain of $\tau$ where $\tau_c$ can be chosen as a local parameter):
\beq
\left\{
\begin{array}{l}
x(z,\tau) = x_c(z) + (\tau-z)\, v(z),   \cr
y(z) = {1-2 v(z)\over 1+2 v(z)}   \virg
{\rm where}\quad v(z) = {dx_c(z)\over dz}.
\end{array}
\right.
\eeq

Therefore, knowing the envelope allows to recover the full spectral curve, i.e. the functions $x(z,\tau)$ and $y(z)$.

\subsubsection{Large size Asymptotic expansion}
\label{secLSAE}

Now, suppose that we have found those functions, and that we have indeed found the correct spectral curve.
The spectral curve is then the pair of functions ${\cal S}_t=(\hat X(z,t),\hat Y(z,t))$, i.e., up to a symplectic transformation:
\beq
{\cal S} = (T u(z), \ln{(y(z))}),
\eeq
and we notice that $u(z)$ and $y(z)$ do not depend on $T$, therefore we have $F_g(Tu,\ln y)=T^{2-2g} F_g(u,\ln y)$.

And thus, from theorem \ref{thZtopchmattasep}, we have the full large $T$ asymptotic expansion:
\begin{conjecture}
$\ln{\cal Z}$ has the large size $T$ expansion
\beq
\ln{\cal Z} \sim \sum_{g=0}^\infty T^{2-2g}\, F_g(u,\ln y).
\eeq
where the meromorphic functions $u(z)$ and $y(z)$ are computed by solving the requirements of section \ref{recipeclassicalspcurve}, and $F_g$ is the $g^{\rm th}$ symplectic invariant defined in \cite{EOFg}.
\end{conjecture}

{\bf sketch of a possible proof:}

A hint to that conjecture, is that the leading large $T$ densities are governed by $F_0(u,\ln y)$, and it can be seen that this coincides with the limit shape found by Kenyon-Okounkov-Sheffield \cite{KOS}.

\smallskip

In order to prove this conjecture, relying on the work of \cite{EPrats}, we only have to prove that we have indeed found the correct spectral curve, and that our guess (that $u$ and $v$ are algebraic functions of smallest possible degree) is correct.

In principle, this means proving that the integration paths in our self-avoiding particles matrix model, are indeed the steepest descent paths for our potentials, but this seems too difficult. For matrix models with polynomial potentials, this is usually proved by the Riemann-Hilbert method of Deift \& co \cite{DKMVZ}.

\smallskip

Another possible proof, is to prove this order by order in some formal parameter, especially if the model tends to a Gaussian matrix integral in the small parameter limit.

\smallskip

One suggestion is to notice that in our matrix model, the potentials $V_t$ may depend on $T$, and our spectral curve can be expected to depend on $T$, and somehow, the Harnack curve gives only the leading term:
\beq
{\cal S}(T) = {\cal S}_{\infty} + O(1/T)
\eeq
where ${\cal S}_\infty$ is given by the Harnack curve of \cite{KOS}.
Indeed, we have not really taken into account the $O(1/T)$ term in \eq{cllimyiyi1} and \eq{cllimyiyik}.

\medskip

However, we have some freedom in the choice of $V_t$, and we can choose any $V_t$ provided  that the spectral curve satisfies \eq{cllimyiyi1} and \eq{cllimyiyik}, and in particular, we may choose the spectral curve ${\cal S}_\infty$, which does satisfy \eq{cllimyiyi1} and \eq{cllimyiyik}, with the $O(1/T)$ term vanishing.
In other words, we use the freedom of choosing the potentials $V_t$, and we define $V_t$ from the spectral curve through \eq{cllimyiyi1}, instead of the contrary.

Somehow, we go backwards. we first construct the spectral curve ${\cal S}_\infty$, and then we construct the potentials $V_t$ which correspond to it.

This very special choice of $V_t$ guarantees that ${\cal S}_\infty$ is the spectral curve of our model, and then it is independent of $T$ $\,\square$.

\subsection{Quantum case}

Now we consider $q\neq 1$.
We also assume for simplicity  that the weights $\alpha(t')=\beta(t')=1$ are constant.

The potential $\td{U}_i(Y)$ appearing in \eq{deftdU} is:
\bea
\ee{-\td U_i(Y)} 
&=&  \ee{- (T_i-T_{i-1}){Y\over 2}}\,\,  {g(-\ee{-Y}\,q^{T_i-T_{i-1}})\over g(-\,\ee{-Y})} \cr
\eea
where $g(y)=\prod_{j=1}^\infty (1-q^j/y)$ is the $q-$product (it is a quantum deformation of the $\Gamma$-function, see appendix \ref{appGammaq}).
And thus:
\beq
\td U_i'(Y) = {T_i-T_{i-1}\over 2} \, +\psi_q(-\ee{-Y}) -\psi_q(-\,\ee{-Y}\,q^{(T_i-T_{i-1})})
\eeq
where $\psi_q(x) = x g'(x)/g(x) = \sum_{j=1}^\infty q^j/(x-q^j)$.

Thus,  \eq{spcurveeqred1} gives:
\beq\label{eqXiYiasympq}
\hat X_i - \hat X_{i-1} = {T_i-T_{i-1}\over 2} + \sum_{j=0}^{T_i-T_{i-1}-1}\, {1 \over 1+q^j\,\ee{-\hat Y_i}}
\eeq
and at intermediate times $T_{i-1}\leq t\leq T_i$:
\beq
\hat X(z,t) = \hat X_i(z)  + {t-T_i\over 2} - \sum_{j=0}^{T_i-t-1}\, {1 \over 1+q^j\,\ee{-\hat Y_i(z)}}
\eeq

\smallskip

We choose the same domain $\domain$ as in the classical case, i.e. a domain which scales with a factor $T=\tmax-\tmin$.

Several possibilities may occur:
\begin{itemize}
\item The regime $T\ll {1\over \ln q}$, is more or less the same as $q=1$, which we have studied in the previous section, i.e. there is a liquid region of typical size $T$.

\item In the regime $T\gg {1\over |\ln q|}$,  there is a liquid region of typical size $1/|\ln q|$, and most of the domain $\domain$ is in a frozen phase.

\item In the intermediate regime $q^T\sim O(1)$, the liquid phase if of typical size $T\sim {1\over \ln q}$. This is the most interesting regime.

\end{itemize}

\subsection{Case $q^T\sim O(1)$}

We consider the regime where $T$ is large and $\ln q$ is small, and $q^T\sim O(1)$.
We define:
\beq
q^T = \qq
\eeq

Then we shall repeat most of the steps of the classical case $q=1$.
First we rescale:
\beq
x_i = q^{\hat X_i}
\virg
y_i=\ee{\hat Y_i}
\virg
\tau = t/T.
\eeq

\subsubsection{Equation of the spectral curve}

The equations \eq{spcurveeqred1} read:
\beq
{y_{i+1}\over y_i} = \qq^{(\tau_{i+1}-\tau_i)}\,\, 
\prod_{j=1}^{m_i}\, {x_i-\qq^{b_{i,j}}\over x_i-\qq^{a_{i,j}}}\,\, (1+ {\rm regular} + O(\ln q))
\eeq
\beq
y_k = \prod_{j=1}^{m_k}\, {x_i-\qq^{a_{i,j}}\over x_i-\qq^{b_{i,j}}}\,\, (1+ {\rm regular} + O(\ln q))
\eeq
And to leading order at small $\ln q$ we have:
\beq
{x_{i-1}\over x_i} = \qq^{-{\tau_i-\tau_{i-1}\over 2}} \,\,{y_i+\qq^{\tau_i-\tau_{i-1}} \over y_i+1} \,\, \left( 1+O(\ln q)) \right)
\eeq

Notice that those equations imply only meromorphic singularities for $y_i$ and $x_i$, and again, it is natural to look for an algebraic spectral curve.
Notice that if $y_i$ has a zero (resp. a pole) at $z$, then we have $x_{i+1}(z)=\qq^{- (\tau_{i+1}-\tau_i)} x_i(z)$ (resp. $x_{i-1}(z)=\qq^{ (\tau_{i-1}-\tau_i)} x_i(z)$).

\subsubsection{Recipe for an algebraic spectral curve}
\label{recipeclassicalspcurveq}

Therefore, let us make the assumption that our functions $x_i(z)$ are algebraic of smallest possible degree satisfying all the constraints.
The consistency of that assumption can be discussed like in section \ref{secLSAE} for $q=1$, and we discuss it again in section \ref{secLSAEq} below.

The problem we have to solve in order to find the spectral curve is then:

\begin{itemize}

\item Find a Riemann surface whose genus is equal to the number of holes in $\domain$.

\item Find 2 meromorphic functions $u(z)$ and $v(z)$, with only simple poles, and let us denote the set of poles as $\infty_{i,j}$, $i=0,\dots,k$, $j=1,\dots,m_i$.

\item We denote 
\beq
x(z,\tau) = \qq^{\tau/2} u(z) + \qq^{-\tau/2} v(z),
\eeq
and $\forall\, i$:
\beq
x_i(z) = x(z,\tau_i)= \qq^{\tau_i/2} u(z) + \qq^{-\tau_i/2} v(z).
\eeq
and we require that $\forall\, i,j$ (cusp condition):
\beq
\Res_{\infty_{i,j}} x_i(z) =0
\eeq

\item The meromorphic functions $u(z)$ and $v(z)$ must be such that $\forall\, i,j$ (tangency conditions):
\beq
{\rm if}\, i<k\quad \exists\, z\qquad {\rm such\, that}\qquad
x_i(z)=\qq^{a_{i,j}}
\qquad {\rm and} \qquad
x_{i+1}(z)=\qq^{a_{i,j}-{\tau_{i+1}-\tau_i\over 2}},
\eeq
\beq
{\rm if}\, i>0\quad \exists\, z\qquad {\rm such\, that}\qquad
x_i(z)=\qq^{a_{i,j}}
\qquad {\rm and} \qquad
x_{i-1}(z)=\qq^{a_{i,j}+{\tau_{i-1}-\tau_i\over 2}},
\eeq
\beq
{\rm if}\, i<k\quad \exists\, z\qquad {\rm such\, that}\qquad
x_i(z)=\qq^{b_{i,j}}
\qquad {\rm and} \qquad
x_{i+1}(z)=\qq^{b_{i,j}+{\tau_{i+1}-\tau_i\over 2}},
\eeq
\beq
{\rm if}\, i>0\quad \exists\, z\qquad {\rm such\, that}\qquad
x_i(z)=\qq^{b_{i,j}}
\qquad {\rm and} \qquad
x_{i-1}(z)=\qq^{b_{i,j}-{\tau_{i-1}-\tau_i\over 2}}.
\eeq

\item and we have the filling fractions:
\beq
{1\over 2i\pi} \, \int_{[a_{i,j},b_{i,j}]} \hat Y_i(z)\, d\hat X_i(z) = n_{i,j}
\eeq
where $x_i=q^{\hat X_i}$ and $\hat Y_i$ is obtained by solving \eq{eqXiYiasympq}.

\end{itemize}

\bigskip

Explicit examples for given domains (hexagon) are treated in section \ref{secexs}.

\subsubsection{Envelope}

The envelope is given by the branchpoints $z_c(t)$ solutions of $dx(z_c,\tau)=0$, and again, it is easier to find $\tau$ as a function of $z_c$ than the contrary. We have:
\beq
q^{t}=\qq^{\tau} = -\,{dv(z)\over du(z)}.
\eeq
The envelope $X_c(t)$ is also better given in a parametric form with the parameter $z_c$ as:
\beq
\left\{
\begin{array}{l}
q^{t} = -\,{dv(z_c)\over du(z_c)} \cr
\cr
q^{X_c} = q^{t/2}\,\, u(z_c)\, +\,  q^{-t/2}\,\, v(z_c)
\end{array}
\right. \, \, .
\eeq
Since $u$ and $v$ are meromorphic, this implies that $q^t$ and $q^{X_c}$ are related by a polynomial equation:
\beq
P(q^{t},q^{X_c})=0.
\eeq
Again, we claim that this polynomial is the same as the Harnack curve of Kenyon-Okounkov-Sheffield \cite{KOS}.

In other words, the plane curve $x_c(q^t)=q^{X_c(t)}$, is an algebraic plane curve inscribed in the image of the domain $\domain$ under the map $(x,t)\to(q^x,q^t)$.

\figureframex{14}{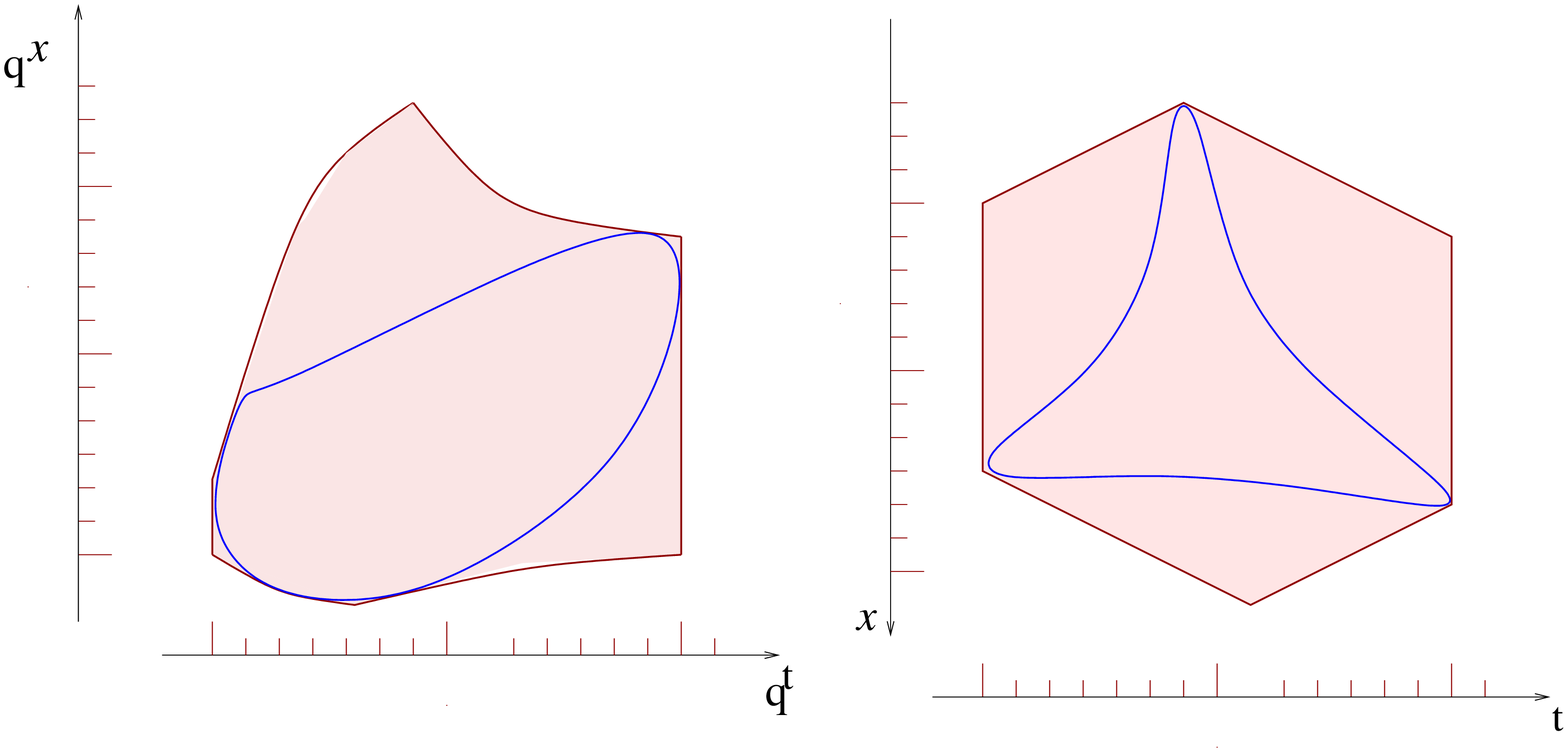}{The envelope of the liquid region, is the smallest degree algebraic plane curve, tangent to the image of the hexagon under $(x,t)\to(q^x,q^t)$. \label{fighexagonq}}

\bigskip
{\bf Recovering the spectral curve from the envelope:}
\medskip

Assume that we know the envelope $X_c(t)$, or, writing $x_c=q^{X_c}$ and $z=q^{t/2}$, assume that we know the plane curve $x_c(z)$, i.e. an algebraic curve tangent to the boundaries of the image of $\domain$ under $(x,t)\to(q^x,q^{t/2})$.
Let us explain how to recover the full spectral curve $x(z,\tau)$.

\medskip

Given the equation of the envelope $x_c(q^{t/2})$ (which is a multivalued algebraic function), one can choose locally $z=q^{t/2}$ as a local parameter.

One finds that the spectral curve is (at least in the domain of $t$ where $q^t$ can be chosen as a local parameter):
\beq
x(z,t) = {q^{t/2}\over 2}\,\left({x_c(z)\over z}+x'_c(z)\right) + {q^{-t/2}\over 2}\,\left(z x_c(z)- z^2 x'_c(z)\right) .
\eeq

Therefore, knowing the envelope allows to recover the full spectral curve, i.e. the functions $x(z,\tau)=q^{\hat X(z,t)}$ and also $\hat Y_i(z)$ from \eq{eqXiYiasympq}.

\subsubsection{Small $\ln q$ Asymptotic expansion}
\label{secLSAEq}

Now, suppose that we have found those functions, and that we have indeed found the correct spectral curve.
The spectral curve is then the pair of functions ${\cal S}_t=(\hat X(z,t),\hat Y(z,t))$, i.e., up to a symplectic transformation:
\beq
{\cal S} = \left({1\over \ln q} \ln{(x(z,\tau))}, \ln{(y(z,\tau))}\right),
\eeq
and we notice that $x(z,\tau)$ and $y(z,\tau)$ do not depend on $q$, they depend only on $\qq$, therefore we have $F_g({1\over \ln q} \ln{x}, \ln{y})= (\ln q)^{2g-2} F_g(\ln x,\ln y)$.

And thus, from theorem \ref{thZtopchmattasep}, we have the full large $T$ asymptotic expansion:
\begin{conjecture}
$\ln{\cal Z}$ has the small $\ln q$ expansion
\beq
\ln{\cal Z} \sim \sum_{g=0}^\infty (\ln q)^{2g-2}\, F_g(\ln x,\ln y).
\eeq
where the meromorphic functions $x(z)$ and $y(z)$ are computed by solving the requirements of section \ref{recipeclassicalspcurveq}, and $F_g$ is the $g^{\rm th}$ symplectic invariant defined in \cite{EOFg}.
\end{conjecture}

A possible proof could follow the same ideas which we discussed in section \ref{secLSAE} for $q=1$.

\subsection{Microscopic asymptotics}

It is a truism to say that since our model is a matrix model, it has all the local universal behaviors of matrix models.

\subsubsection{Zoom near a point}

Let us choose a point $(x_0,t_0)$ anywhere in the plane (it can be in or outside the domain, or on the border).

Let us rescale $x$ and $t$ with some scaling parameter $s$, with some exponents $\alpha$ and $\delta$, i.e. we write:
\beq
x = x_0 + s^{\alpha}\, \xi
\virg
t=t_0+s^\delta\, \tau.
\eeq
Also, on the spectral curve, we choose a point $z_0$ such that $\hat X(z_0,t_0)=x_0$, and we rescale it by choosing a rescaled local parameter $\zeta$:
\beq
z=z_0+s^\gamma\,\zeta.
\eeq

We rewrite the asymptotics of the functions $\hat X(z,t)$ and $\hat Y(z,t)$ in terms of those rescaled variables:
\beq
\hat X(z,t) = x_0 + s^\alpha \hat \xi(\zeta,\tau) + o(s^\alpha)
\virg 
\hat Y(z,t) = y_0 + s^{\beta} \hat \eta(\zeta,\tau) + o(s^\beta).
\eeq

If the exponents $\alpha,\beta,\gamma,\delta$ are well chosen, then the spectral curve
${\cal S}_{\rm zoom} = (\xi(\zeta,\tau),\eta(\zeta,\tau))$, is a regular spectral curve (it has only branchpoints of square-root types).
We call this curve, the {\bf blow up} of the spectral curve $(\hat X,\hat Y)$ near the point $(x_0,t_0)$.
In practice, finding the blow up is a rather trivial task, it merely consists in finding the first non-vanishing terms in the Taylor expansion near a point.

\medskip

From \cite{EOFg}, we have that the correlation functions $W_n^{(g)}(x_1,\dots,x_n;t)$:
\beq
\sum_g W_n^{(g)}(x_1,\dots,x_n;t) = \left< \tr {1\over x_1-M_t}\tr {1\over x_2-M_t} \dots \tr {1\over x_n-M_t}\right>_c
\eeq
behave at small $s$ like:
\beq
W_n^{(g)}(x_1,\dots,x_n;t) \sim s^{(2-2g-n)\,(\alpha+\beta)-n\alpha}\,\, \om_n^{(g)}(\xi_1,\dots,\xi_n;\tau)\,\, (1+o(s^0)),
\eeq
where $\om_n^{(g)}$ are the ``symplectic invariants'' correlation functions of \cite{EOFg} associated to the spectral curve ${\cal S}_{\rm zoom}$.

\medskip

In other words, since the correlation functions are symplectic invariant correlators of a spectral curve ${\cal S}$, their local behavior is, to leading order, given by the symplectic invariant correlators of the blown up spectral curve. This theorem found in \cite{EOFg} is very easy to prove by recursion on $n$ and $g$ (see appedix \ref{appspinv}).

\smallskip

Therefore, it suffices to find the blown up spectral curve to characterize the leading behavior of the correlation functions near a point.

\subsubsection{Airy kernel near regular boundaries of the liquid region}

Near a regular point of the envelope $X_c(t_0)$, we have $d\hat X/dz=0$ at $t=t_0$ and thus, we have a Taylor expansion in $z$ of the form:
\beq
\hat X(z,t) \sim X_0(t) + (t-t_0)X_1(t)\, z  + X_2(t) z^2 + \dots,
\eeq
where each $X_i(t)$ has a regular Taylor expansion in $t-t_0$.
Let us rescale:
\beq
z=s \, \zeta
\virg
t=t_0+s\tau.
\eeq
We have to the first few orders in $s$:
\beq
\hat X(z,t) = X_0(t_0) + s\,\left( \dot X_0 \tau \right) + s^2\,\left(  X_1 \tau \zeta + X_2 \zeta^2\right) + O(s^3).
\eeq
We also have
\beq
\dot{\hat X}(z,t) = \dot X_0 + s\,\left( X_1 \zeta \tau\right) + O(s^2),
\eeq
and therefore
\beq
\hat Y(z,t) \sim Y_0 + s\, X_1 \zeta  + O(s^2).
\eeq
The blown up curve is thus:
\beq
{\cal S}_{Airy}=\,\,
\left\{\begin{array}{l}
x(\zeta,\tau) = \zeta \tau + \zeta^2 \,\,  \cr
y(\zeta,\tau) = \zeta\,
\end{array}\right. \,\, ,
\eeq
which satisfies:
\beq
(y+\tau/2)^2 = x+\tau^2/4.
\eeq
The spectral curves such that ${\rm Pol}_q(y)={\rm Pol}_p(x)$ where ${\rm Pol}_q$ is a polynomial of degree $q$ and ${\rm Pol}_p$ is a polynomial of degree $p$, appear in the so-called $(p,q)$-minimal models, i.e. in the classification of conformal field theories \cite{KazakovRMTcrit, ZJDFG}.
It has central charge $c=1-6(p-q)^2/pq$.
Here, this is the $(1,2)$ minimal model, with central charge $c=-2$, which is well known to be generated by the Airy differential system $Ai''=x Ai$, and the correlation functions are determinants of the Airy kernel.
It is also well known to be related to the Tracy-Widom law of extreme eigenvalues statistics \cite{TW}.

\subsubsection{Pearcey kernel near cusps}

We have a cusp each time a pole disappears, generically a simple pole. Thus we have a Laurent expansion in $z$ starting at $z^{-1}$, and such that the residue of the pole vanishes at $t=t_0$, i.e.:
\beq
\hat X(z,t) \sim{(t-t_0) v(t)\over z}+x_0(t)+u(t) z+\dots,
\eeq
and all the coefficients $v(t),x_0(t),u(t),\dots$ have a regular Taylor expansion near $t=t_0$.
Let us rescale $z=s \zeta/u(t_0)$ and $t=t_0+s^2 \tau/u(t_0)v(t_0)$, to order $s$, and up to constant terms, we have:
\beq
\hat X(z,t)  = x_0 + s\, \left(\zeta+{\tau\over \zeta} \right) + O(s^2),
\eeq
and generically, $\hat Y$ behaves like:
\beq
\hat Y(z,t) = y_0- {z\over v_0} + \dots .
\eeq
The blown up curve is thus:
\beq
{\cal S}_{Pearcey}=\,\,
\left\{\begin{array}{l}
x(\zeta,\tau) = \zeta+{\tau\over \zeta} \,\,  \cr
y(\zeta,\tau) = \zeta\,
\end{array}\right. \,\, ,
\eeq
and the local behaviors of correlation functions, are the correlation functions of that universal curve.

This is the spectral curve whose correlators are generated by the Pearcey kernel.

We thus recover the well known Pearcey kernel behaviour \cite{TWP}.

\subsubsection{Critical points}

If we consider a point on the boundary, such that more derivatives vanish, typically we find that the spectral curve behaves locally like:
\beq
y(x)\sim y(a)+(x-x(a))^{p\over q} + \dots
\eeq
It was shown in \cite{EOFg}, that the correlation functions tend towards those of the $(p,q)$ reduction of the K-P hierarchy, i.e. the $(p,q)$ minimal model of central charge $c=1-6(p-q)^2/pq$.

The model $(3,2)$ of central charge $c=0$ is called "pure gravity", the model $(5,2)$ of central charge $c=-22/5$ is called Lee-Yang, The model $(4,3)$ of central charge $c={1\over 2}$ is called the Ising model,...
Their correlators are generated by determinantal formulae from a kernel involving the $\psi$-system associated to a non-linear equation of Gelfand-Dikii type (Painlev\'e I is the Gelfand-Dikii equation for pure gravity $(3,2)$). Some details can be found in \cite{ZJDFG, EBerg}.

\subsubsection{Other local behaviors}

Also, we expect that locally in the bulk of the liquid region, the behavior is given by the sine-kernel \cite{BOO, BDJ, johansson},
and near vertical boundaries of the liquid region, we have $y(x)\sim y(a)+(x-a)^2$, i.e. we expect the $(2,1)$ model, described by the Hermit kernel of the ``birth of a cut'' (see \cite{BrezHik, eynbirthcut}).
And we also expect to find the "Bead model" limit in the tentacles of the amoeba when $|T \ln q|$ is large, see \cite{boutillier}.

However, these cases are such that the blown up curve is not regular, and we cannot directly apply the method of \cite{EOFg}.

\subsubsection{Arbitrary local behaviors}

Then, one could easily invent some domains, for which a local blown up curve could be any spectral curve specified in advance, and by choosing sufficiently complicated domains one can obtain any limit law.

The classifications of all possible laws is more or less the classification of singularities of spectral curves, and it is more or less the classification of spectral curves themselves.

\section{Examples}
\label{secexs}

In this section, we illustrate our method by applying it to several classical examples, which were already studied in the literature with other methods.
Here, however, we have a method to obtain not only the large size limit, but also all corrections to all orders.

\subsection{The hexagon}

Our domain $\domain$ is the hexagon of  figure \ref{fighexagon}, with slopes $\pm {1\over 2}$.
\figureframex{9}{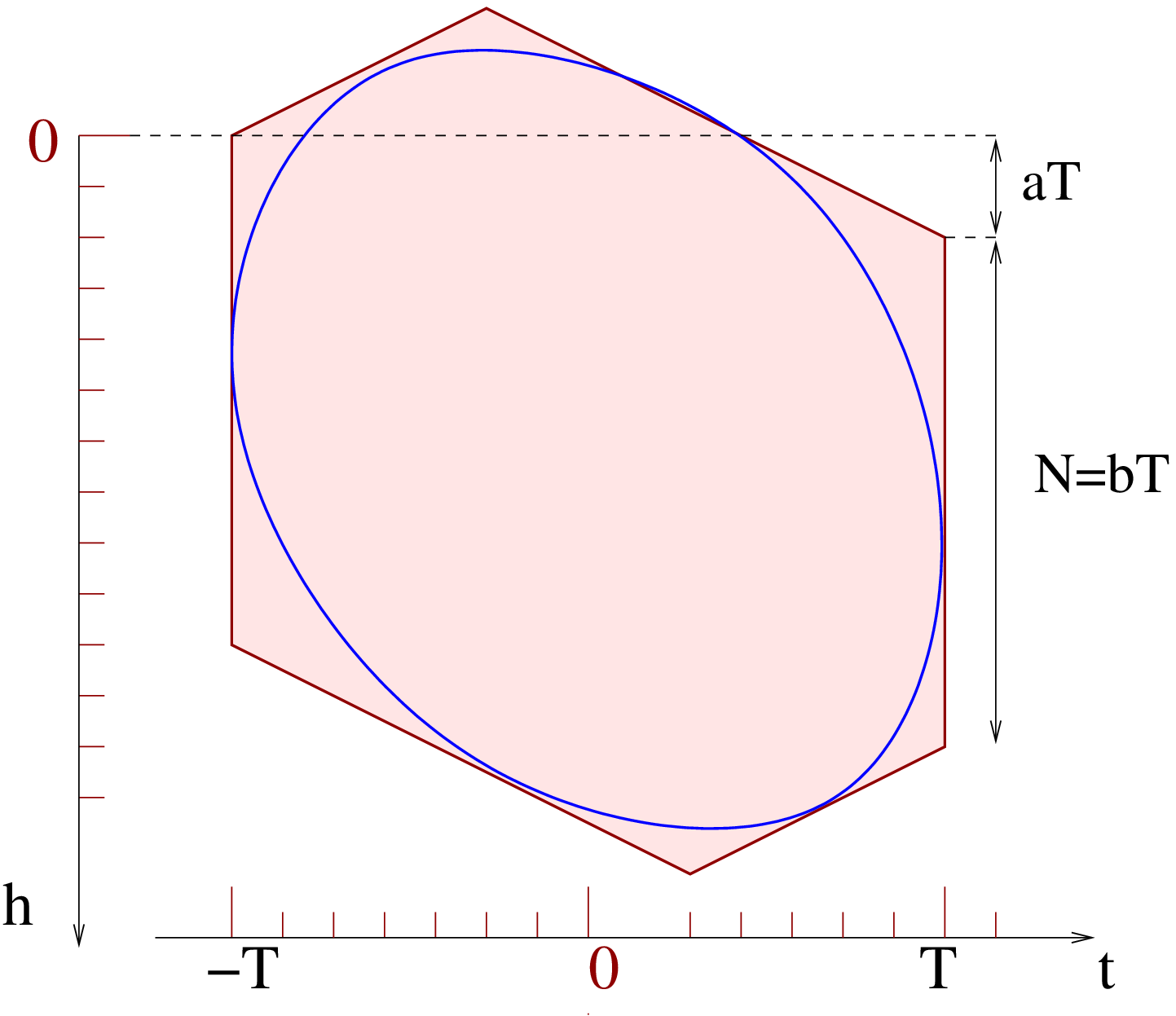}{\label{fighexagon} Domain for the hexagon.}

We choose $\tmax=-\tmin=T$. We choose the weighs $\alpha(t)=\beta(t)=1$ at all times.
We write $N=bT$.

There is no defect, therefore we have $k=1$, the reduced matrix integral is a 2-matrix model with an external field:
\beq
 {\cal Z}_{\rm hexagon}  
= \int_{H_{N}} dM_0\,\, \int_{i H_{N}}  dR_1\,\,\, 
 \, \ee{-\Tr V_{0}(M_0)}\, \,
 \, \ee{-\Tr \td{U}_{1}(R_1)}\,\,  \ee{\Tr R_1\,(M_1-M_0 )}
\eeq
where $M_1$ is the fixed following matrix:
\beq
M_1 = {\rm diag}(Ta+1,Ta+2,\dots, Ta+N) = T\,{\rm diag}(a+{1\over T},a+{2\over T},\dots, a+b).
\eeq
For $V_0$, we can choose
\beq
\ee{-V_0(X)} = {1\over \Gamma(X)\,\Gamma(N+1-X)}.
\eeq

\subsubsection{The classical hexagon $q=1$}

We apply the recipe of section \ref{recipeclassicalspcurve}.

First, the domain has no hole, and we look for an algebraic curve of genus $0$, therefore it can be parametrized by a uniformizing variable in the complex plane $z\in\mathbb C$.
Let us look for 2 rational fractions $u(z)$ and $v(z)$, with 2 poles, and we write:
\beq
x(z,\tau) = u(z) + \tau v(z).
\eeq

Up to a reparametrization of $z$, we choose the poles to be at $z=0$ and $z=\infty$.
Moreover, we require that at $\tau=-1$ the pole at $z=0$ disappears, 
and that at $\tau=+1$ the pole at $z=\infty$ disappears, this implies that $u$ and $v$ and $x$ are of the form:
\beq
x(z,\tau) = c+r\,\tau + \gamma\,\left( (1-\tau)z + {1+\tau\over z}\right),
\eeq
\beq
u(z) = c+ \gamma\,\left( z + {1\over z}\right)
\virg
v(z) = r+ \gamma\,\left( -z + {1\over z}\right).
\eeq
We define:
\beq
x_0(z) = x(z,-1) = c-r+2\gamma z
\virg
x_1(z) = x(z,1) = c+r+2\gamma /z.
\eeq
Then we find the coefficients $r,c,\gamma$ by solving the following system:
\beq
\left\{\begin{array}{l}
\exists\, z\qquad {\rm such\, that}\qquad
x_0(z)=0 \qquad {\rm and} \qquad v(z)=-{1\over 2}, \cr
\exists\, z\qquad {\rm such\, that}\qquad
x_0(z)=b
\qquad {\rm and} \qquad
v(z)={1\over 2}, \cr
\exists\, z\qquad {\rm such\, that}\qquad
x_1(z)=a
\qquad {\rm and} \qquad
v(z)={1\over 2}, \cr
\exists\, z\qquad {\rm such\, that}\qquad
x_1(z)=a+b
\qquad {\rm and} \qquad
v(z)=-{1\over 2}.
\end{array}\right.
\eeq
We have 3 unknowns and 4 equations, but one can check that there are only 3 independent equations, and the system has a solution.

We easily find:
\beq
c={a+b\over 2}
\virg
r={a(1+b)\over 2}
\virg
16\,\gamma^2 = (1-a)(1+a)b(b+2).
\eeq

\bigskip
{\bf Envelope}
\medskip

The branchpoints $z_c(\tau)$ are found from $x'(z_c,\tau)=0$, and we find that there are two branchpoints:
\beq
z_c(\tau) = \pm \sqrt{1+\tau\over 1-\tau}.
\eeq
That gives:
\beq
x_c(\tau) = c+r\tau \pm 2\gamma \sqrt{1-\tau^2},
\eeq
or explicitely
\beq
x_c(\tau) = {1\over 2} \left( a+b+\tau\, a(1+b) \pm \sqrt{(1-\tau^2)\,(1-a^2)\,b(b+2)}\right).
\eeq
One can easily check that this is the equation of the ellipse tangent to all sides of the hexagon.
See the ellipse in fig.\ref{fighexagon}.

\subsubsection{The quantum hexagon $q\neq 1$}

Now consider $q^T=O(1)$, and write:
\beq
\qq = q^T
\eeq
and we shall define the $\qq$-numbers:
\beq
[\tau] = {\qq^{-\,{\tau\over 2}}-\qq^{\tau\over 2}\over \qq^{-\,{1\over 2}}-\qq^{1\over 2}}.
\eeq

Let us now apply the recipe of section \ref{recipeclassicalspcurveq}.


We write:
\beq
x(z,\tau) = \qq^{\tau/2} u(z) + \qq^{-\tau/2} v(z)
\eeq
where $u$ and $v$ are rational fractions with two poles. 
Up to a reparametrization of $z$, we choose the poles to be at $z=0$ and $z=\infty$.
Moreover, we require that at $\tau=-1$ the pole at $z=0$ disappears, 
and that at $\tau=+1$ the pole at $z=\infty$ disappears, this implies that $u$ and $v$ and $x$ are of the form:
\beq
x(z,\tau) = {[1-\tau]\over [2]}\,(c+z) + {[1+\tau]\over [2]}\,(r+{d\over z}) ,
\eeq
We define:
\beq
x_0(z) = x(z,-1) = c+z
\virg
x_1(z) = x(z,1) = r+{d\over z}.
\eeq
Then we find the coefficients $r,c,d$ by solving the following system:
\beq
\left\{\begin{array}{l}
\exists\, z\qquad {\rm such\, that}\qquad
x_0(z)=1 \qquad {\rm and} \qquad x_1(z)=\qq^{-1}, \cr
\exists\, z\qquad {\rm such\, that}\qquad
x_0(z)=\qq^b \qquad {\rm and} \qquad x_1(z)=\qq^{b+1}, \cr
\exists\, z\qquad {\rm such\, that}\qquad
x_1(z)=\qq^a \qquad {\rm and} \qquad x_0(z)=\qq^{a-1}, \cr
\exists\, z\qquad {\rm such\, that}\qquad
x_1(z)=\qq^{a+b} \qquad {\rm and} \qquad x_0(z)=\qq^{a+b+1}.
\end{array}\right.
\eeq
We have 3 unknowns and 4 equations, but one can check that there are only 3 independent equations, and the system has a solution.
We find:
\beq
c = {\qq^{-1}+\qq^{a+b}-\qq^a-\qq^{b+1}\over \qq^{-1}-\qq}
\virg
r= -\,{1+\qq^{a+b+1}-\qq^{a-1}-\qq^{b}\over \qq^{-1}-\qq},
\eeq
\beq
d= {(1-\qq^b)(\qq^a-\qq)(\qq^a-\qq^{-1})(\qq^{b+1}-\qq^{-1})\over (\qq^{-1}-\qq)^2} \, .
\eeq

\begin{figure}[bth]
\hrule\hbox{\vrule\kern8pt 
\vbox{\kern8pt \vbox{
\begin{center}
{\mbox{\epsfxsize=5.truecm\epsfbox{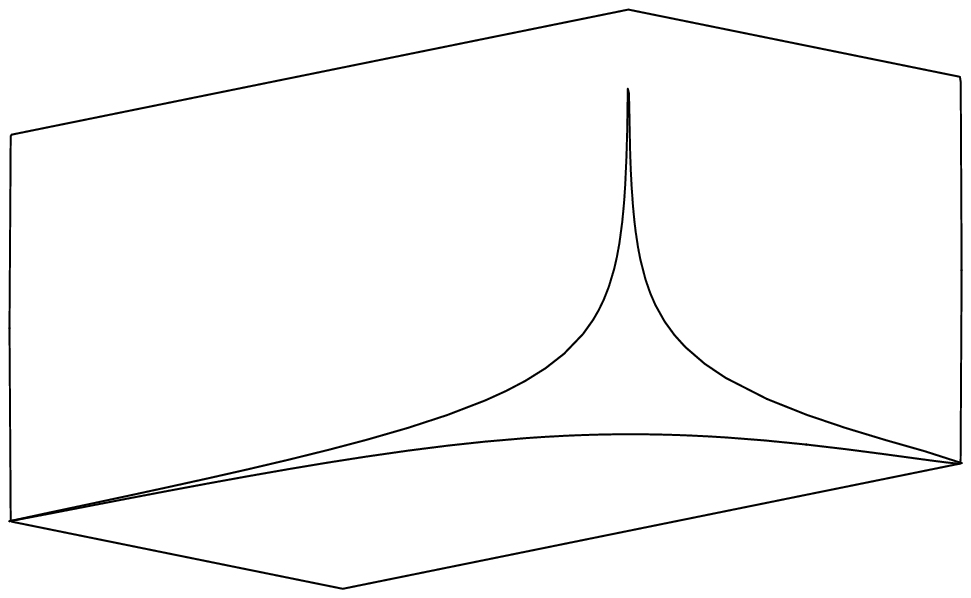}}}
\hspace{1cm} {\mbox{\epsfxsize=5.truecm\epsfbox{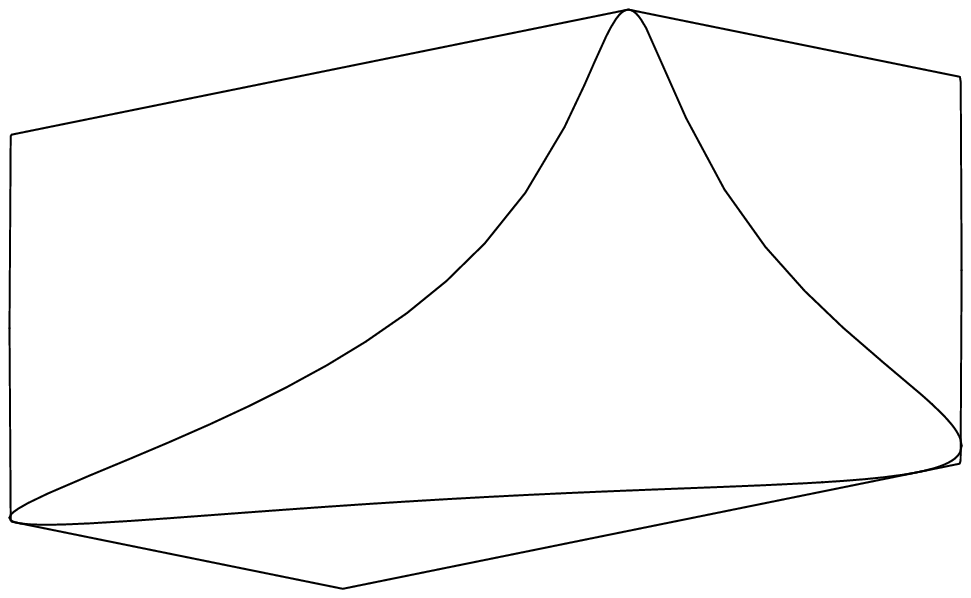}}}
\hspace{1cm} {\mbox{\epsfxsize=5.truecm\epsfbox{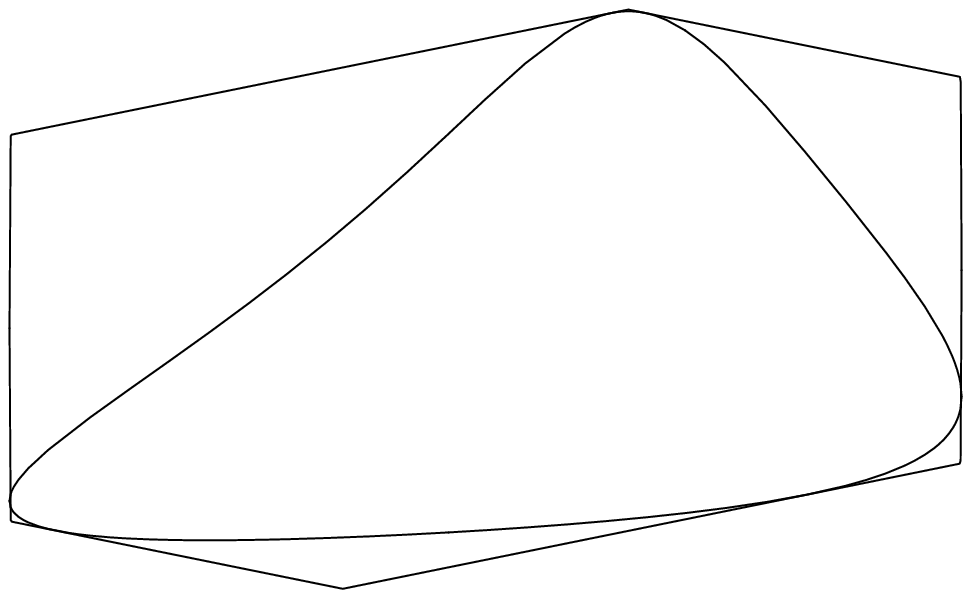}}}
\hspace{1cm} {\mbox{\epsfxsize=5.truecm\epsfbox{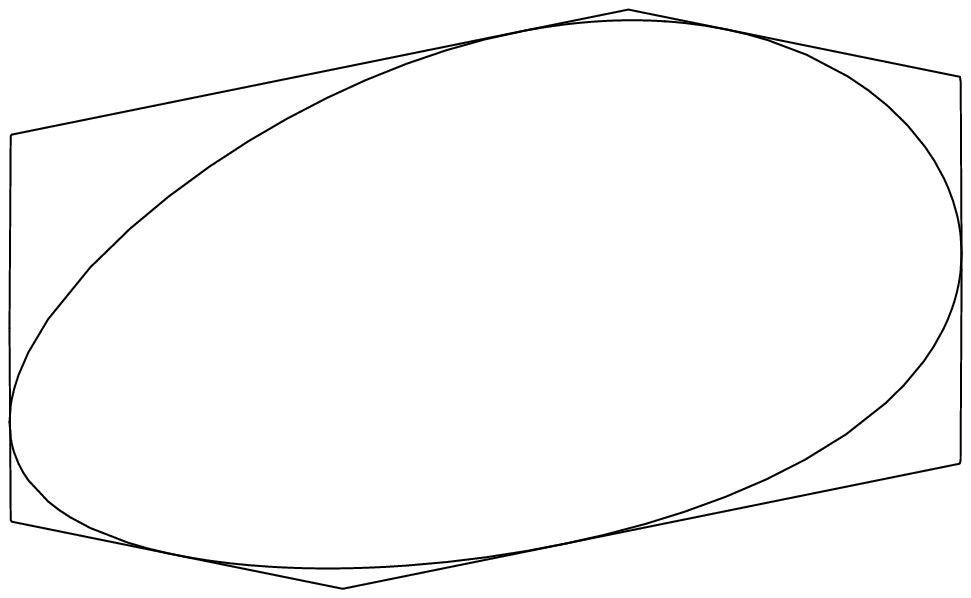}}}
\hspace{1cm} {\mbox{\epsfxsize=5.truecm\epsfbox{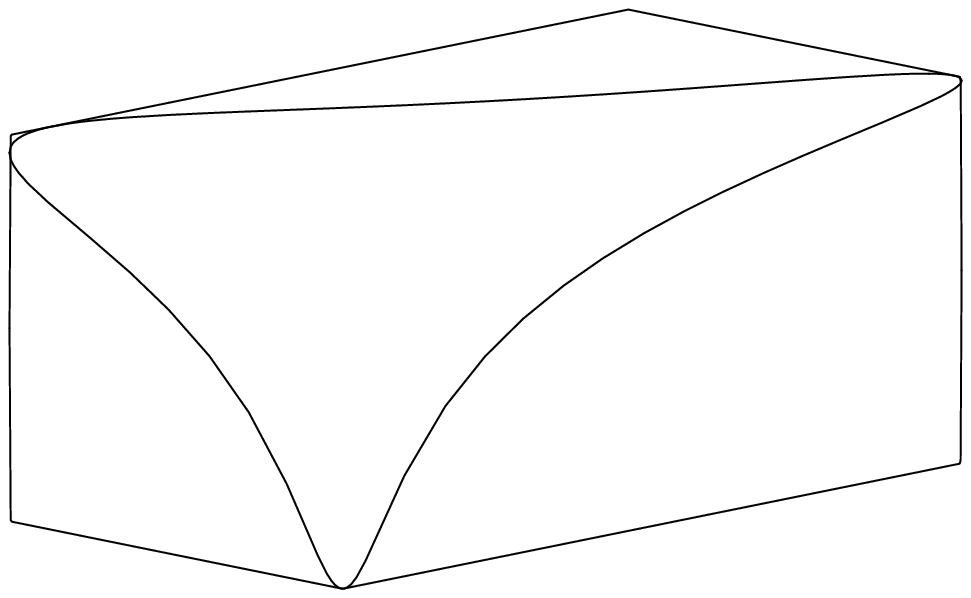}}}
\hspace{1cm} {\mbox{\epsfxsize=5.truecm\epsfbox{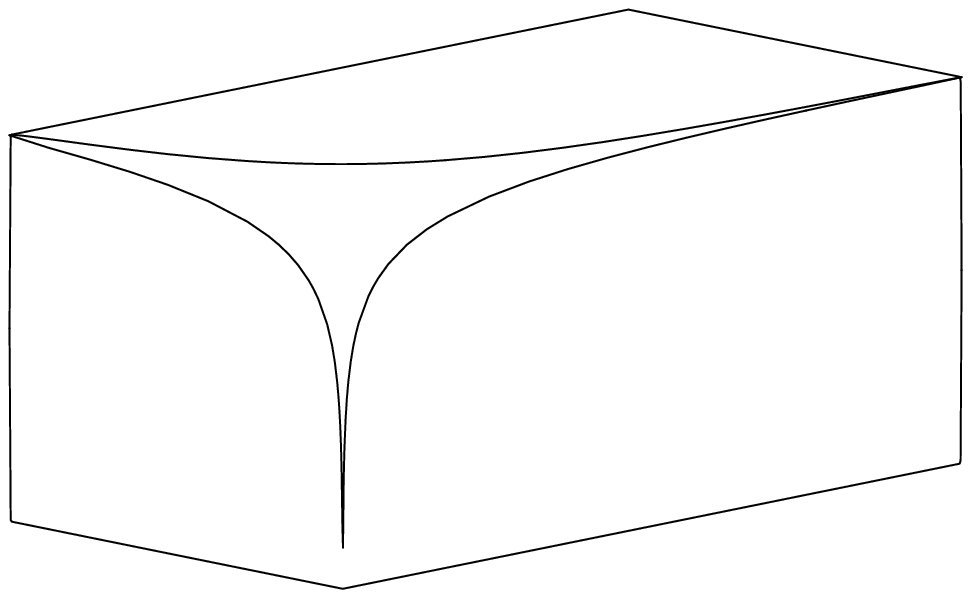}}}
\end{center}
\caption{The envelope of the hexagonal domain. The plots of \eq{eqenvhexq} are for the hexagon $a=0.3$, $b=2$, and for values of $q$ respectively $q=0.001,0.1,0.3,0.8,10,1000$.
Notice that the liquid region is convex only for $q$ close to $1$.\label{figexhexq}}
}\kern8pt} 
\kern8pt\vrule}\hrule
\end{figure}

\bigskip
{\bf Envelope}
\medskip

The branchpoints are at $z_c(\tau)=\pm\sqrt{d\,[1+\tau]/[1-\tau]}$, and thus the envelope is:
\beq\label{eqenvhexq}
x_c(\tau) = {[1-\tau]c+[1+\tau] r \pm 2\sqrt{d\,[1+\tau]\,[1-\tau]}\over [2]}.
\eeq
See some examples plotted in figure \ref{figexhexq}.

\subsection{The cardiod}
 
 Consider the classical case $q=1$. The domain is the one represented in figure \ref{figcardioid}.
 We assume $0<a<1$, and $b>0$.
\figureframex{9}{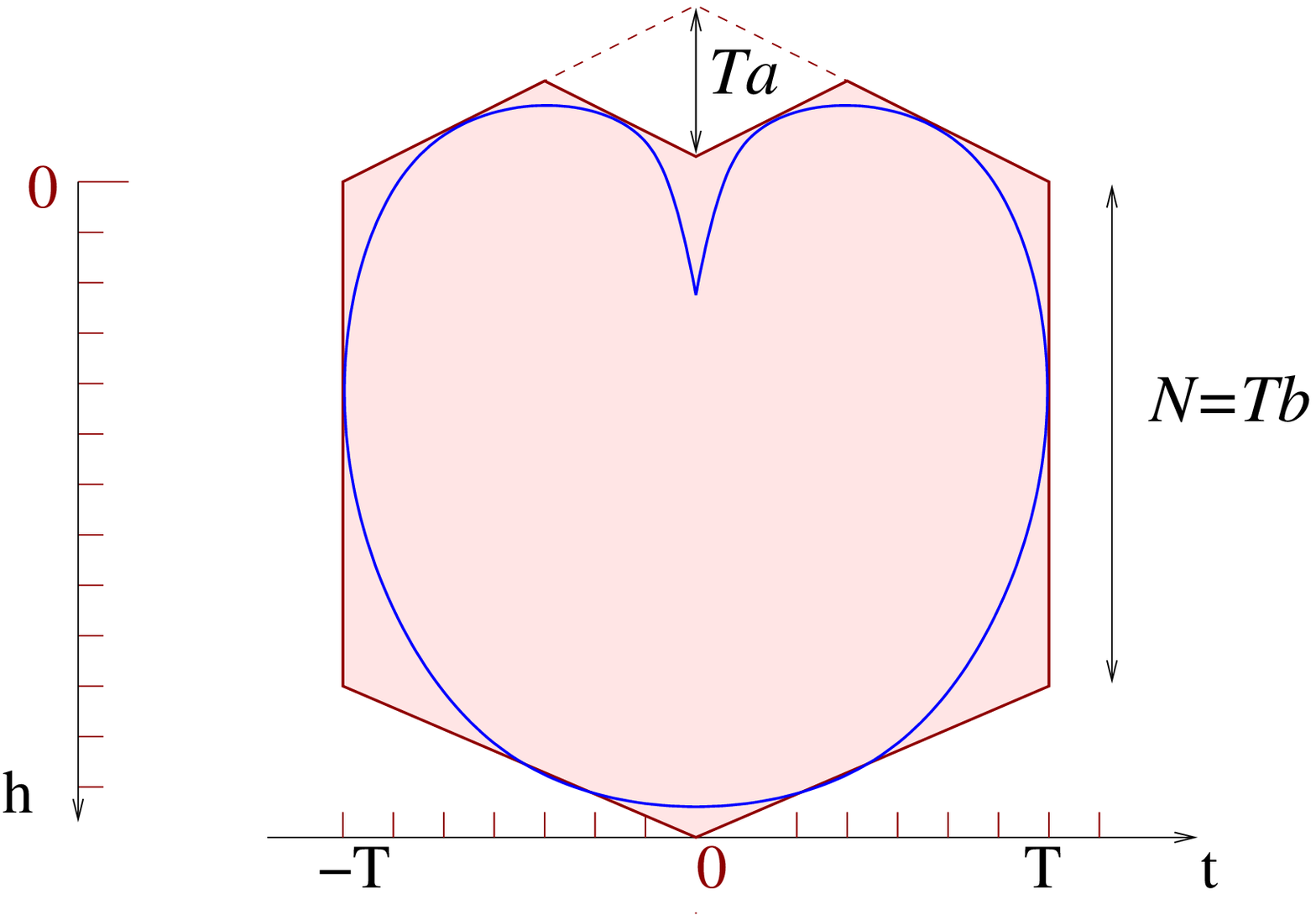}{The domain of the cardioid. The envelope is a cardioid.\label{figcardioid}}

\medskip

We find the spectral curve by applying the recipe of section \ref{recipeclassicalspcurve}.

Consider
\beq
x(z,\tau) = u(z)+\tau v(z),
\eeq
where $u$ and $v$ are rational fractions with 3 poles.

Up to a reparametrization of $z$, we choose the poles to be at $z=-1,0,1$.
Moreover, we require that at $\tau=-1$ the pole at $z=-1$ disappears, at $\tau=0$ the pole at $z=0$ disappears, and at $\tau=+1$ the pole at $z=+1$ disappears. Moreover, since the domain has a symmetry $\tau\to -\tau$, we choose $u$ and $v$ and $x$ of the form:
\beq
x(z,\tau) = a+b-{1\over 2}-2 w+{\gamma (1-\tau) \over 1-z}+{s\tau\over z}+{\gamma(1+\tau)\over 1+z},
\eeq
\beq
u(z)=a+b-{1\over 2}-2 w+{2\gamma \over 1-z^2}
\virg
v(z) = {2\gamma z\over z^2-1} + {s\over z}.
\eeq
The symmetry $\tau\to -\tau$ is such that $x(-z,-\tau)=x(z,\tau)$.

We write:
\beq
x_0(z)=x(z,-1) = a+b-{1\over 2}-2 w+{2\gamma\over 1-z}-{u\over z},
\eeq
\beq
x_1(z)=x(z,0) = a+b-{1\over 2}-2 w+{2\gamma \over 1-z^2},
\eeq
\beq
x_2(z)=x(z,1) = a+b-{1\over 2}-2 w+{u\over z}+{2\gamma\over 1+z}.
\eeq

The coefficients $w,s,\gamma$ are determined from the following system (we have written only the independent equations):
\beq
\left\{\begin{array}{l}
\exists\, z\qquad {\rm such\, that}\qquad
x_2(z)=0 \qquad {\rm and} \qquad v(z)={1\over 2}, \cr
\exists\, z\qquad {\rm such\, that}\qquad
x_0(z)=b
\qquad {\rm and} \qquad
v(z)={1\over 2}, \cr
\exists\, z\qquad {\rm such\, that}\qquad
x_1(z)=a-{1\over 2}
\qquad {\rm and} \qquad
v(z)={1\over 2}.
\end{array}\right.
\eeq

Those equations imply that:
\beq
3w^2 - w(2a+2b-1)+b(a-1)=0,
\eeq
\beq
\gamma = 2 w (b+1-w)(a-w)=2(w+1)(b-w)(a-w-1),
\eeq
\beq
2s = -(2w+1)(2b+1-2w)(2a-1-2w).
\eeq
And we have:
\beq
3w=a+b-{1\over 2}+2\gamma+s.
\eeq

\bigskip
{\bf Envelope}
\medskip

Let $s/2\gamma = e$.

Parametrically

\beq
\tau = {2\, z^3\over z^2(1+z^2)+e(z^2-1)^2}
\eeq
\beq
X_c = a+b-{1\over 2}-2w+{2\gamma}\,\,{z^2+e(1+z^2)\over z^2(z^2+1)+e (1-z^2)^2}
\eeq

One can check that it is the equation of the cardioid inscribed in the domain, see fig.\ref{figcardioid}.

\bigskip

The same domain can also be considered in the quantum case, but we don't do it here.

\subsection{The trapezoid}
\label{sectrapezoid}

Take $N=T$,
$\domain$ is the domain $\domain(\tmin)=[0,2T]$ at $\tmin=0$ and $\domain(\tmax)=[{T\over 2},{3T\over 2}]$ at $\tmax=T$.
We choose the weights $\beta=1$ and  $\alpha\neq 1$ with $\alpha$  constant in time.
We study the classical case $q=1$.

Notice that $\# \domain(\tmin)=2N \neq N=\#\domain(\tmax)$, therefore the initial matrix $M_\tmin$ is not fixed, it must have a potential $V_\tmin$ satisfying conditions \eq{defVtchardefectt} instead of \eq{defVtchardefecttmin}, but in fact it suffices to choose:
\beq
V_\tmin=0.
\eeq

\begin{figure}[bth]
\hrule\hbox{\vrule\kern8pt 
\vbox{\kern8pt \vbox{
\begin{center}
{\mbox{\epsfxsize=5.truecm\epsfbox{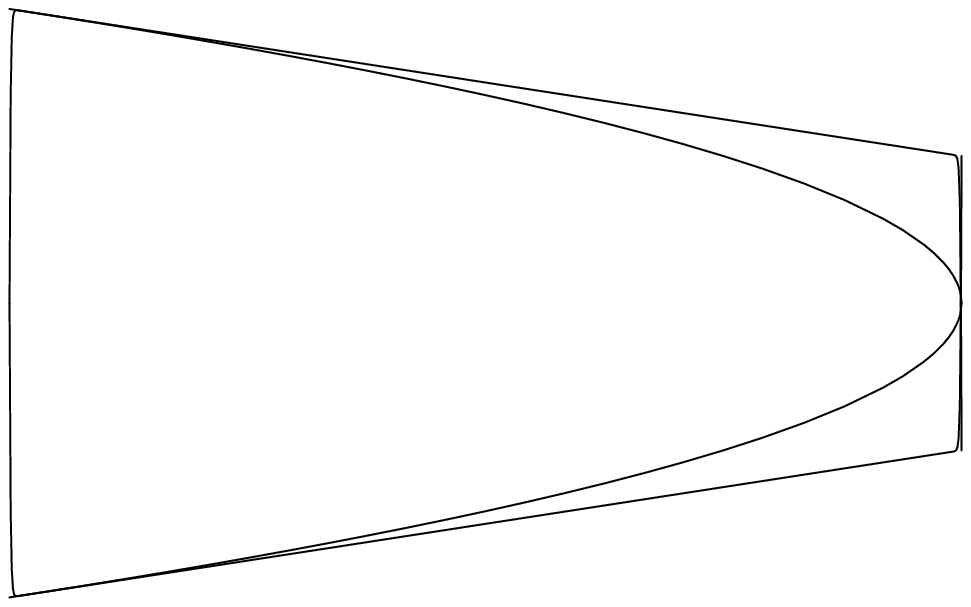}}}
\hspace{1cm} {\mbox{\epsfxsize=5.truecm\epsfbox{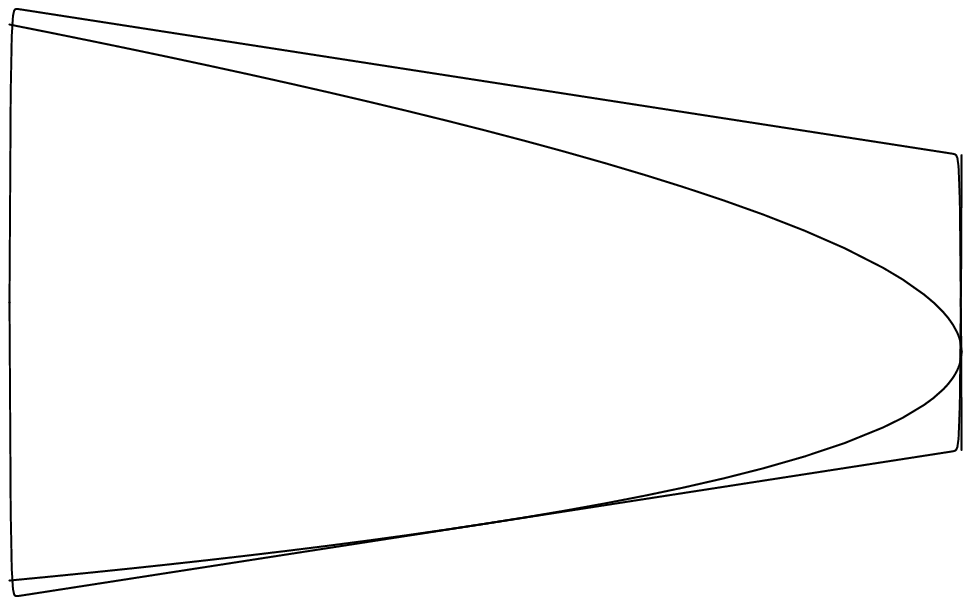}}}
\hspace{1cm} {\mbox{\epsfxsize=5.truecm\epsfbox{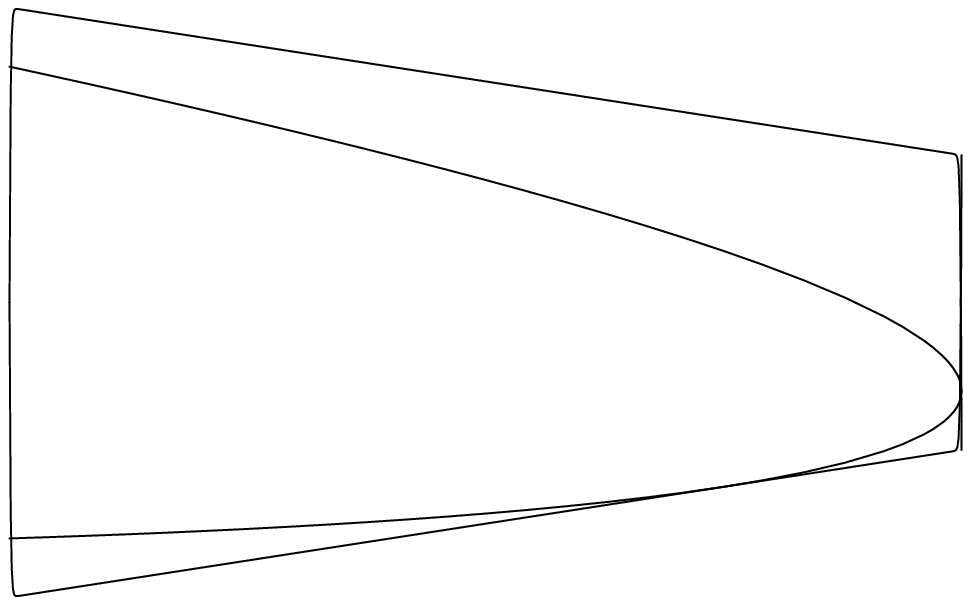}}}
\hspace{1cm} {\mbox{\epsfxsize=5.truecm\epsfbox{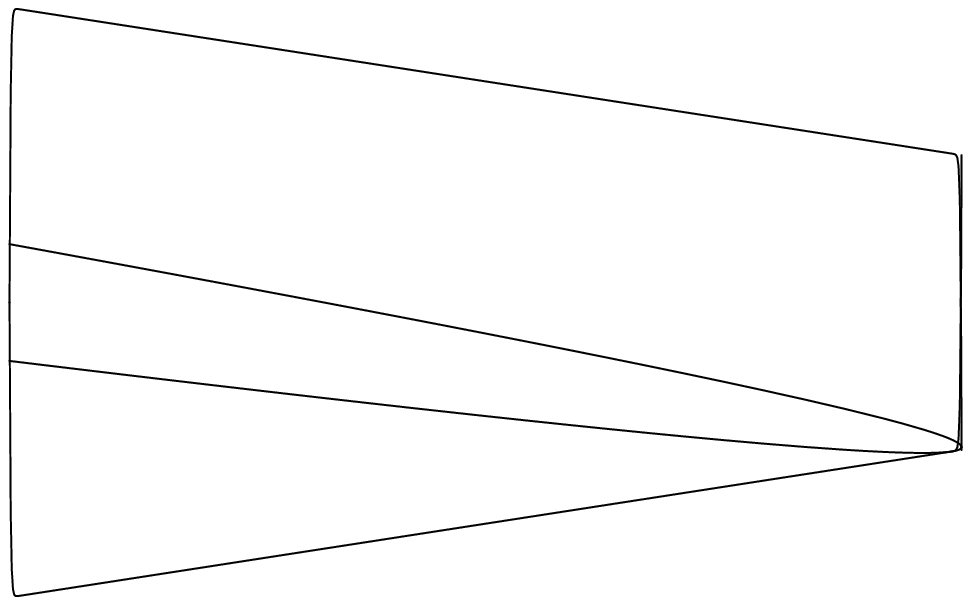}}}
\hspace{1cm} {\mbox{\epsfxsize=5.truecm\epsfbox{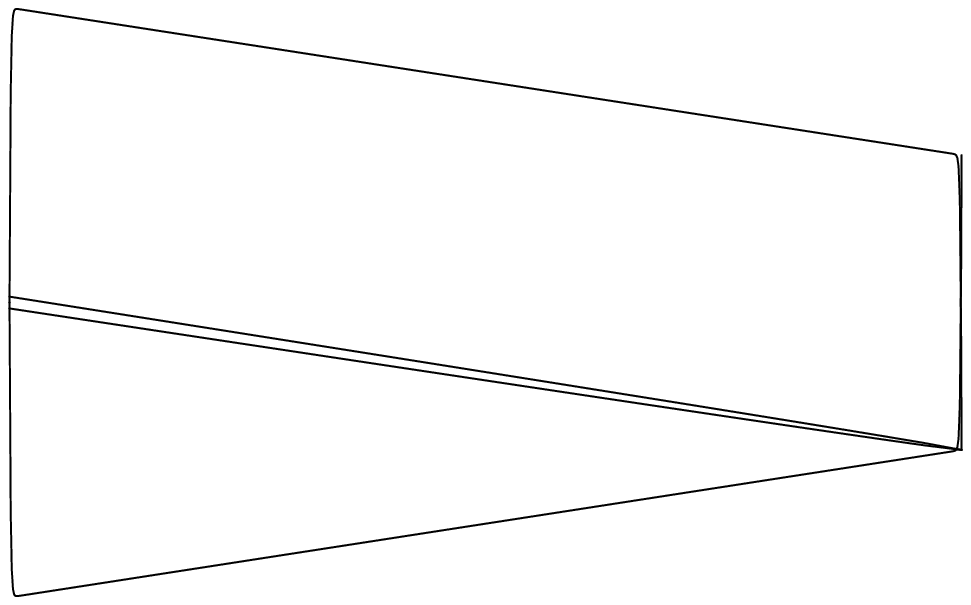}}}
\hspace{1cm} {\mbox{\epsfxsize=5.truecm\epsfbox{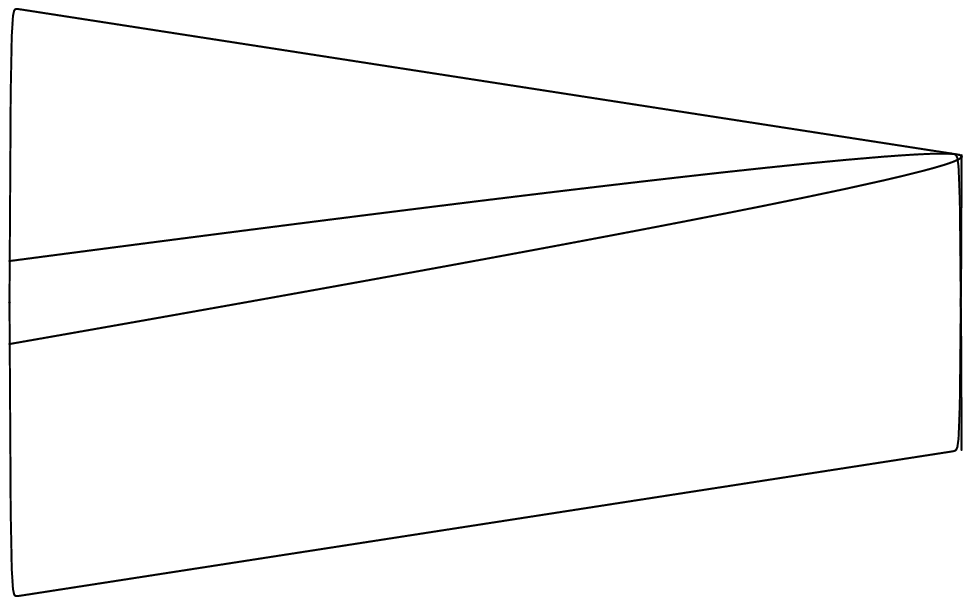}}}
\end{center}
\caption{The envelope of the trapezoidal domain. The plots are for $\alpha=1,0.5,0.25,0.01,0.0001,200$.\label{figextrapez}}
}\kern8pt} 
\kern8pt\vrule}\hrule
\end{figure}

Let us look for an algebraic genus zero spectral curve of minimal degree.
First, notice that we have no cusp condition at $\tmin$. We thus need to have only 2 poles, let us say at $z=0$ and $z=\infty$, and the pole at $z=0$ disappears at $t=T$.
Thus we write (with $\tau = t/T$):
\beq
x(z,\tau) = 1+c+ r \tau + (u\tau+v) z  - {1-\tau\over z}.
\eeq
That means
\beq
x_0(z) = x(z,0) = 1+c+v z  -{1\over z}
\virg
x_1(z) = x(z,1) = 1+c+r+(u+v) z.
\eeq
and
\beq
y_0(z)=y_1(z) = \alpha\,\, {{1\over 2} - (r+uz+{1\over z})\over  {1\over 2} + (r+uz+{1\over z})}.
\eeq

The coefficients $c,r,u,v$ can be determined by requiring that:

$\bullet$ $\exists z$ such that $x_0(z)=0$ and $x_1(z)={1\over 2}$.

$\bullet$ $\exists z$ such that $x_0(z)=2$ and $x_1(z)={3\over 2}$.

$\bullet$ $\ln{(y_0(z))} \sim 1/x_0(z)$ at $z\to \infty$.

The last condition implies $u=0$ and $r={1\over 2}\,{\alpha-1\over \alpha+1}$.
Then, the 2 other conditions imply $c=0$ and $v=(r-{1\over 2})(r+{1\over 2})$.

Finally we find the following spectral curve:
\beq
x_0(z) = 1 -{\alpha z\over (\alpha+1)^2} - {1\over z}
\virg
x_1(z) = 1+{1\over 2}\,{\alpha-1\over \alpha+1} -{\alpha z\over (\alpha+1)^2} 
\eeq
\beq
x(z,t) = 1+{t\over 2}\,{\alpha-1\over \alpha+1} -{\alpha z\over (\alpha+1)^2} - {1-t\over z}
\eeq

\beq
y_1(z) = {z -1 -\alpha\over  z + 1+{1\over \alpha}}
\eeq
The envelope of the liquid region is:
\beq
x_c(t) = 1+{t\over 2}\,{\alpha-1\over \alpha+1} \pm {2\over \alpha+1}\,\sqrt{(1-t)\alpha}
\eeq
this is a parabola shifted by a straight line, and tangent to at least two of the trapezoid boundaries.

\medskip

\begin{figure}[bth]
\hrule\hbox{\vrule\kern8pt 
\vbox{\kern8pt \vbox{
\begin{center}
{{\mbox{\epsfxsize=10truecm\epsfbox{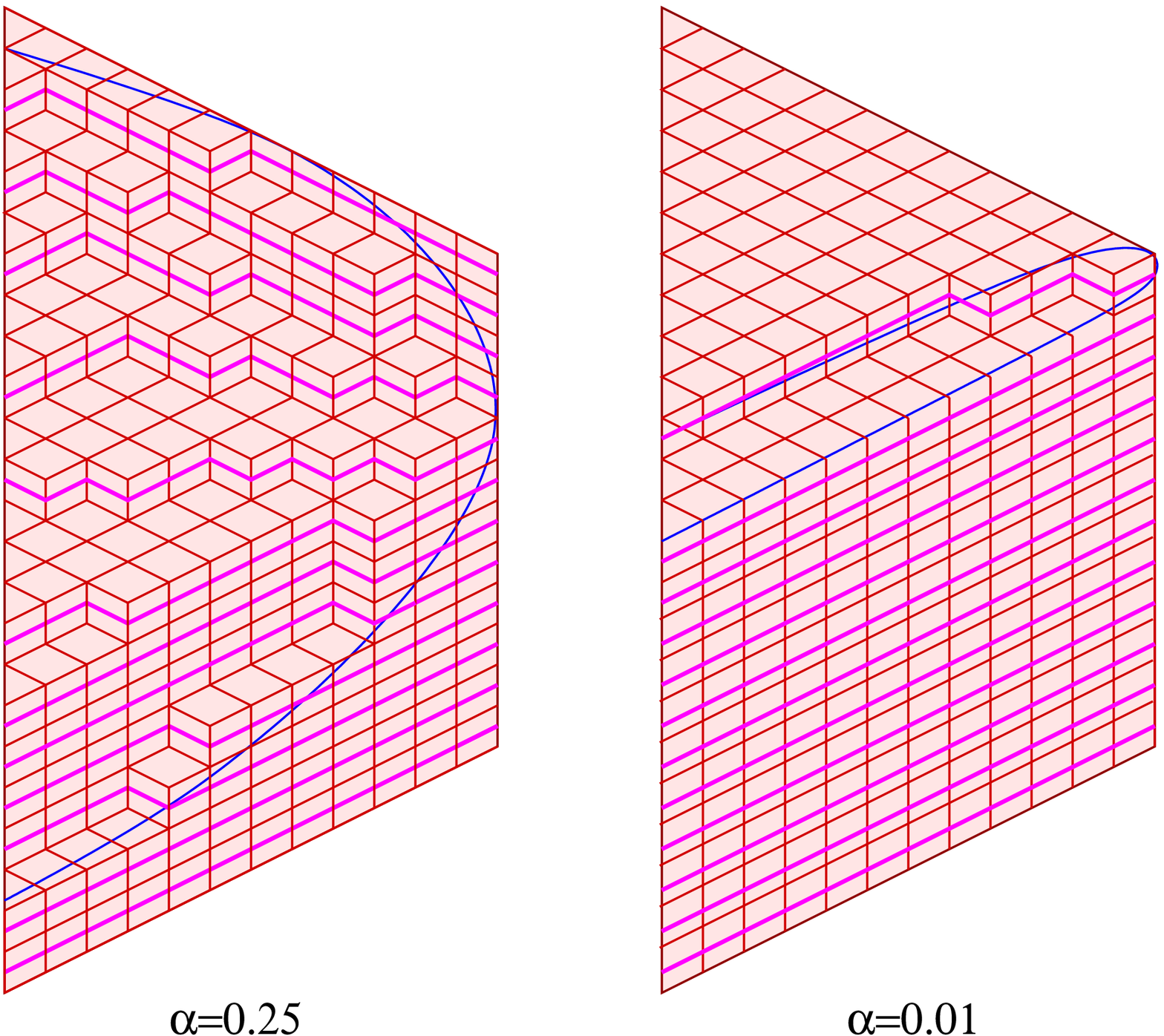}}}}\end{center}
\caption{Typical self-avoiding particles model, or typical plane partitions in the trapezoid.
The tiling outside the liquid region is regular, it is frozen.
For small $\alpha$, the probability to go downward is very small, and therefore almost all the particles are in the solid region going upward, and the liquid region becomes a narrow region around the line $x=1-{t\over 2}$.\label{figextrapeztasep}}
}\kern8pt} 
\kern8pt\vrule}\hrule
\end{figure}

Notice that when $\alpha=1$, particles have the same probability to go upward $h_i\to h_i+{1\over2}$ or downward $h_i\to h_i-{1\over2}$, and the spectral curve is symmetric with respect to $x\to 2-x$.
When $\alpha$ is very small, the probability to go downward is very small, and therefore almost all the particles are in the solid region going upward, and the liquid region becomes a narrow region around the line $x=1-{t\over 2}$.
On the countrary, when $\alpha$ is very large, the probability to go downward is very large, and  almost all the particles are in the solid region going downward, and the liquid region becomes a narrow region around the line $x=1+{t\over 2}$.
See fig.\ref{figextrapeztasep}.

\subsection{The Plancherel law}
\label{secplancherel}

Let us choose a partition $\mu$, and $N\geq n(\mu)$. we write:
\beq
h_i(\mu) = \mu_i-i+N
\eeq
we have:
\beq
h_1(\mu)>h_2(\mu)>\dots>h_N(\mu)\geq 0
\eeq
Consider the domain $\domain$, comprised between $\tmin=0$, and $\tmax=T$, 
and such that at $t=T$ the particles are at $h_i(T)={T\over 2}+N-i$ (in some sense the boundary is the trivial partition shifted by $T\over 2$), and at $t=0$, the particles are at $h_i(0)=h_i(\mu)$ (the boundary is the partition $\mu$).
See fig \ref{figdomainplancherel}.

\figureframey{7}{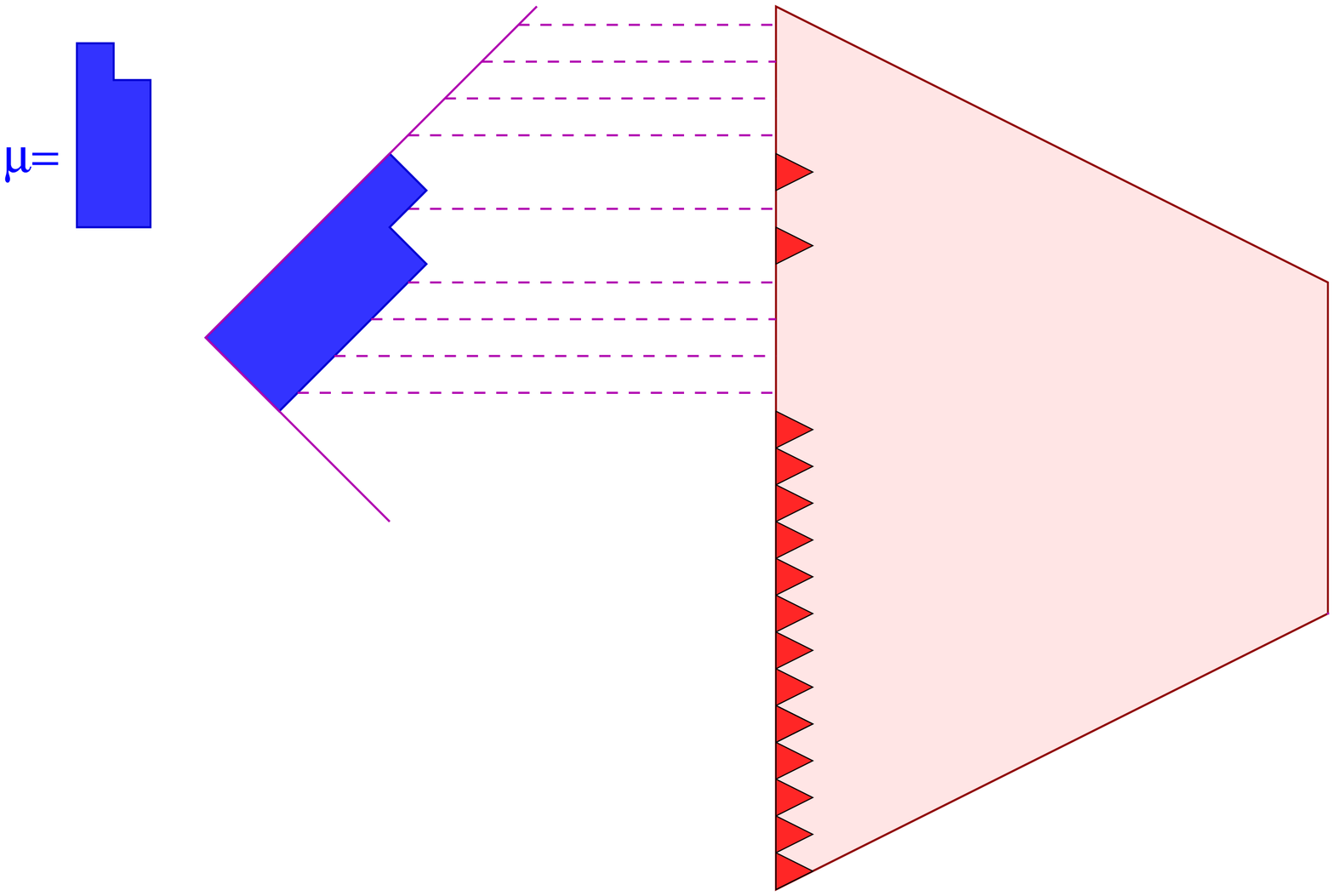}{\label{figdomainplancherel} The domain for the Plancherel law. At time $t=\tmax=T$, we have $h_i(\tmax)={T\over 2}+N-i$, and at time $t=0=\tmin$ we have $h_i(\tmin) = h_i(\mu)$.}

Let us call ${\cal P}_T(\mu)$ the plane partitions generating function in this domain:
\beq
{\cal P}_T(\mu) = Z_T(\mu) = \sum_{\pi,\partial\pi |_0 = \mu, \partial \pi |_T=\emptyset}\,\, q^{|\pi|}
\eeq
Let us define the "Plancherel law"  as the the limit $T\to\infty$:
\beq
{\rm Plancherel}(\mu) = {\cal P}(\mu) = {\cal P}_\infty(\mu).
\eeq
It has been well known from a really long time \cite{Planchrel1},
that:
\beq
{\cal P}(\mu) = {\prod_{i>j} \left(q^{h_i(\mu)-h_j(\mu)\over 2} -q^{h_j(\mu)-h_i(\mu)\over 2}\right)\over \prod_{i=1}^N \prod_{j=1}^{h_i(\mu)} (q^{-h_i(\mu)\over 2} - q^{h_i(\mu)\over 2})}
= {\prod_{i>j} [h_i(\mu)-h_j(\mu)]\over \prod_i [h_i(\mu)]!},
\eeq
and if $q=1$, that reduces to the classical Plancherel law:
\beq
{\cal P}(\mu) = {\Delta(h_i(\mu))\over \prod_{i=1}^N h_i(\mu)!}.
\eeq
As a check of our matrix model approach, 
let us recover this classical result from the matrix model.

\subsubsection{The classical Plancherel law $q=1$}

As presented in section \ref{secMM}, the domain $\domain$ is characterized by:

$\bullet$ the matrix $M_T$:
\beq
M_T = {T\over 2}\,{\rm Id}\,\, + {\rm diag}(0,1,2,\dots,N-1),
\eeq

$\bullet$ a potential at $t=0$, which satisfies the conditions of \eq{defVtchardefecttmin}, we choose:
\bea
\ee{-V_0(x)}
&=& {\ee{i\pi x}\over \Gamma(-x)}\,\,\prod_{i=1}^{N} {1\over x-h_i(\mu)}
\eea
Notice that we have:
\beq
\ee{-V_0(h_i(\mu))} = { h_i(\mu) ! \over \prod_{j\neq i} (h_i(\mu)-h_j(\mu))}
\eeq
and thus:
\beq
\ee{-\sum_i V_0(h_i(\mu))} = (-1)^{N(N-1)/2}\,\,\, {\prod_i h_i(\mu)!\over \Delta^2(h_i(\mu))}
\eeq

$\bullet$ and
\bea
\td{U}(y) = -T\,\ln{(2\cosh{y\over 2})}.
\eea

\bigskip

Theorem \ref{thTASEPMM}, or more precisely its reduced version theorem \ref{thTASEPMMreduced}, gives the relationship between the self-avoiding particles model partition function ${\cal P}_T(\mu)$ and the matrix model:
\bea
{\cal Z}
&=& \int dM_0 dR_1\,\, \ee{-\Tr V_0(M_0)} \ee{-\Tr \td{U}(R)}\,\, \ee{\Tr R_1(M_T-M_0)}  \cr
&=& C_{N,T,2}\,\,{\Delta(h_i(\mu))\over \Delta(M_T)}\,\, \ee{-\sum_i V_0(h_i(\mu))}\,\, {\cal P}_T(\mu)
\eea
i.e.:
\beq
{\cal P}_T(\mu) = C\,\,{\Delta(h_i(\mu))\over \prod_i h_i(\mu)!}\,\,\, {\cal Z}
\eeq
And we can already guess, that in the large $T$ limit, the role of the $h_i(\mu)$'s in the spectral curve is going to be subleading, i.e., to large $T$ leading order ${\cal Z}$ is going to be independent of $h_i(\mu)$.

\smallskip
In principle, our matrix model could be used to find the asymptotics of the Plancherel measure \cite{Okounkov2}.

\section{Obliged places and TSSCPPs}
\label{secTSSCPP}

So far, we have considered a self-avoiding particles model with defects, i.e. forbidden places for the particles. Defects were introduced by choosing $\ee{-V_t}(x)=0$ at the corresponding place $(x,t)$.

\smallskip
One could also be interested in constrained self-avoiding particles model, where we want to oblige some places to be visitted at some given times.
This cannot be achieved directly by tuning the potentials $V_t$, but this can be achieved as follows.

\subsection{Obliged places}

We choose a potential $V_t$ such that $\ee{-V_t(x)}=1-\eta$ at the place $(x,t)$, i.e. we enforce a probability $1-\eta$ that the place $(x,t)$ can be visited.
In other words, the contribution of self-avoiding particles  configurations such that one particle visits $(x,t)$ will be proportional to $1-\eta$, and  the contribution of self-avoiding particles  configurations such that no particle visits $(x,t)$ will be independent of $\eta$ (notice that no more than one particle can visit $(x,t)$).
Therefore the partition function ${\cal Z}$ is made of two terms:
\beq
{\cal Z} = {\cal Z}_0 + (1-\eta) {\cal Z}_1
\eeq 
where ${\cal Z}_1$ is the partition function of self-avoiding particles  configurations which visit $(x,t)$.
We have:
\beq
{\cal Z}_1 = -\,\left.{d\over d\eta}\, {\cal Z}\right|_{\eta=0} =  {\cal Z}\,\,\left< {d\over d\eta} \Tr V_t(M_t)\right>_{\eta=0}
\eeq
In other words, ${\cal Z}_1$ can be realized as some expectation value in our matrix integral.
We write:
\beq
A(x) = \left.{d\over d\eta} \Tr V_t(M_t)\right|_{\eta=0}
\eeq

We have that:
\beq
{{\cal Z}_1\over {\cal Z}} = <\Tr A(M_t)>.
\eeq

More generally, if we want to have several obliged places $(x_i,t_i)$, $i=1,\dots,s$, we introduce a function $A(x,t_i)$ for each of them.
\beq
{{\cal Z}_{\{(x_i,t_i)\}}\over {\cal Z}} = <\prod_{i=1}^s\,\Tr A(M_{t_i},t_i)>.
\eeq

\bigskip

Again, we have some arbitrariness in the choice of $A(x,t)$. The only requirements are:

$\bullet$ and for $t$ integer, $\tmin<t<\tmax$, we choose  $A(x,t)$ such that:
\beq
\forall x\in \domain(t)/\widehat\defect\,\, , \qquad \,\,
A(x,t) = 
\left\{\begin{array}{l}
1\,\, {\rm if}\,\, \exists\,i \,\,\,\, (x,t)=(x_i,t_i) \cr
0\,\, {\rm otherwise,\, in\,} \domain(t)/\widehat\defect,  \cr
\end{array}\right.
\eeq
and arbitrary values everywhere else.

A possible choice could be:
\beq
A(x,t) = \sum_i \delta_{t,t_i}\,\,{\sin{(\pi (x-x_i))}\over \pi\, (x-x_i)},
\eeq
but many other choices could also be made.

\subsection{TSSCPP}

An application of what precedes is the partition function for counting TSSCPP (Totally symmetric self-complementary plane partitions) see fig \ref{figTSSCPP}.
Counting TSSCPP has become a famous combinatorics problem, due to its link with alternating sign matrices, Razumov-Stroganoff conjecture, Hecke algebras and qKZ relations \cite{PDFTSSCPP, andrew}.

\figureframex{10}{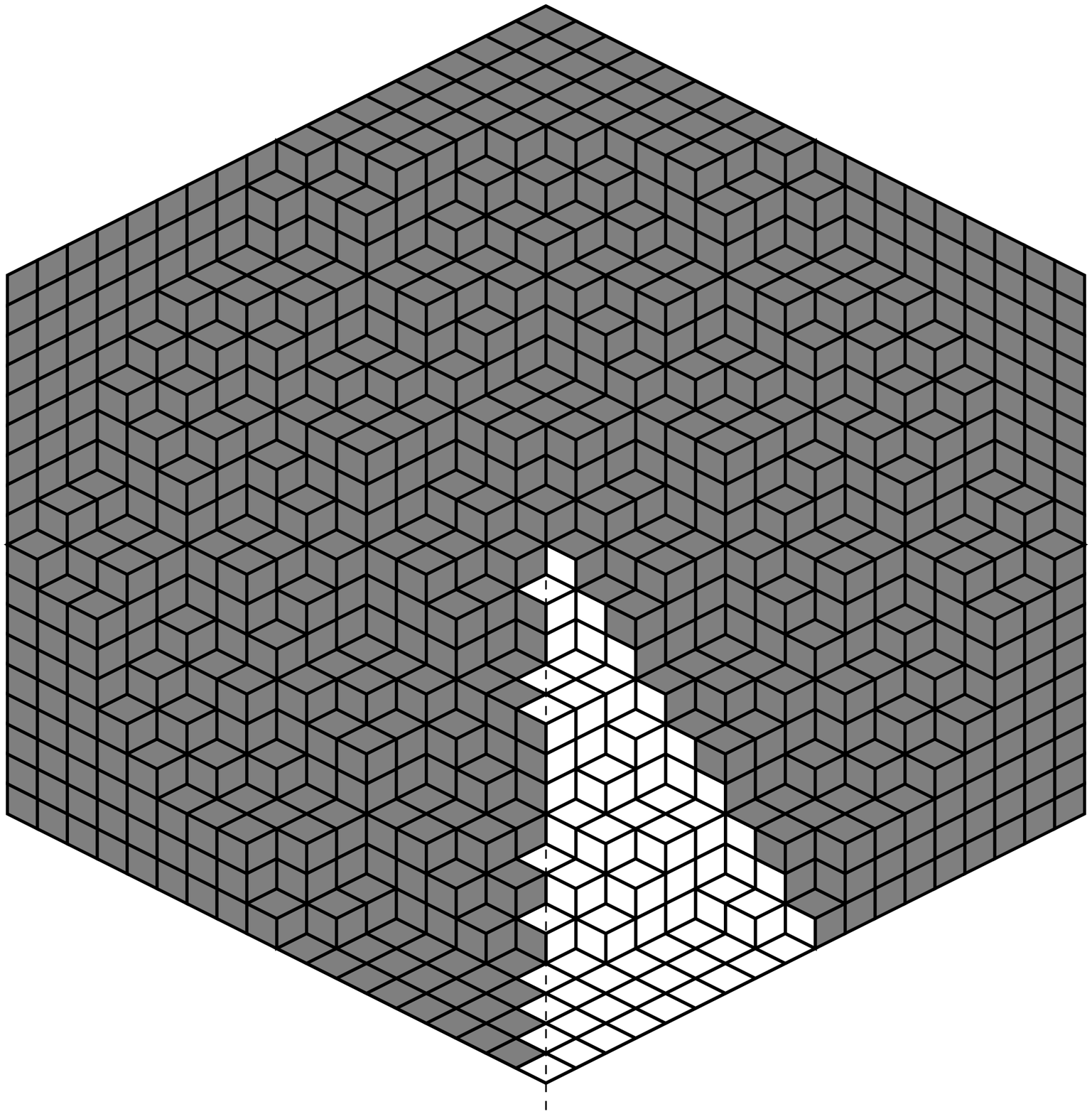}{A TSSCPP, is a plane partition with all symmetries of the hexagon+ self complementarity (i.e. imagine that it is a pile of cubes within a big cube, then it must be equal to its complement).
A TSSCPP configuration is completely determined by a partition of $1/12$th of the hexagon (the white region).\label{figTSSCPP}}

A TSSCPP configuration is completely determined by a partition of $1/12$th of the hexagon.

In terms of a self-avoiding particles process, see fig. \ref{figTSSCPP12}, this can be viewed as $N$ self avoiding particles $h_i(t)$ jumping by $\pm {1\over 2}$, and such that particle $i$ has to follow a straight line after time $t\geq N-i$:
\beq
h_i(t) = N-i+{t\over 2} \qquad {\rm if} \quad t\geq N-i.
\eeq

\figureframex{5}{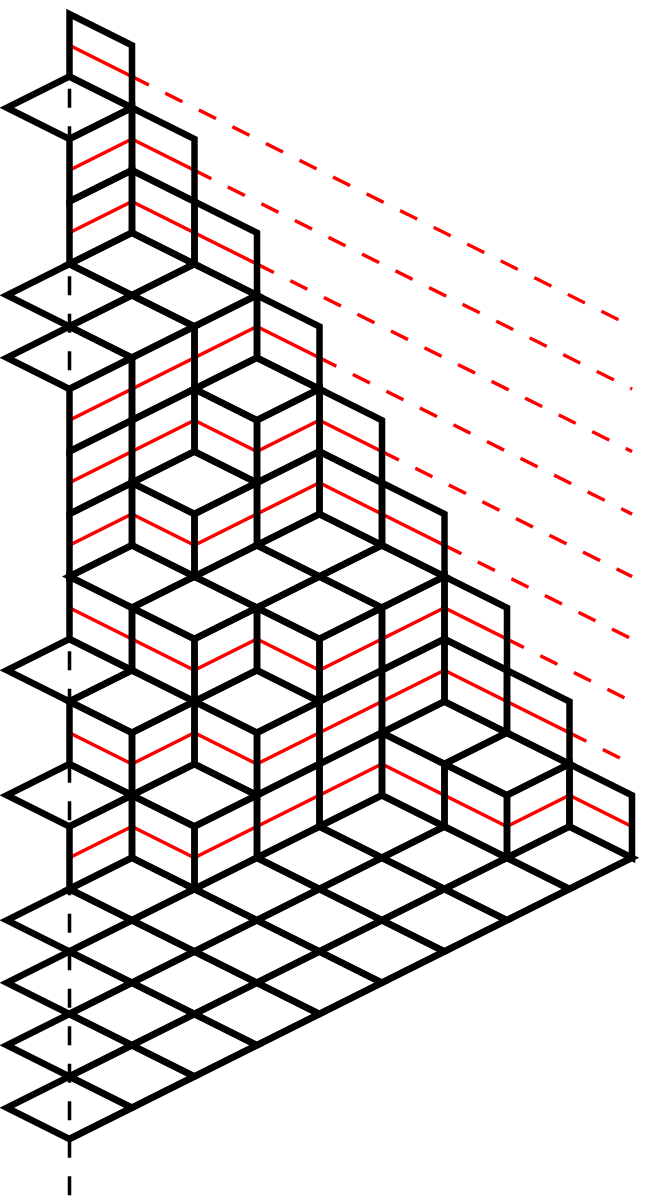}{A plane partition of 1/12th of the hexagon, can be realized as a self-avoiding particles  process such that $h_i(t) = N-i+{t\over 2}$ if $t\geq N-i$.\label{figTSSCPP12}}

In other words, we have a self-avoiding particles  process with some obliged positions.
Also, we don't fix the positions of the particles at time $t=0$ (although we could easily do it in our matrix model) .

Notice that if we fix $h_i(N-1) = N-i+{N-1\over 2}$ at time $t=N-1$, and if we oblige $h_i(N-i)={3\over 2}\,(N-i)$ only at time $t=N-i$, then the self-avoiding particles  process necessarily evolves in a way such that 
$h_i(t) = N-i+{t\over 2}$ if $t\geq N-i$.
In other words, it is sufficient to oblige only 1 position at each time: we oblige the position $({3t\over 2},t)$ at time $t$.

\subsubsection{The matrix model}

We choose for $t=0,\dots,N-1$:
\beq
A_t(x) = {C_t\over x-{3t\over 2}}\,\prod_{j=0}^{2N-2-t}\, (x-{t\over 2}-j)
\eeq
where the constant $C_t=1/t!(2N-2-2t)!$ is such that $A_t({3t\over 2})=1$.

We may also choose:
\beq
A_t(x) = {C_t\over (x-{3t\over 2})\,\Gamma(x-{t\over 2})}\, 
\eeq
where the constant $C_t=(-1)^t/t!$ is such that $A_t({3t\over 2})=1$.

\medskip

For our matrix model, we choose the potential $V_t=0$ for all $t$ (there is no defect).

The TSSCPP partition function is thus realized by the matrix model:
\bea
{\cal Z}_{\rm TSSCPP}(N;q) 
&=& \int \prod_{t=0}^{N-2} dM_t \prod_{t'={1\over 2}}^{N-{3\over 2}} dR_{t'}\,\,\,\,\left(\prod_{t=0}^{N-1}  \Tr A_t(M_t) \right)\,\,\, \cr
&& \quad \prod_{t=0}^{N-2} q^{\Tr M_t}\,\, \prod_{t'={1\over 2}}^{N-{3\over 2}}\ee{-\tr U_{t'}(R_{t'})}\,\,  \ee{\tr R_{t'}(M_{t'+{1\over 2}}-M_{t'-{1\over 2}})}
\eea
where we integrate $M_0,\dots,M_{N-2}$ an $R_{1\over 2},\dots, R_{N-{3\over 2}}$,
and $M_{N-1} = {\rm diag}({N-1\over 2},\dots,{N-1\over 2}+N-1)$ is not integrated upon.
All matrices are of size $N\times N$.


\subsubsection{Examples}

$\bullet$ $N=2$:
\beq
{\cal Z}_{\rm TSSCPP}(2) = \int dM_0 dR_{1\over 2}\,\, \Tr A_0(M_0)\,\,\,  q^{\Tr M_0}\,\ee{-\tr U_{1\over 2}(R_{1\over 2})}\,\, \ee{\tr R_{1\over 2}(M_{1}-M_0)}
\eeq
\beq
M_1 = {\rm diag}({1\over 2},{3\over 2})
\virg
A_0(x) = {1\over 2}\,(x-1)(x-2)
\eeq

\medskip

$\bullet$ $N=3$:
\bea
{\cal Z}_{\rm TSSCPP}(3) 
&=& \int dM_0 dM_1 dR_{1\over 2} dR_{3\over 2}\,\, \Tr A_0(M_0)\,\, \Tr A_1(M_1)\,\,\,  q^{\Tr M_0}\,q^{\Tr M_1}\cr
&& \,\ee{-\tr U_{1\over 2}(R_{1\over 2})}\,\, \ee{\tr R_{1\over 2}(M_{1}-M_0)} \,\ee{-\tr U_{3\over 2}(R_{3\over 2})}\,\, \ee{\tr R_{3\over 2}(M_{2}-M_1)}
\eea
\beq
M_2 = {\rm diag}(1,2,3)
\virg
A_0(x) = {1\over 24}\,(x-1)(x-2)(x-3)(x-4)
\eeq
\beq
A_1(x) = {1\over 2}\,(x-{1\over 2})(x-{5\over 2})(x-{7\over 2})
\eeq

\subsubsection{The spectral curve}

Since we have chosen all $V_t=0$, we have the same spectral curve as for the Trapezoid of section \ref{sectrapezoid}.
Our partition function is computed as an expectation value.

\bea
{\cal Z}_{\rm TSSCPP}(N) 
&=& \sum_{g} (\ln q)^{2g-2+N}\,\, \Res_{\infty_1}\dots\Res_{\infty_N} \cr
&& \qquad W_N^{(g)}(z_1,\dots,z_N) \,\, A_1(\hat X_1(z_1))\dots A_N(\hat X_N(z_N))
\eea

\subsubsection{Fixed position at time t=0}

If we want to fix the positions of particles at time $t=0$, it suffices to introduce a potential
\beq
\ee{-V_0(x)} = {\prod_{i=0}^{2N-2} (x-i)\over \prod_{i=1}^N (x-h_i(0))}.
\eeq

It would be interesting to relate those partition functions with fixed initial conditions for the particles, and see if we have some qKZ equations, and if they can be related to Alternating sign matrices partition functions.

It was already noted \cite{PZJ} that there is a matrix model formulation for the 6-vertex model counting Alternating sign matrices.
It would be interesting to compare the spectral curve of the 6-vertex matrix model, and the spectral curve of the matrix model we introduced here.

\section{Application to topological strings}

Regarding the applications to algebraic geometry and topological strings, 
our method allows to see that Gromov-Witten invariants of the toric Calabi-Yau 3-folds $\mathfrak X=\mathbb C^3$ with branes (called the topological vertex), are the symplectic invariants of their mirror's spectral curve ${\cal S}_{\td{\mathfrak X}}$. In other words, we confirm that the "remodelling the B-model" proposal of BKMP \cite{reBmodel} holds for those $\mathbb C^3$ toric CY 3-folds:
\beq
\hbox{BKMP's claim:} \qquad GW_g(\mathfrak X)=F_g({\cal S}_{\td{\mathfrak X}}).
\eeq
This claim is in the spirit of Dijkgraaf-Vafa conjecture, saying that B-model topological strings should be equivalent to some matrix model.
Here, we have proved, that the topological vertex itself can be written as a matrix model.

In order to prove BKMP for general toric CY 3 folds, It remains to find a "matrix-model way" to glue topological vertices together.

\section{Conclusion}

We have introduced a new formulation of self-avoiding particles model in statistical physics, or lozenge tilings, dimer models, or plane partitions, in terms of matrix models.
Then we briefly discussed how to apply the rich technology of matrix models.

\medskip

However, this is only a starting point, it is only a proposed framework, and we expect that exploiting the immense knowledge of matrix integrals developped since Wigner-Dyson-Mehta, we can find new consequences for self-avoiding particles model.

\medskip

It should be possible to look at many extensions of this approach, for instance taking appropriate limits, it  should be possible to apply this formalism to Dyson processes, or vicious walkers, and recover many results in the matrix model framework.

\medskip

As an application to algebraic geometry, we have only found the "topological vertex", i.e. the building block for computing Gromov-Witten invariants of all toric CY 3-folds, but unfortunately, it is at the present time, not known how to perform the gluing of vertices, in the language of symplectic invariants of \cite{EOFg}. This question needs to be addressed.

\bigskip

\setcounter{section}{0}

\section*{Acknowledgments}
We would like to thank Marcos Mari\~ no for numerous discussions.
We would also like to thank G. Borot, P. Di Francesco, A. Kashani-Poor, P. Ferrari, K. Mallick, M. Mulase, N. Orantin, A. Prats-Ferrer, S. Sheffield, C. Tracy, H. Widom, P. Zinn-Justin for useful and fruitful discussions on this subject.
This work is partly supported by the Enigma European network MRT-CT-2004-5652, 
by the ANR project GIMP (G\'eom\'etrie et int\'egrabilit\'e en physique math\'ematique) ANR-05-BLAN-0029-01, 
by the ANR project GranMa (Grandes Matrices Al\'eatoires) ANR-08-BLAN-0311-01,
by the Enrage European network MRTN-CT-2004-005616,
by the European Science Foundation through the Misgam program,
by the Quebec government with the FQRNT.

\appendix

\section{Gamma-function and q-product}
\label{appGammaq}

\subsection{Gamma-function}

The Gamma-function $\Gamma(x)$ is such that:
\beq
\Gamma(x) = \int_0^\infty \,t^{x-1}\, dt\,\, \ee{-t}
\eeq
If $x=n+1$ is a positive  integer:
\beq
\Gamma(n+1)=n! = n (n-1)(n-2)\dots
\eeq
$\Gamma$ has poles at all negative integers:
\beq
\Gamma(-n+\epsilon ) \mathop{{\sim}}_{\epsilon\to 0}  {(-1)^n\over n!\,\,\epsilon}
\eeq
We have:
\beq
\Gamma(x)\Gamma(-x) = {-\,\pi\over x\sin{(\pi x)}}
\eeq

Stirling formula:
\beq
{\Gamma'(x)\over \Gamma(x)}  = \psi(x)
\sim \ln{x} -{1\over 2x}  - \sum_{n=1}^\infty {B_{2n}\over 2n\, x^{2n}} 
\eeq
where $B_n$ are the Bernouilli numbers.
\beq
B_1=-{1\over 2}
\virg
B_2={1\over 6}
\virg
B_4=-{1\over 30}
\,\, ,\, \dots
\eeq

\subsection{q-product}

The q-product is defined as an infinite product:
\beq
g(x) = \prod_{n=1}^\infty (1-{1\over x}\, q^n) \,\, \stackrel{{\rm def}}{=}\, {1\over (1-{1\over x})\,\Gamma_q({1\over x})}
\eeq
The function $\Gamma_q(x)$ shares many similarities with the function $\Gamma(x)$, by replacing integers with $q-$numbers. It is a quantum $\Gamma$-function.
$q-$numbers are:
\beq
[n] = q^{-n\over 2}-q^{n\over 2}
\eeq
They tend to usual numbers in the limit $q\to 1$.

If $n$ is an integer we have:
\beq
g(q^n) = 0
\virg
g'(q^n) = g(1)\,q^{-{n(n+1)\over 2}}\,\, \prod_{m=1}^{n-1} [m]
 = g(1)\,q^{-{n(n+1)\over 2}}\,\, [n-1]!
\eeq

Stirling formula at small $\ln q$:
\beq
\ln{g(x)} = {1\over \ln q}\,\sum_{n=0}^\infty {(-1)^n\,B_n\over n!}\,(\ln q)^n\,\, Li_{2-n}(1/x) 
\eeq
\beq
x\,{g'(x)\over g(x)} = -\,{1\over \ln q}\,\sum_{n=0}^\infty {(-1)^n\,B_n\over n!}\,(\ln q)^n\,\, Li_{1-n}(1/x) 
\eeq
where $Li_n$ is the Polylogarithme:
\beq
Li_n(x) = \sum_{k=1}^\infty\, {x^k\over k^n}
\eeq
\beq
Li_1(x) = -\ln{(1-x)}
\virg
Li_0(x) = {x\over 1-x}
\virg
Li_n'(x) = {1\over x}\,Li_{n-1}(x)
\eeq
That is:
\beq
{x\,g'(x)\over g(x)} \sim {1\over \ln{q}}\,\Big[
\ln{(1-{1\over x})} -{\ln q\over 2(x-1)}
-\sum_{n=1}^\infty {B_{2n}\over (2n)!}\,\, (\ln q)^{2n}\,Li_{1-2n}(x) \Big]
\eeq

\bigskip

\section{Introduction to symplectic invariants}
\label{appspinv}

Here, we briefly recall the definition of symplectic invariants, but we refer the reader to \cite{EOFg, EOreview} for more details.

\smallskip
They were first introduced in \cite{eynloop1mat, CE} and further formalized in \cite{EOFg}, as a solution of the so called "topological expansion" of matrix integrals.
But then, in \cite{EOFg}, they were defined as algebraic quantities for spectral curves,  independently of the existence of an underlying matrix model.

\subsection{Definition of symplectic invariants}

\medskip
Consider a spectral curve ${\cal S}=({\cal L},x,y)$, where ${\cal L}$ is a compact Riemann surface, and $x$ and $y$ are two analytical functions, such that $dx$ and $dy$ are meromorphic forms on ${\cal L}$.

The branchpoints $a_i$ are defined as the zeroes of $dx$:
\beq
dx(a_i)=0
\eeq
We assume that the spectral curve is regular, i.e. we have a finite number of branchpoints, and they are simple branchpoints, which means that each $a_i$ is a simple zero of $dx$, and $dy(a_i)\neq 0$.
This is equivalent to say that near $a_i$, we have a square root behaviour:
\beq
y(z) \sim y(a_i) + C\,\sqrt{x(z)-x(a_i)}.
\eeq

Since we have a square root branchpoint near $a_i$, this means that in the vicinity of $a_i$, there exists $\bar z\neq z$ on the other branch, i.e. $x(\bar z)=x(z)$:
\beq
{\rm near}\, a_i\, , \qquad \exists !\, \bar z\neq z\, , \quad x(\bar z)=x(z).
\eeq
$\bar z$ is called the conjugate of $z$ near $a_i$. It is defined only locally near branchpoints, and it is not necessarily defined globally.

\medskip

On a Riemann surface ${\cal L}$, there exists a "Bergman kernel", i.e. a symmetric meromorphic 2-form, having a double pole, with no residue, on the diagonal, i.e. which behaves like:
\beq
B(z_1,z_2) \mathop{{\sim}}_{z_1\to z_2}\,\, {dz_1 dz_2\over (z_1-z_2)^2} + O(1)
\eeq
where $z$ is any local parameter.

The Bergman kernel is unique if we also fix its periods $\oint_{{\cal A}_i} B$, but this is not necessary for the definition of symplectic invariants.
Let us assume that we have choosen one Bergman kernel $B(z_1,z_2)$.

\medskip

\bd
We define the following $n-$forms:
\beq
W_1^{(0)}(z) = -y(z)dx(z)
\eeq
\beq
W_2^{(0)}(z_1,z_2) = B(z_1,z_2)
\eeq
and by recursion for $2-2g-n<0$ and $n\geq 1$, and if we denote collectively $J=\{z_2,\dots,z_n\}$:
\bea
W_n^{(g)}(z_1,J) 
&=& \sum_i \Res_{z\to a_i} K(z_1,z)\,\Big[ W_{n+1}^{(g-1)}(z,\bar z,J) \cr
&& +\sum_{h=0}^g \sum'_{I\subset J} W_{1+|I|}^{(h)}(z,I)\,W_{n-|I|}^{(g-h)}(\bar z,J\setminus I) \Big].
\eea
where:
\beq
K(z_1,z) = {\int_{z}^{\bar z} B(z_1,z')\over 2(y(z)-y(\bar z))\,dx(z)}.
\eeq

For $n=0$, and $g\geq 2$, we define:
\beq
F_g=W_0^{(g)} = {1\over 2-2g}\,\,\sum_i \Res_{z\to a_i} W_1^{(g)}(z)\,\Phi(z)
\eeq
where $\Phi$ is any function such that $d\Phi = y dx$.
\ed

The $W_n^{(g)}$'s defined in this way are always symmetric meromorphic $n$-forms, having poles only at branchpoints (except $W_1^{(0)}$ and $W_2^{(0)}$). And $F_g$ does not depend on the choice of integration constant for $\Phi$.

\smallskip

There is also a definition for $F_0$ and $F_1$, but we refer the reader to \cite{EOFg}.

\subsection{Symplectic invariants of genus zero spectral curves}

An interesting example is when ${\cal L}$ is the complex plane $\mathbb C$.
$x$ and $y$ are functions of a complex variable $z$.
In that case we have:
\beq
B(z_1,z_2) = {dz_1\,dz_2\over (z_1-z_2)^2}.
\eeq

For example we have:
\beq
W_3^{(0)}(z_1,z_2,z_3) = \sum_i {1\over x''(a_i)\,y'(a_i)}\,\, {dz_1\,dz_2\,dz_3\over (z_1-a_i)^2\,(z_2-a_i)^2\,(z_3-a_i)^2}.
\eeq


\subsection{Examples of spectral curves and their symplectic invariants}




Let us give a few examples of spectral curves:

$\bullet$ The spectral curve ${\cal S}_{WP}$ defined by the functions $x(z)=z^2, y(z)={1\over 2\pi}\sin{(2\pi z)}$, i.e. $y={1\over 2\pi}\sin{(2\pi\sqrt x)}$ appears in the Weil-Petersson volumes of moduli spaces, its $F_g$'s are the Weil-Petersson volumes:
$ F_g({\cal S}_{WP}) = {\rm Vol}({\cal M}_{g})$, see \cite{EMvol}.

$\bullet$ The spectral curve ${\cal S}_{\rm Airy}$ defined by $y^2-x=0$, i.e. $y=\sqrt{x}$ is associated to the Airy kernel law, and to the universal behavior of extreme eigenvalues distribution, i.e. to the Tracy-Widom law \cite{TW}. The $F_g$'s of the Airy curve are equal to zero: $F_g({\cal S}_{\rm Airy})=0$.

$\bullet$ The spectral curve $y=\sqrt{{\rm Pol}_{2m+1}(x)}$, where ${\rm Pol}_{2m+1}$ is a polynomial of degree $2m+1$, is associated to the $(2m+1,2)$ minimal model in conformal field theory (with central charge $c=1-3\,{(2m-1)^2\over 2m+1}$), i.e. a reduction of the KdV hierarchy. Its $F_g$'s are related to the KdV Tau-function, and they can be computed by the asymptotic expansion of the solution of a Painlev\'e type equation (in fact the $m+1^{\rm st}$ Gelfand-Dikii equation $R_{m+1}(u(t))=t$).
The case $m=0$ i.e. the $(1,2)$ minimal model is the Airy case.
The case $m=1$ i.e. the $(3,2)$ minimal model is called pure gravity, and the case $m=2$ i.e. the $(5,2)$ minimal model is called Lee-Yang singularity.

$\bullet$ Spectral curves of type ${\rm Pol}(\ee{x},\ee{y})=0$ appear in topological strings, where ${\rm Pol}$ is a polynomial.
More precisely, consider a Toric Calabi-Yau 3-fold $\mathfrak X$. Through mirror symmetry \cite{bookmirror}, it has a mirror $\td{\mathfrak X}$, which is also a toric Calabi-Yau 3-fold, and which is given by a submanifold of $\mathbb C^4$ of equation $H(\ee{x},\ee{y})=uv$ where $H$ is a polynomial. The spectral curve ${\cal S}_{\td{\mathfrak X}}$, is defined as the singular locus of  $\td{\mathfrak  X}$, which satisfies the equation $H(\ee{x},\ee{y})=0$.
The "remodelling the B-model" conjecture of BKMP \cite{reBmodel}, is that the generating function of Gromov-Witten invariants $GW_g(\mathfrak X)$ of genus $g$ of $\mathfrak X$, is equal to the symplectic invariant $F_g({\cal S}_{\td X})$ of the spectral curve ${\cal S}_{\td{\mathfrak  X}}$ of the mirror:
\beq
GW_g({\mathfrak  X}) \stackrel{?}{=} F_g({\cal S}_{\td{\mathfrak  X}})
\eeq
This conjecture was proved in few cases, and in particular for the family of Hirzebruch manifolds ${\mathfrak X}_p= O(-p)\oplus O(p-2) \to \mathbb P^1$ (which include the connifold), see  \cite{eynLP}.
It was also partially proved for $SU(n)$ Seiberg-Witten theories \cite{KSSW}, and for $2^*$ geometries \cite{Sulk}.

\subsection{Some properties of symplectic invariants}

Let us now give a few properties of symplectic invariants. We refer the reader to \cite{EOFg, EOreview} for the general theory, and for proofs and details.

\medskip
\underline{\bf Properties:}
\smallskip

$\bullet$ {\bf Homogeneity:} If we rescale $y\to \l y$, which we note $\l {\cal S} = (x,\l y)$, we have:
\beq\label{FgHomogeneity}
F_g(\l{\cal S})=\l^{2-2g}\,F_g({\cal S})
\virg
W_n^{(g)}(\l{\cal S})=\l^{2-2g-n}\,W_n^{(g)}({\cal S}).
\eeq
In particular that implies $F_g(-{\cal S})=F_g({\cal S})$.

\smallskip

$\bullet$ {\bf Symplectic invariance:}
if two spectral curves ${\cal S}=\{y(x)\}$ and $\td{\cal S}=\{\td{y}(\td{x})\}$ are symplectically equivalent, i.e. if there is a map from $\mathbb C\times \mathbb C$ to $\mathbb C\times \mathbb C$ which sends one spectral curve to the other and conserves the symplectic form $dx\wedge dy=d\td{x}\wedge d\td{y}$, then:
\beq
F_g({\cal S}) = F_g(\td{\cal S})
\eeq
For instance we can change $y\to y+R(x)$ where $R(x)$ is a rational function, or $(y\to x, x\to -y)$, or $(x\to \ln x, y\to xy)$, ... etc, without changing $F_g$.

The $W_n^{(g)}$ with $n\geq 1$ are not symplectic invariants, they change under symplectic transformations, but they change in a covariant way. For example $W_1^{(g)}({\cal S}) - W_1^{(g)}(\td{\cal S})$ is an exact form.

\smallskip

$\bullet$ {\bf Variations:}
Consider an infinitesimal deformation of the spectral curve  
$(x,y)\to (x+\epsilon\delta x,y+\epsilon \delta y)$, such that $\Omega = \delta y \, dx - \delta x \, dy$ is a meromorphic differential form on the spectral curve. Any meromorphic form $\Omega$ is dual to a cycle $\Omega^*$ on the curve, the duality pairing is realized through the Bergman kernel:
\beq
\Omega(z) = \oint_{\Omega^*} B(z,z')
\eeq
Then, the infinitesimal variation of $F_g$ and $W_n^{(g)}$ is given by:
\beq\label{eqWngvar}
{d\over d\epsilon}\, W_n^{(g)}(z_1,\dots,z_n) = \oint_{\Omega^*}\, W_{n+1}^{(g)}(z_1,\dots,z_n,z')
\eeq
The infinitesimal variation of $F_g$ is the case $n=0$: ${d F_g\over d\epsilon}= \oint_{\Omega^*}\, W_{1}^{(g)}(z')$.

\smallskip

$\bullet$ {\bf Limits:}
Consider a one parameter family of spectral curves ${\cal S}(t)$, such that at $t=t_c$, the spectral curve ${\cal S}(t_c)$ becomes singular. Consider its blow up: ${\cal S}(t) \sim (t-t_c)^{\nu}\, {\cal S}_c + o((t-t_c)^\nu)$, where the exponent $\nu$ is chosen such that the curve ${\cal S}_c$ is regular. Then we have:
\beq
F_g({\cal S}(t)) \sim (t-t_c)^{(2-2g)\nu}\, F_g({\cal S}_c) + o((t-t_c)^{(2-2g)\nu})
\eeq
and more generally
\beq
W_n^{(g)}({\cal S}(t)) \sim (t-t_c)^{(2-2g-n)\nu}\, W_n^{(g)}({\cal S}_c) + o((t-t_c)^{(2-2g-n)\nu}).
\eeq
This theorem is very useful, it gives the asymptotic behaviors. For instance, if we zoom near a regular branch point, the spectral curve always behaves like $y=\sqrt{x}$, i.e. ${\cal S}_c={\cal S}_{Airy}$, and we find the Airy law, and Tracy-Widom.
Near an algebraic cusp singularity $y\sim x^{p/q}$, we find the spectral curve ${\cal S}_c={\cal S}_{(p,q)}$ of the $(p,q)$ minimal model of conformal field theories (of central charge $c=1-6\,{(p-q)^2\over pq}$), which is a reduction of the KP hierarchy.

\bigskip

Thes are just a few of the properties satisfied by the $F_g$'s. See \cite{EOFg} for more.
For instance there are also modular properties and holomorphic anomaly equations, Hirota equations, and diagrammatic representations.

\section{Solution of matrix models}
\label{appchainmat}

Let us present a brief review about matrix models.

\subsection{Generalities on the solution of matrix models}

Consider a matrix integral of type:
\beq\label{Zgenchain}
{\cal Z}=\int_{\prod H_N({\cal C}_i)} \prod_{i=1}^{p} dM_i\,\,\, \ee{-Q \Tr [\sum_{i=1}^p V_i(M_i) + \sum_i c_i M_i M_{i+1}]}
\eeq
where $H_N({\cal C}_i)$ is the set of normal matrices having their eigenvalues on contour ${\cal C}_i$, and where $M_{p+1}$ is not integrated upon.

If one wants to find a large $Q$ expansion of $\ln {\cal Z}$, the answer, is that one has first to compute the "spectral curve" ${\cal S}$ (defined below), then, in \cite{EPrats}, it is proved that:

\bt\label{thZFgMM}
If ${\cal S}$ is the spectral curve associated to potentials $V_i$ and paths ${\cal C}_i$, and if ${\cal Z}$ has a large $Q$ expansion of the form
\beq\label{ThZFg}
\ln{{\cal Z}} = \sum_{g=0}^\infty Q^{2-2g}\, F_g({\cal S})
\eeq
then, the coefficient  $F_g=F_g({\cal S})$ is the symplectic invariant of degree $2-2g$ of the spectral curve ${\cal S}$.
Symplectic invariants $F_g({\cal S})$ of a spectral curve ${\cal S}$ were defined and introduced in \cite{EOFg}.

\et

\medskip

We shall explain the meaning of this theorem below in further details.
We shall explain how to compute the spectral curve ${\cal S}$ of a matrix model, and how to compute its symplectic invariants $F_g$. We recall the definition of symplectic invariants in appendix \ref{appspinv}.

Let us just mention that finding the spectral curve is rather automatic, it is mostly an algebraic task, and for most examples the spectral curve is a rather simple object.
Here, we have the the spectral curves of Kenyon-Okounkov-Sheffield \cite{KOS}.

\smallskip

Then, computing the symplectic invariants of a spectral curve, is rather easy. The definition of symplectic invariants \cite{EOFg} involves computing residues of rather standard functions (exponentials, logs, rational functions, ...), and can be completely automatized, see appendix \ref{appspinv}.
It is really an efficient method.
For example $F_1$ can often be computed by hand.
Moreover, symplectic invariants satisfy many properties, which make them really convenient to use, for example there are formulae for computing their derivatives with respect to any parameter, and formulae for finding their limits near singularities.

\subsection{Generalities about loop equations}

The theorem \ref{thZFgMM} for the chain of matrices was proved from loop equations \cite{EPrats}.

Loop equations are obtained by integration by parts in the matrix integral, or equivalently, by writing that an integral is invariant under change of variable.
For example, assume that we make an infinitesimal change of variable in \eq{Zgenchain}:
\beq
M_3 \to M_3 + \epsilon M_2^5 M_3^3 + O(\epsilon^2)
\eeq
one finds the Jacobian to first order in $\epsilon$:
\beq
dM_3\to dM_3\,\Big(1+\epsilon(\Tr M_2^5\Tr M_3^2 + \Tr M_2^5 M_3 \Tr M_3 + \Tr M_2^5 M_3^2 \Tr {\rm Id} )+O(\epsilon^2)\Big)
\eeq
Writing that the integral is invariant to first order in $\epsilon$ implies:
\bea\label{exloopeq}
&& <\Tr M_2^5\Tr M_3^2 + \Tr M_2^5 M_3 \Tr M_3 + \Tr M_2^5 M_3^2 \Tr {\rm Id}> \cr
&=& Q\,<\Tr V'_3(M_3)M_2^5 M_3^3 + c_2 \Tr M_2^6 M_3^3 + c_3 \Tr M_2^5 M_3^2 M_4>
\eea
Similarly, by considering other appropriate changes of variables, one can write a family of relationships of the type \eq{exloopeq} between expectation values, this was done systematically in \cite{EPrats}.

\smallskip

Instead of considering expectation values of powers $<\Tr M^k>$, is is more convenient to group them in a formal generating function $\bar W_1(x) = <\Tr {1\over x-M}> = \sum_{k=0}^\infty {<\Tr M^k>\over x^{k+1}}$.
This leads to introduce:
\beq\label{defWngbarchaingen}
\bar W_n(x_1,\dots,x_n) = <\Tr {1\over x_1-M_1}\Tr {1\over x_2-M_1} \dots \Tr {1\over x_n-M_1}>_c
\eeq
(where the subscript $<.>_c$ means the cumulant, or connected part).
Since we made the hypothesis that there exists a large $Q$ expansion we write:
\beq\label{genMMtopexp}
\bar W_n(x_1,\dots,x_n) = \sum_{g=0}^\infty  Q^{2-2g-n}\,\, \bar W_n^{(g)}(x_1,\dots,x_n).
\eeq

The problem consists in finding a family of relationships among those quantities, and which allow to find the solution, i.e. compute all of them.
This was done in \cite{EPrats,eynmultimat}.

The result of \cite{EPrats}, is that the $W_n^{(g)}(x_1,\dots,x_n)=\bar W_n^{(g)}(x_1,\dots,x_n)\,dx_1\dots dx_n$'s are the n-forms of theorem \ref{thZFgMM}, i.e. the $W_n^{(g)}$'s of \cite{EOFg}, and where the spectral curve ${\cal S}=(x,y)$ is  $y=-\bar W_1^{(0)}(x)$.

In some sense, $\bar W_1^{(0)}(x)$ can be viewed as the "large $Q$" limit of the resolvent of the matrix $M_1$ (the first matrix in the chain \eq{Zgenchain}), it is often called the "equilibrium distribution of eigenvalues of $M_1$", but one should take this denomination with some care.
Indeed, $\bar W_1^{(0)}$ can be shown to be the "weak" large $Q$ limit of the resolvent, for some potentials (in particular in the potentials don't depend on $Q$), and some integration paths ${\cal C}_i$, but this is probably wrong in general.
In fact, in our case, the potentials do depend on $Q$, and saying that the spectral curve is the "large $Q$" limit of the resolvant is slightly wrong (and at least one needs to make precise the notion of large $Q$ limit).

The only correct definition of the spectral curve, which corresponds to theorem \ref{thZFgMM}, is that $y=\bar W_1^{(0)}(x)$, is the first term in the formal large $Q$ expansion, doing as if the potentials were independent of $Q$.

\medskip

Working to order $g=0$ in the expansion \eq{genMMtopexp} within the loop equations, is formally equivalent to replacing expectation values of product of traces, by product of expectation values of each trace (indeed, the connected part comes with a factor $Q^{2-2g-n}$, whereas the non-connected term, i.e. the factorized term comes with a factor $Q^{2-2g+n}$, i.e. all terms which are not disconnected don't contribute to the highest power of $Q$):
\beq
<\Tr A_1 \Tr A_2  \dots \Tr A_k> \,\longrightarrow\, <\Tr A_1><\Tr A_2>\dots<\Tr A_k>
\eeq
This formal manipulation allows to find an equation which determines the spectral curve $\bar W_1^{(0)}(x)$.


\subsection{Spectral curve of the chain of matrices}

Here, we shall just give a "ready to use"  recipe of how to find the spectral curve for a chain of matrices matrix model of type \eq{Zgenchain}. This recipe is extracted from \cite{EPrats}, and it is really technical to explain how to obtain it. However, it is easy to use.

\subsection{General case}
\label{secspcurvegen}

The spectral curve of a matrix integral of type \eq{Zgenchain}:
\beq\label{Zgenchainbis}
{\cal Z}=\int_{\prod H({\cal C}_i)} \prod_{i=1}^{p} dM_i \ee{-Q \Tr [\sum_{i=1}^p V_i(M_i) + \sum_i c_i M_i M_{i+1}]}
\eeq
is characterized by a set  (see \cite{eynmultimat, EPrats}) of $p+2$ analytical functions of a variable $z$ ($z$ belongs to a Riemann surface ${\curve}$). There is one such analytical function for each matrix $M_i$, $i=1,\dots,p+1$ of the chain, plus one additional function at the end of the chain. Let us call them:
\beq
\hat X_i(z)\quad , \quad i=0,\dots,p+1.
\eeq
Those functions are completely determined by the following system of equations:
\beq\label{spcurveeqmotgen}
\forall\, i=2,\dots,p\, , \qquad \quad c_{i-1} \hat X_{i-1}(z)+ c_i \hat X_{i+1}(z) + V'_i(\hat X_i(z)) = 0
\eeq
and
\beq
\bar W_1^{(0)}(\hat X_1(z)) = \hat X_0(z) = V'_1(\hat X_1(z)) - c_1 \hat X_{2}(z),
\eeq
together with the conditions:

$\bullet$ 
$\hat X_p(z) $ has simple poles at the values of $z$ such that $\hat X_{p+1}(z)$ is an eigenvalue of $M_{p+1}$.
Let us call $\zeta_i$, the value of $z$ such that $\hat X_{p+1}(z)=\lambda_i$ the $i^{\rm th}$ eigenvalue of $M_{p+1}$.
Then we have in the vicinity of $z\to \zeta_i$:
\beq
\hat X_p(z) \sim {1\over \hat X_{p+1}(z)-\lambda_i}
\eeq

$\bullet$ The function $\bar W_1^{(0)}(x)$,  is analytical outside of the cuts $[a_i,b_i]$, the endpoints of the cuts being zeroes of $d\hat X_1$.
Near $\hat{X}_1\to\infty$ (only in the physical sheet), we have:
\beq
\hat X_0(z) \sim {N/Q\over \hat X_1(z)} + O(1/\hat X_1(z)^2)
\eeq

$\bullet$ the genus of the Riemann surface ${\cal L}$, and the period integrals $\oint_{{\cal A}_i} \hat X_0 d\hat X_1$ on non-contractible cycles ${\cal A}_i$ of ${\cal L}$, are related to the choice of contours ${\cal C}_i$ in the matrix integral \eq{Zgenchainbis}.
The following quantity is called "filling fraction"
\beq
\epsilon_i = {1\over 2i\pi}\,\oint_{{\cal A}_i} \hat X_0 d\hat X_1,
\eeq
and it is such that $Q\epsilon_i$ is the number of eigenvalues of $M_1$ contained in the domain surrounded by the projection in $\mathbb C$ of the cycle ${\cal A}_i$ by the application $\hat X_1$.

More generally, ${Q\over 2i\pi}\,\oint_{{\cal A}_i} \hat X_{j-1} d\hat X_{j}$ is the number of eigenvalues of $M_{j}$ contained in the projection in $\mathbb C$ of the cycle ${\cal A}_i$ by the application $\hat X_j$.

\bigskip

Finding the spectral curve for arbitrary potentials $V_i$ and arbitrary contours ${\cal C}_i$ can be extremely tedious.
However, here we are interested in formal matrix integrals, which are defined as formal power series expansions in some formal parameter, and in particular, the spectral curve we are looking for, must also be found as a formal power series. In particular, the genus is the genus of the spectral curve when we send the formal parameter to $0$, and in general that simplifies the computation of the spectral curve a lot. Very often the genus is in fact zero.

\subsubsection{Topological expansion}
\label{sectopexpgen}

When we have determined the spectral curve ${\cal S}$, i.e. the two functions $x(z)=\hat X_1(z)$ and $y(z)=\hat X_0(z)$:
\beq
{\cal S}=(x,y)={\cal S}_{1,0}=(\hat X_1,\hat X_0)
\virg
x(z)=\hat X_1(z), \,\, y(z)=\hat X_0(z),
\eeq
we have:
\beq
\ln{{\mathcal Z}} = \sum_{g=0}^\infty F_g({\cal S}).
\eeq

It is then useful to use the symplectic invariance of the $F_g$'s.

Consider the following spectral curves:
\beq
{\cal S}_{i,j}=(\hat X_i,\hat X_j)
\eeq
Due to \eq{spcurveeqmotgen}, we have:
\beq
c_i d\hat X_{i+1} \wedge d\hat X_i = -c_{i-1} d\hat X_{i-1} \wedge d\hat X_i = c_{i-1} d\hat X_{i} \wedge d\hat X_{i-1}
\eeq
i.e. all the spectral curves ${\cal S}_{i,i+1}$ are symplectically equivalent:
\beq
c_i {\cal S}_{i,i+1} \equiv c_{i-1} {\cal S}_{i-1,i} \equiv - c_{i-1} {\cal S}_{i,i-1}
\eeq
and therefore, the theorem of symplectic invariance of the $F_g$'s \cite{EOFg, EOsym} gives $\forall i$:
\beq
c_{i}^{2-2g}\,\, F_g({\cal S}_{i,i+1}) = c_{i-1}^{2-2g}\,\, F_g({\cal S}_{i-1,i}) = c_{i-1}^{2-2g}\,\, F_g({\cal S}_{i,i-1})
\eeq
In other words, the $F_g$'s can be computed with the spectral curve ${\cal S}=(x,y)$ where $x$ and $y$ are any two consecutive $\hat X_i, \hat X_{i+1}$, we don't need to choose the pair $\hat X_1,\hat X_0$.

\subsubsection{Densities and correlation functions}
\label{secdensitygen}

For the general chain of matrices, the $W_n^{(g)}$'s were defined as the Stieltjes transforms of the density correlation functions of the first matrix $M_1$ in the chain, i.e. the resolvents \eq{defWngbarchaingen}.
The densities can be recovered by taking the discontinuities:
\beq
\rho(h) = \sum_g Q^{1-2g}\,\rho^{(g)}(h) = \left< \sum_i \delta(h-h_i) \right>
\eeq
where
\beq
\rho^{(g)}(h) = {1\over 2i\pi}\,\, \left( \bar W_1^{(g)}(h-i0)-\bar W_1^{(g)}(h+i0) \right).
\eeq
We recall that the spectral curve used to compute $W_n^{(g)}$ here, is the spectral curve ${\cal S}_{1,0}=(\hat X_1,\hat X_0)$, and the $W_n^{(g)}$'s are not invariant under symplectic transformations.

However, just by applying \eq{eqWngvar}, one can see that, for any matrix $M_j$ of the chain, one has that the density of eigenvalues of $M_j$:
\beq
\rho_j(h) = \sum_g Q^{1-2g}\,\rho_j^{(g)}(h) = \left< \sum_i \Tr \delta(h-M_j) \right>
\eeq
is given by:
\beq
\rho_j^{(g)}(h) = {1\over 2i\pi\,\, d\hat X_j(h)}\,\, \left( W_1^{(g)}(h-i0)-W_1^{(g)}(h+i0) \right)_{{\cal S}_{j,j-1}}.
\eeq

Therefore, the density of eienvalues of each matrix of the chain, can be computed to all orders in the large $Q$ expansion, using the $W_n^{(g)}$'s defined in \cite{EOFg}.

\bigskip

In the general chain of matrices, the support of the densities must be related to the integration paths ${\cal C}_i$'s:
One defines the support of densities as the loci where densities are real.
The supports of densities, together with their analytical continuations, must be homologicaly equivalent to the integration paths ${\cal C}_i$'s.
And saying that the paths ${\cal C}_i$ are steepest descent paths, is more or less equivalent to saying that the densities $\rho_i$ must be real and positive on their supports, and $i$ times  the analytical continuation of the densities, along the analytical continuations of the supports, must be real and strictly positive (stability condition).
This is a highly non trivial condition in general.

Fortunately, for formal matrix integrals, defined as formal power series in a formal parameter, the small limit of the parameter often gives a very simple matrix model, with a very simple spectral curve, and it is often easy to check positivity to leading order, and then, that the formal power series expansion does not destroy the supports and positivity.


\begin{thebibliography}{99}
\bibliographystyle{plain}

\bibitem{AVM} M Adler, P van Moerbeke,  ``The spectrum of coupled random matrices'',
 Annals of Mathematics, 1999.

\bibitem{Aldous} D.J. Aldous, P. Diaconis, Longest increasing subsequences: from patience sorting to the Baik-Deift-Johansson theorem, Bull. Amer. Math. Soc. 36 (1999), 413-432.


\bibitem{topvertex} M. Aganagic, A. Klemm, M. Mari\~ no, C. Vafa, The topological vertex, hep-th 0305132.


\bibitem{andrew} G. Andrews, Plane partitions. V. The TSSCPP conjecture, J. Combin. Theory Ser. A 66 (1994), no. 1, 28Ð39.


\bibitem{baik} J. Baik, E.M. Rains, Symmetrized random permutations, Random Matrix Models and Their Applications, vol. 40, Cambridge University Press, 2001, pp. 1-19.

\bibitem{BDJ} J. Baik, P. Deift, K. Johansson, On the distribution of the length of 
the longest increasing subsequence of random permutations, Journal of 
AMS, 12 (1999), 1119Ð1178.

\bibitem{Behrend} K. Behrend, Gromov-Witten invariants in algebraic geometry, Invent. 
Math. 127 (1997), 601Ð617. 

\bibitem{EBerg} M. Berg\`ere, B. Eynard, Determinantal formulae and loop equations,
  math-ph: arxiv.0901.3273.


\bibitem{bogo} N. M. Bogoliubov, Determinantal Representation of the Time-Dependent Stationary Correlation Function for the Totally Asymmetric Simple Exclusion Model,
 SIGMA 5 (2009), 052, 
arXiv:0904.3680.

\bibitem{BOO}
A Borodin, A Okounkov, G Olshanski, Asymptotics of Plancherel measures for symmetric groups,
Journal of the American Mathematical Society, 2000 - ams.org.

\bibitem{reBmodel} V. Bouchard, A. Klemm, M. Mari\~ no, S. Pasquetti, Remodelling the B-model, hep-th 0709.1453.

\bibitem{boutillier} C. Boutillier, The bead model and limit behaviors of dimer models, 
	arXiv:math/0607162.

\bibitem{BrezHik} E. Brezin and S. Hikami, Universal singularity at the closure of a gap in a 
random matrix theory, cond-mat/9804023. Phys. Rev. E (3) 57 (1998), no. 
4, 4140Ð4149. 
	

\bibitem{bryan} J. Bryan and R. Pandharipande, The local Gromov-Witten theory of curves. 
math.AG/0411037. 


\bibitem{CE} L. Chekhov and B. Eynard, ÒHermitean matrix model free energy: Feynman graph 
technique for all genera,Ó JHEP 0603, 014 (2006) [arXiv:hep-th/0504116]. 

\bibitem{Cohn} H. Cohn, M. Larsen, and J. Propp, The shape of a typical boxed plane partition, New York J. Math. 4 (1998), 
137Ð165, arXiv:math/9801059.



\bibitem{DKMVZ} P. Deift, T. Kriecherbauer, K. T-R McLaughlin, S. Venakides, X. Zhou,
``Uniform asymptotics for polynomials orthogonal with respect to varying exponential weights and applications to universality questions in random matrix theory'', 
Comm. Pure and applied Maths., Vol 52, 1335-1425, 1999.

\bibitem{derrida} Derrida, An exactly soluble non-equilibrium system: the asymmetric simple exclusion process, Phys. Rep. 301, 65, (1998).

\bibitem{derrida2} B. Derrida, M. R. Evans, V. Hakim, V. Pasquier, 1993, Exact solution of a 1D asymmetric exclusion model using a matrix formulation, J. Phys. A: Math. Gen. 26, 1493.


\bibitem{ZJDFG} P. Di Francesco, P. Ginsparg, J. Zinn-Justin, 2D Gravity and Random Matrices, Phys. Rep. 254, 1 (1995).

\bibitem{PDFTSSCPP}
P. Di Francesco, Totally symmetric self-complementary plane partitions and the quantum Knizhnik-Zamolodchikov equation: a conjecture, 
J. Stat. Mech. Theory Exp. (2006), no. 9, P09008, 14 pp., 
arXiv:cond- mat/0607499.

\bibitem{ZJ3} P. Di Francesco, P. Zinn-Justin,  Around the Razumov-Stroganov conjecture: proof of a multi-parameter sum rule,
arXiv:math-ph/0410061.

\bibitem{DFZJZ} P. Di Francesco, P. Zinn-Justin, J.-B. Zuber, 
Determinant Formulae for some Tiling Problems and Application to Fully Packed Loops,
 arXiv:math-ph/0410002.
    
        
  
\bibitem{eynloop1mat}
 B.~Eynard,
``Topological expansion for the 1-hermitian matrix model correlation
functions,''
arXiv:hep-th/0407261.
 
 \bibitem{eynform}
 B.~Eynard, ``Formal matrix integrals and combinatorics of maps,''
  arXiv:math-ph/0611087.
 
\bibitem{eynLP} B. Eynard, All orders asymptotic expansion of large partitions,\\ 
math-ph: arxiv.0804.0381.
  
\bibitem{EOFg}
B. Eynard and N. Orantin, ``Invariants of algebraic curves and topological expansion'',
arXiv:math-ph/0702045.

\bibitem{EOreview}
B. Eynard and N. Orantin, Algebraic methods in random matrices and enumerative geometry,
  math-ph: arxiv.0811.3531.


\bibitem{EMvol}
B. Eynard, Recursion between Mumford volumes of moduli spaces, math-ph: arXiv:0706.4403.
  
\bibitem{EOsym} B. Eynard and N. Orantin, Topological expansion of mixed correlations in the hermitian 2 Matrix Model and $x-y$ symmetry of the $F_g$ algebraic invariants, math-ph/arXiv:0705.0958, to appear in J.Phys A.



\bibitem{eynMehta} B. Eynard, M.L. Mehta, Matrices coupled in a chain: eigenvalue correlation, 
Journal of Physics A 19 (1998) 4449,
arxiv: cond-mat/9710230.

\bibitem{eynmultimat} B. Eynard, Master loop equations, free energy and correlations for the chain of matrices,  JHEP11(2003)018,
hep-th/0309036.

\bibitem{EPrats}  B. Eynard, A. Prats Ferrer, Topological expansion of the chain of matrices, 
math-ph: arxiv.0805.1368.

\bibitem{eynbirthcut} B. Eynard, Universal distribution of random matrix eigenvalues near the ``birth of a cut'' transition, math-ph/0605064, JSTAT, Theory and Experiment P07005 (2006) P07005.



\bibitem{ferrari1}  P. L. Ferrari, M. Praehofer, One-dimensional stochastic growth and Gaussian ensembles of random matrices,
 proceedings of ''Inhomogeneous Random Systems 2005'', Markov Processes Relat. Fields 12 (2006) 203-234, arXiv:math-ph/0505038.

 \bibitem{ferrari2} Patrik L. Ferrari, Polynuclear growth on a flat substrate and edge scaling of GOE eigenvalues, Comm. Math. Phys., 252 (2004), 77-109,     arXiv:math-ph/0402053.

    
\bibitem{GV} IM. Gessel, X. Viennot, Determinants, paths, and plane partitions, preprint, 1989 - 132.180.197.15.


\bibitem{Mallick} O. Golinelli, K. Mallick, 2006, The asymmetric simple exclusion process : an integrable model for non-equilibrium statistical mechanics, J. Phys. A: Math. Gen. 39 12679.

\bibitem{HC} Harish-Chandra, Amer. J. Math. 79 (1957) 87-120.
  
 
 \bibitem{bookmirror} K. Hori, S. Katz, A. Klemm, R. Pandharipande, R. Thomas, C. Vafa, R. Vakil, E. 
Zaslow,   AMS, Providence, 2003. 

\bibitem{IZ} C. Itzykson and J.-B. Zuber, Jour. Math. Phys. 21 (1980) 411.

\bibitem{Jarctic} K. Johansson, The arctic circle boundary and the Airy process, Ann. Probab. 33 (2005), 1-30.

\bibitem{johansson} K. Johansson, The longest increasing subsequence in a random permuta- 
tion and a unitary random matrix model, Math. Res. Lett., 5 (1998), 
no. 1-2, 63Ð82.

\bibitem{kasteleyn}
Kasteleyn, P. W. (1961), "The statistics of dimers on a lattice. I. The number of dimer arrangements on a quadratic lattice", Physica 27 (12): 1209Ð1225, doi:10.1016/0031-8914(61)90063-5.

\bibitem{kasteleyn1}
P. Kasteleyn, Graph theory and crystal physics, Graph Theory and Theoretical Physics, Academic Press, London, 1967, pp. 43Ð110.

\bibitem{KSSW} A. Klemm, P. Sulkowski,  Seiberg-Witten theory and matrix models,
eprint arXiv:0810.4944.


\bibitem{Kenyon} R. Kenyon, Local statistics of lattice dimers, Ann. Inst. H. Poincar\'e Probab. 
Statist. 33 (1997), no. 5, 591Ð618.

    
\bibitem{KOS}
R. Kenyon, A. Okounkov, S. Sheffield,
Dimers and Amoebae,
Annals of mathematics, Vol. 163, N¼ 3,2006 ,1019-1056,
arXiv:math-ph/0311005.





\bibitem{KazakovRMTcrit} V.A. Kazakov, I. Kostov, A.A. Migdal,
"Critical properties of randomly triangulated planar random surfaces", {\em Physics Letters} {\bf B}, 1985.

\bibitem{Krat1} C. Krattenthaler, ``Generating functions for plane partitions of a given shape'', manuscripta mathematica, vol 69, 1, 1432-1785, 1990.

\bibitem{liggett} T.M. Liggett, Coupling the simple exclusion process, Ann. Probab. 4 (1976), 339-356.


\bibitem{mmhouches}
  M.~Mari\~no, ``Les Houches lectures on matrix models and topological strings,''
  arXiv:hep-th/0410165.


\bibitem{Mehtamultimat} M. L. Mehta, ``A method of integration over matrix variables'', Comm. Math. Phys., vol 79, 3, 327-340, 1981.

\bibitem{MehtaBook}
M.L. Mehta, Random Matrices, 3rd edition, (Academic Press, New York, 2004).

\bibitem{MOO} D. Maulik, A. Oblomkov, A. Okounkov, R. Pandharipande,
 Gromov-Witten/Donaldson-Thomas correspondence for toric 3-folds,
 arXiv:0809.3976.


 \bibitem{nv}
 A.~Neitzke and C.~Vafa, 
 ``Topological strings and their physical applications,''
  arXiv:hep-th/0410178.


\bibitem{nekrasov2}  N. Nekrasov, A. Okounkov, Seiberg-Witten Theory and Random Partitions, hep-th/0306238.

\bibitem{ORV} A. Okounkov, N. Reshetikhin, C. Vafa, 
 Quantum Calabi-Yau and Classical Crystals,
 arXiv:hep-th/0309208.

\bibitem{OR} A. Okounkov and N. Reshetikhin, Correlation function of Schur process with application to local geometry of a random 3-dimensional Young diagram, J. Amer. Math. Soc. 16 (2003), no. 3, 581Ð603, arXiv:math.CO/0107056.


\bibitem{Okounkov2} A. Okounkov, Asymptotics of Plancherel measures for symmetric groups, Journal of American Mathematical Society, Vol 13, 3, 481-515.

\bibitem{Okounkov2bis} A. Okounkov,  Random surfaces enumerating algebraic curves, hep-th 0412008.

\bibitem{Okounkov3} A. Okounkov and R. Pandharipande, Gromov-Witten theory, Hurwitz 
numbers, and matrix models, I, math.AG/0101147. 


\bibitem{spohn} M. Pr\"ahofer and H. Spohn, Universal Distributions for Growth Processes 
in 1+1 Dimensions and Random Matrices, Phys. Rev. Lett. 84, 4882 
(2000). 


\bibitem{Robinson} G. de B. Robinson, "On representations of the symmetric group," Amer. J. Math. 60 (1938), 745Ð760.

\bibitem{ssuites1} C. Schensted, Longest increasing and decreasing subsequences, Canad. J. Math., 13, 1961,  179-191. MR 22:12047 

\bibitem{Sulk} P. Sulkowski, ``Matrix models for 2* theories'', arXiv:0904.3064.


\bibitem{Takasaki} K. Takasaki, 
Integrable structure of melting crystal model  with two q-parameters, arxiv:0903.2607 [math-ph].


\bibitem{Planchrel1} A. Vershik and S. Kerov, Asymptotics of the Plancherel measure of the symmetric group and  the limit form of Young tableaux, Soviet Math. Dokl., 18, 1977, 527Ð531. 

\bibitem{vershik} A. Vershik, Statistical mechanics of combinatorial partitions and their 
limit configurations, Func. Anal. Appl., 30, no. 2, 1996, 90Ð105.


\bibitem{TW} C. A. Tracy and H. Widom, Level-spacing distributions and the Airy 
kernel, Commun. Math. Phys., 159, 1994, 151Ð174.

\bibitem{TWP} C. Tracy and H. Widom, The Pearcey Process, math.PR/0412005. 

\bibitem{PZJ} P. Zinn-Justin, Six-vertex model with domain wal l boundary conditions and one-matrix model, Phys. Rev. E 62 (2000), 
no. 3, part A, 3411Ð3418, arXiv:math- ph/0005008 .



\end{thebibliography}
\end{document}